\documentclass{aastex61}

\received{\today}
\revised{}
\accepted{}
\submitjournal{ApJ}

\shorttitle{Smoke and Mirrors}
\shortauthors{Hakkila et al.}

\begin{document}

\title{Smoke and Mirrors: Signal-to-Noise and Time-Reversed Structures \\ in Gamma-Ray Burst Pulse Light Curves}

\correspondingauthor{Jon Hakkila}
\email{hakkilaj@cofc.edu}

\author{Jon Hakkila}
\affiliation{The Graduate School, University of Charleston, SC at the College of Charleston, 66 George St., Charleston, SC 29424-0001, USA}
\affiliation{Department of Physics and Astronomy, College of Charleston, 66 George St. Charleston, SC 29424-0001, USA}

\author{Stephen Lesage}
\affiliation{Department of Physics and Astronomy, College of Charleston, 66 George St. Charleston, SC 29424-0001, USA}

\author{Stanley McAfee}
\affiliation{Department of Physics and Astronomy, College of Charleston, 66 George St. Charleston, SC 29424-0001, USA}

\author{Eric Hofesmann}
\affiliation{Department of Physics and Astronomy, College of Charleston, 66 George St. Charleston, SC 29424-0001, USA}

\author{Corinne Maly Taylor}
\affiliation{Department of Physics and Astronomy, College of Charleston, 66 George St. Charleston, SC 29424-0001, USA}
\affiliation{Department of Physics and Astronomy, Clemson University, Clemson, SC 29631, USA}

\author{Thomas Cannon}
\affiliation{Department of Physics and Astronomy, College of Charleston, 66 George St. Charleston, SC 29424-0001, USA}



\begin{abstract}

We demonstrate that the `smoke' of limited instrumental sensitivity smears
out structure in gamma-ray burst (GRB) pulse light curves,
giving each a triple-peaked appearance at moderate signal-to-noise
and a simple monotonic appearance at low signal-to-noise.
We minimize this effect by studying six very bright GRB pulses (signal-to-noise generally $> 100$),
discovering surprisingly that each exhibits {\em complex time-reversible} 
wavelike residual structures.
These `mirrored' wavelike structures can have large amplitudes,
occur on short timescales, begin/end
long before/after the onset of the monotonic pulse component,
and have pulse spectra that generally evolve hard to soft,
re-hardening at the time of each structural peak. Among 
other insights, these observations help explain the existence of negative 
pulse spectral lags, and allow us to conclude that GRB pulses are 
less common, more complex, and have longer durations
than previously thought.
Because structured emission mechanisms that can operate forwards
and backwards in time seem unlikely, we look to {\em kinematic}
behaviors to explain the time-reversed light curve structures. We
conclude that each GRB pulse involves a single
impactor interacting with an independent medium. Either the material is distributed
in a bilaterally symmetric fashion, the impactor is structured in a
bilaterally symmetric fashion, or the impactor's motion is
reversed such that it returns along its original path of motion.
The wavelike structure of the time-reversible component
suggests that radiation is being both
produced and absorbed/deflected dramatically, repeatedly, and 
abruptly from the monotonic component.


\end{abstract}




\section{Introduction} \label{sec:intro}

It has become apparent that pulses are the basic radiative units of gamma-ray burst (GRB) prompt emission
({\em e.g.,} see \cite{hak11,hak15,hak18} and references contained therein), containing key information about the physical mechanisms of GRBs
and the environments in which they occur. 
However, GRB pulses do not conform to the standard definition of ``single vibrations or short bursts of sound, electric current, light, or other waves"  (Oxford English Dictionary).

Instead, a variety of observed properties characterize GRB pulses, including durations ranging from milliseconds to hundreds of seconds \citep{nor96}, longer decay times than rise times (pulse asymmetry) \citep{nor96}, longer durations at lower energies than at higher energies \citep{ric96}, durations that anti-correlate with their peak fluxes \citep{hak11}, fluences that correlate with their durations \citep{hak11}, emissions that start nearly simultaneously in different energy bands \citep{hak09}, hard-to-soft spectral evolution \citep{nor05, hak15}, more pronounced hard-to-soft evolution for hard asymmetric pulses than for soft symmetric ones \citep{hak15,hak18}, similar correlated behaviors regardless of the burst class to which they belong (long, intermediate, or short; \cite{hak11, hak15, hak18}), and more symmetric, longer duration light curves at lower energies than at higher energies \citep{hak15}. Additionally, the interpulse separations in short GRBs increase with the durations of both the initial and subsequent pulses \citep{hak18}. This effect has not been apparent in long GRB pulses \citep{rrm00}.

GRB pulses are non-monotonic in nature \citep{hak14,hak15,hak18}. They undergo
brighter intensity fluctuations than are expected from only Poisson noise, and these fluctuations 
are in phase with the underlying monotonic pulse structure. 
The fluctuations are thus not indicative of noise, nor of isolated pulses, 
but rather of pulse components.
This non-monotonicity is critically important in characterizing GRB pulses, because otherwise 
every observed non-stochastic fluctuation in a pulse light curve can be interpreted as being
a separate pulse, leading to improper conclusions about the number of pulses,
pulse characteristics, and inter-pulse separations.
The most basic non-monotonic pulse fluctuations are smoothly-varying, triple-peaked structures 
that can be viewed by fitting a GRB pulse light curve with a monotonic function, subtracting the
fit from the data, and examining the remaining pulse residuals. 
Although pulses generally exhibit hard-to-soft evolution, those with noticeable triple-peaked
structures tend to undergo spectral re-hardening just before or at the time of a peak,
and this can give some pulses (particularly those that are softer and more symmetric)
the appearance of having intensity-tracking spectral evolution. Although some pulses 
exhibit smooth monotonic or triple-peaked structures, others exhibit more
complex, chaotic structures that are often particularly noticeable at higher energies.
Pulses observed in noisy data show little structure and are best described as monotonic.

We use the word {\em structure} to describe variations
exceeding those commonly expected from 
a monotonic GRB pulse light curve.
This structure definition refers to any intensity variation that 
results in an inadequate monotonic fit to an otherwise distinct emission episode.
As such, it is instrument-dependent, because bright GRB pulses often exhibit structure whereas faint pulses rarely do.
We have recently developed a classification scheme 
for Short GRB pulses \citep{hak18} which demonstrates
that the amount of structure in GRB pulses increases
as the signal-to-noise ($S/N$) increases, and as the 
temporal bin size decreases. In other words, structure
seems to be an inherent feature of GRB pulses that
is not always apparent in data because it
can get washed out by instrumental effects.

Structure is less pronounced in faint GRB pulses than in bright ones \citep{hak14,hak15,hak18},
and $S/N$ plays an important role in this.
Signal-to-noise can decrease for a variety of instrumental reasons, including
inefficient photon detection,
small detector surface area, decreased temporal bin size, decreased spectral range,
increased spectral resolution, and detection at lower (noisier) energies.
It seems intuitively obvious that structure and noise should become indistinguishable
from one another when they have comparable amplitudes.  
However, structure and an underlying smooth pulse profile are both present in
light curves of GRB pulses. As signal-to-noise decreases, does structure disappear before or after the smoothly-varying
remainder of the pulse disappears? How does this affect the properties of pulses measured
at low signal-to-noise relative to those at high signal-to-noise? Is it possible that other pulse properties 
might accompany these changes at low signal-to-noise ({\em e.g.,} duration, 
asymmetry, and spectral evolution), and what might such changes, if observed, tell us about 
the influence of instrumental biases on GRB pulse measurements?

We wish to explore these questions further
using pulses from bright Long GRBs observed by
BATSE (the Burst And Transient Source Experiment; \cite{hor91})
on NASA's Compton Gamma Ray Observatory (CGRO).
BATSE had one of the largest surface areas of any
GRB experiment and thus was able to observe many
GRBs at large $S/N$ ratios.

\section{Empirical Pulse and Residual Model} \label{sec:pulse}

\subsection{The Pulse Model}\label{sec:pulsemodel}

In order to explore structure as a function of $S/N$, we must first 
define and characterize structureless pulses.
We do this by using the monotonic asymmetric intensity function of \cite{nor05}:

\begin{equation}\label{eqn:function} 
I(t) = A \lambda e^{[-\tau_1/(t - t_s) - (t - t_s)/\tau_2]},
\end{equation}
where $t$ is time since trigger, $A$ is the pulse amplitude, $t_s$ is the pulse start time, $\tau_1$ is the pulse rise parameter, $\tau_2$ is the pulse decay parameter, and the normalization constant $\lambda$ is $\lambda = \exp{[2 (\tau_1/\tau_2)^{1/2}]}$. 
The intensity is the count rate obtained from the total counts observed in BATSE energy channels 1 ($20 - 50$ keV),
2 ($50 - 100$ keV), 3 ($100 - 300$ keV), and 4 (300 keV $- 1$ MeV) at 64 ms temporal resolution.
Poisson statistics and a two-parameter background counts model of the form $B=B_0+BS \times t$ are assumed (where $B$ is the background counts in each bin and $B_0$ and $BS$ are constants denoting the mean background (counts) and the rate of change of the mean background (counts/s)). This model can be used to produce observable pulse parameters such as pulse peak time ($\tau_{\rm peak} = t_s + \sqrt{\tau_1 \tau_2}$), pulse duration ($w$) and pulse asymmetry ($\kappa$). As a result of the smooth rapid rise and more gradual fall of the pulse model, $w$ and $\kappa$ are measured relative to some fraction of the peak intensity. Following \citep{hak11}, we measure $w$ and $\kappa$ at $I_{\rm meas}/I_{\rm peak}=e^{-3}$ (corresponding to $0.05 I_{\rm peak}$), or

\begin{equation}\label{eqn:duration}
w = \tau_2 [9 + 12\mu]^{1/2},
\end{equation}
where $\mu = \sqrt{\tau_1/\tau_2}$, and

\begin{equation}\label{eqn:asymmetry}
\kappa \equiv [1 + 4\mu/3]^{-1/2}.
\end{equation}
Asymmetries range from symmetric ($\kappa=0$ when $\tau_1 \gg \tau_2$ and $\mu \rightarrow \infty$)
to asymmetric ($\kappa=1$ when $\tau_1 \ll \tau_2$ and $\mu \rightarrow 0$) with longer decay than rise times.

\subsection{The Residual Model}\label{sec:resmodel}

Residuals to the \cite{nor05} model are generated by obtaining the best monotonic pulse fit
for the observed pulse light curve and subtracting this fit from the data. 
Small yet distinct deviations in the residuals are found to be systematically
in phase with the {\bf monotonic} light curve \citep{hak14}, and these 
deviations are needed to accurately describe GRB pulse shapes. 
Although the deviations are closely aligned with the pulse duration, they are not always contained within it.
Thus we have defined a larger {\em fiducial time interval} $w_{\rm f}$ as
\begin{equation}\label{eqn:res}
w_{\rm f} = t_{\rm end} - t_{\rm start} = 4.4 \tau_2 (\sqrt{1+\mu/2}+1) + \sqrt{\tau_1 \tau_2},
\end{equation}
with the fiducial end time $t_{\rm end}$ given by 
\begin{equation}\label{eqn:tend}
t_{\rm end} = \frac{w}{2} (1+\kappa)+t_s+\tau_{\rm peak}
\end{equation}
and the fiducial start time $t_{\rm start}$ given by
\begin{equation}\label{eqn:tstart}
t_{\rm start} = t_s - 0.1 [\frac{w}{2} (1+\kappa) - \tau_{\rm peak} ].
\end{equation}
The fiducial time interval is a unitless interval introduced
to demonstrate the generic nature of GRB pulse residual structures, regardless
of the durations and asymmetries of the underlying pulses containing them. Residuals
are obtained using this time interval so that normalized pulse residual structures can be directly
compared with one another, and so that goodness-of-fit measurements
used in making these comparisons are not based on the amount of background sampled.
The residuals exhibit variations that can be fitted with an empirical function \citep{hak14}
in the fiducial time interval {\em f} where $0 \le t' \le 1$:
\begin{equation}\label{eqn:residual}
\textrm{res($t'$)} = \left\{ \begin{array}{ll}
a  J_0( \sqrt{\Omega_{\rm f} [t_{\rm 0; f} - t' - 0.005]}) & \textrm{if $t' < t_{\rm 0; f}  - 0.005$} \\
a & \textrm{if $t_{\rm 0; f}  - 0.005 \le t' \le t_{\rm 0; f}  + 0.005$} \\
a  J_0(\sqrt{s_{\rm f} \Omega_{\rm f} [t' - t_{\rm 0; f}  - 0.005]}) & \textrm{if $t' > t_{\rm 0; f}  + 0.005$.}
\end{array} \right. 
\end{equation}
Here, $J_0(x)$ is an integer Bessel function of the first kind, $t_{\rm 0; f}$ is the unitless time of the residual peak, $a_{\rm f}$ is the amplitude of the residual peak, $\Omega_{\rm f}$ is the unitless Bessel function's angular frequency that defines the timescales of the residual wave (a large $\Omega_{\rm f}$ corresponds to a rapid rise and fall), and $s_{\rm f}$ is a dimensionless scaling factor that relates the fraction of time which the function pre-$t_{\rm 0; f}$ has been compressed relative to its time-inverted form post-$t_{\rm 0; f}$. The time during which the pulse intensity is a maximum is required to be a plateau instead of a peak because otherwise the width of the Bessel function's first lobe would not be broad enough to capture the extent of the central peak. Therefore, a duration of $w_{\rm plateau} \approx 0.010 w_{\rm fid}$ is hard-coded into the fitting routine to create a small plateau structure in the function. Without evidence in the pulse shape that the Bessel function continued beyond the third zero (following the second half-wave), \cite{hak14} truncated the function at the third zero $J_0 (x = \pm 8.654)$, giving the fitting function a triple-peaked appearance. Because the central plateau most commonly aligns with the peak of the monotonic pulse, the three residual function peaks were named the precursor peak, the central peak, and the decay peak, respectively \citep{hak14}. 

In some pulses the central peak of the residual function does not closely align with the peak of the monotonic pulse (see \cite{hak18} and Section \ref{sec:hr_evol} of this manuscript). The reason for this misalignment is still not well-understood but it appears to signify that the residual function and the monotonic function are separate pulse components.

After fitting, the unitless fiducial values are converted back to the observed time interval using:
\begin{equation}
a=a_{\rm f}
\end{equation}
\begin{equation}
s=s_{\rm f},
\end{equation}
\begin{equation}
t_0=t_{\rm 0; f} (t_{\rm end}-t_{\rm start}) +t_{\rm start},
\end{equation}
and
\begin{equation}
\Omega=\Omega_{\rm f}/(t_{\rm end}-t_{\rm start}),
\end{equation}
where $t_{\rm start}$ and $t_{\rm end}$ are the real time values corresponding to the start and end of the fiducial time interval.
The Bessel function frequency $\Omega$ is now given in units of $s^{-1}$ and the residual peak time $t_0$ is now given in units of $s$.

The flux contribution that the residual function makes to the combined pulse fit, as well as its relative temporal alignment, is capable of significantly altering a pulse shape from that of monotonicity. Although the precursor peak and the decay peak of the residual function typically fall during the rise and decay portions of the monotonic fitting function, in some cases the precursor peak can occur before the formal rise of the monotonic pulse. When the precursor peak is harder than the central peak, the hard-to-soft pulse evolution is pronounced. When the precursor peak is softer than the central peak, the pulse begins evolving hard-to-soft but then re-hardens to a maximum value around the time of the central peak, and the pulse hardness appears to track with intensity. Additionally, the central residual peak and the peak of the monotonic pulse do not always align with one another, producing an extra smooth pulse structure in some cases.


The 64 ms $S/N$ is a measure of the peak brightness of each GRB,
relative to its background count measured with 64 ms temporal resolution. The $S/N$ is
\begin{equation}\label{eqn:SN}
S/N = (P_{64}-B)/\sqrt P_{64},
\end{equation}
where $P_{64}$ is the 64 ms peak counts and $B$ is the mean background count.
This signal-to-noise ratio is primarily appropriate when analyzing
GRB pulses fit on the 64 ms time scale. 

Using the approach of \cite{hak18}, we classify GRB pulses based on the 
amount of structure they exhibit, as defined relative to the $\chi^2_\nu$ $p-$values
of the pulse fits. This method first fits each pulse with the monotonic \cite{nor05} fitting function and analyzes the goodness-of-fit with a $\chi^2-$test. We define $\chi^2$ over the fiducial timescale and $\nu$ from the \cite{nor05} model as the number of temporal bins minus the number of pulse- and background-fit parameters -- two for the background and four for a single pulse. The $p-$value associated with $\chi^2_\nu$ is the probability that a value of $\chi^2$ or less having $\nu$ degrees of freedom will be measured in a random distribution. We consider good fits (indicating relatively smooth light curves) to be those having standard best-fit $p-$values of $p_{\rm best} \ge 5 \times 10^{-3}$. Next, this {\bf monotonic} pulse fit is subtracted from the light curve leaving a residual light curve. The residual light curve is then fitted with the \cite{hak14} residual fitting model. Finally the residual fit is added to the monotonic pulse fit to produce a total fit, and this is tested to see whether or not adding the residual fitting function significantly improved the overall pulse fit. A $\Delta \chi^2$ test is used to indicate whether or not the residual function should be included in the fit: $\Delta \chi^2$ is the difference in $\chi^2$ obtained from the \cite{nor05} model minus that obtained from the \cite{nor05} model combined with the \cite{hak14} residual model. The difference in the number of degrees of freedom between these fits is four for a single pulse. If adding the residual fit significantly improved the pulse fit, then the combined models act as the pulse fit. But if the addition of the residual fit does not improve the overall pulse fit, then only the original monotonic pulse fit is used to represent the final pulse fit. We require a $\Delta \chi^2$ $p-$value of $p_{\Delta} \le 10^{-3}$ for the model to be considered significantly improved and therefore use the combined pulse models to represent the final pulse fit.

We use this approach to classify GRB pulses into one of four groups. {\em Simple} pulses are those best characterized by 
Equation \ref{eqn:function} alone ($p_{\rm best} \ge 5 \times 10^{-3}$). {\em Blended} pulses are fits improved using Equation \ref{eqn:residual} ($p_{\Delta} \le 10^{-3}$ and $p_{\rm best} \ge 5 \times 10^{-3}$). {\em Structured} pulses have many characteristics that can be explained by the pulse and residual models, but also have significant deviations from these structures ($p_{\Delta} \le 10^{-3}$ and $10^{-5} \le p_{\rm best} < 5 \times 10^{-3}$). {\em Complex} pulses have complicated light curves ($p_{\rm best} < 10^{-5}$). 

\section{Analysis} \label{sec:analyze}

We hypothesize that bright, complex GRB emission episodes
are really structured pulses that can be reduced to triple-peaked
and monotonic shapes with the loss of $S/N$. We test this
hypothesis by analyzing four of BATSE's brightest GRBs.
Specifically, we study BATSE triggers 143 (GRB 910503), 
249 (GRB 910601), 5614 (GRB 960924), and 7301 (GRB 990104B) \citep{pac00}.
All four bursts have been observed in BATSE energy channels $1 - 4$ (20 keV to 1 MeV)
with 64 ms temporal resolution.
We initially limited our sample to five pulses with $S/N > 100$. However,
because BATSE 7301 contains two bright pulses, we also chose to analyze the 
second pulse of BATSE 143 ($S/N=55.9$) for completeness.
As we will show in Section \ref{sec:structure}, any pulse with $S/N > 25$
can be considered to be ``bright" for the purposes of this analysis.

BATSE trigger 143 contains two distinct emission episodes
peaking at $1.7$ seconds and 48 seconds after the trigger. The brighter first event
(143p1) has a $S/N=131.5$, a duration of $w=9.9$ seconds, an asymmetry of $\kappa=0.75$,
and has considerable temporal structure, 
while the smoother and fainter second event (143p2) has $S/N=55.9$, $w=11.2$ seconds,
$\kappa=0.59$, and shows clear evidence of the triple-peaked structure.

BATSE trigger 249 is a complex, time-symmetric emission episode with $S/N=108.0$ 
peaking $21.9$ seconds after the trigger and containing at least 5 separate emission peaks. 
The entire episode lasts some 38 seconds, although the fitted pulse lasts only $w=20.8$ seconds
and has an asymmetry of $\kappa=0.07$.

BATSE trigger 5614 is a bright ($S/N=177.2$) single-peaked asymmetric emission episode
($\kappa=0.69$) 
peaking $8.4$ seconds after the trigger with a duration of $w=6.7$ seconds.
Although the burst has the appearance of a fairly smooth single pulse,
the main part of the pulse rise does not occur until roughly 7 seconds after the trigger.
Of the six pulses in the sample, this one most clearly exhibits
the triple-peaked pulse shape described in previous papers.

BATSE trigger 7301 is a complex burst with what appear to be four
separate complex emission episodes. For reasons we will discuss later, we
identify only the two brightest episodes as pulses 7301p1 and 7302p2 (these peak $15.8$ seconds 
and $165.1$ seconds after the trigger, respectively) while excluding the fainter
emission episodes starting at the trigger ($t = 0$ seconds) and at around $t = 32$ seconds.
Pulse 7301p1 is a bright, complex episode with $S/N=100.9$, $w=10.7$ seconds and $\kappa=0.51$.
Pulse 7301p2 is a bright, complex episode with $S/N=144.3$, $w=29.3$ seconds and $\kappa=0.52$;
the pulse contains three bright peaks each
lasting a few seconds which occur during fainter emission lasting roughly 50 seconds.

We initially assume that each of the bright, complex emission episodes
in BATSE triggers 143, 249, 5614, and 7301 is composed
of a single structured pulse. This is completely antithetical to the standard pulse-fitting approach in which each 
pulse would be treated as a light-curve variation that could not be attributed to noise. 
Using such an approach, 143p1, 249, 7301p1, and 7301p2
would be best fit by dozens of pulses. Even the apparently smoothly-varying
143p2 and 5614 would consist of at least half a dozen pulses because 
the burst is so structured at this $S/N$ level.

\begin{figure}
\plottwo{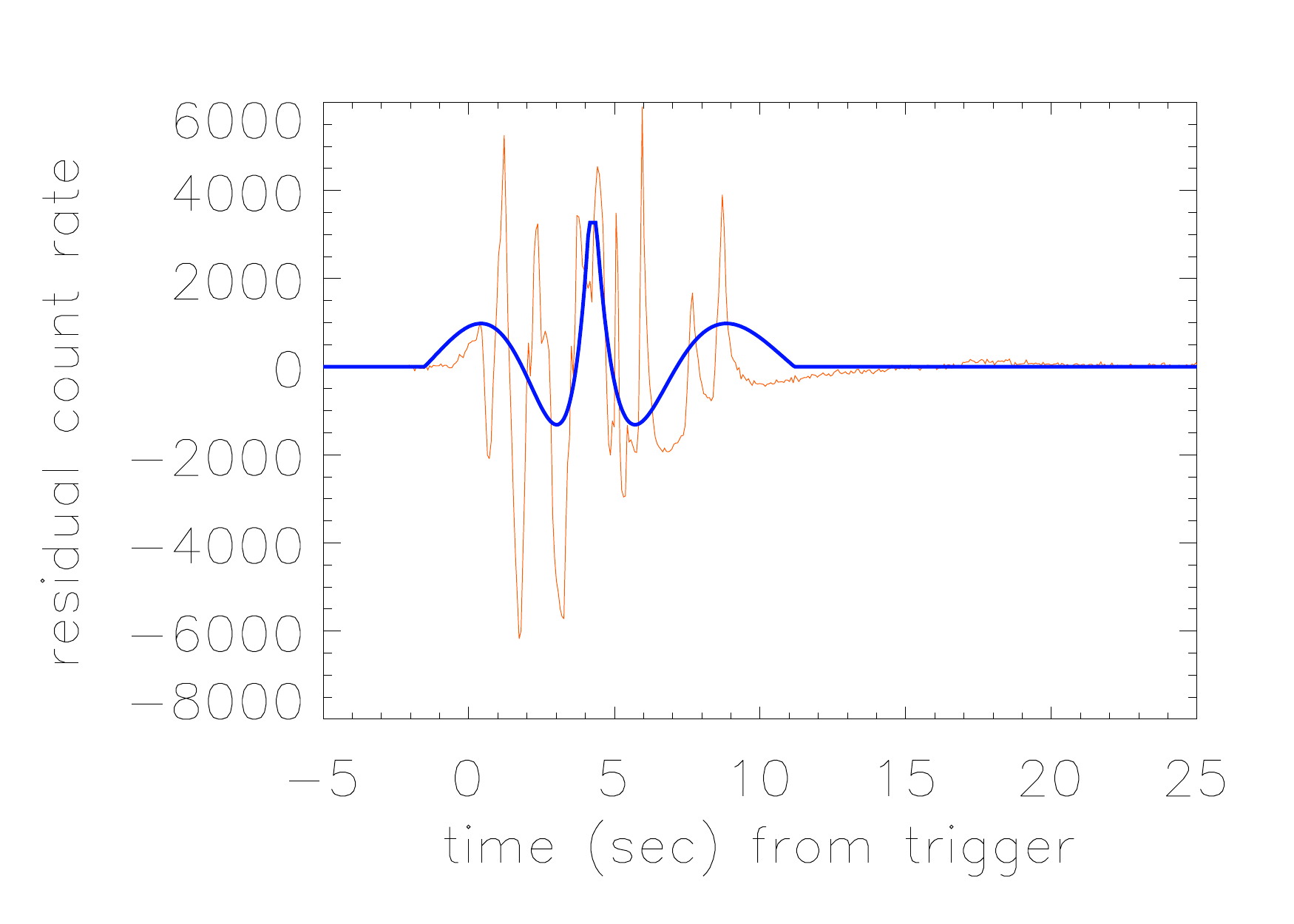}{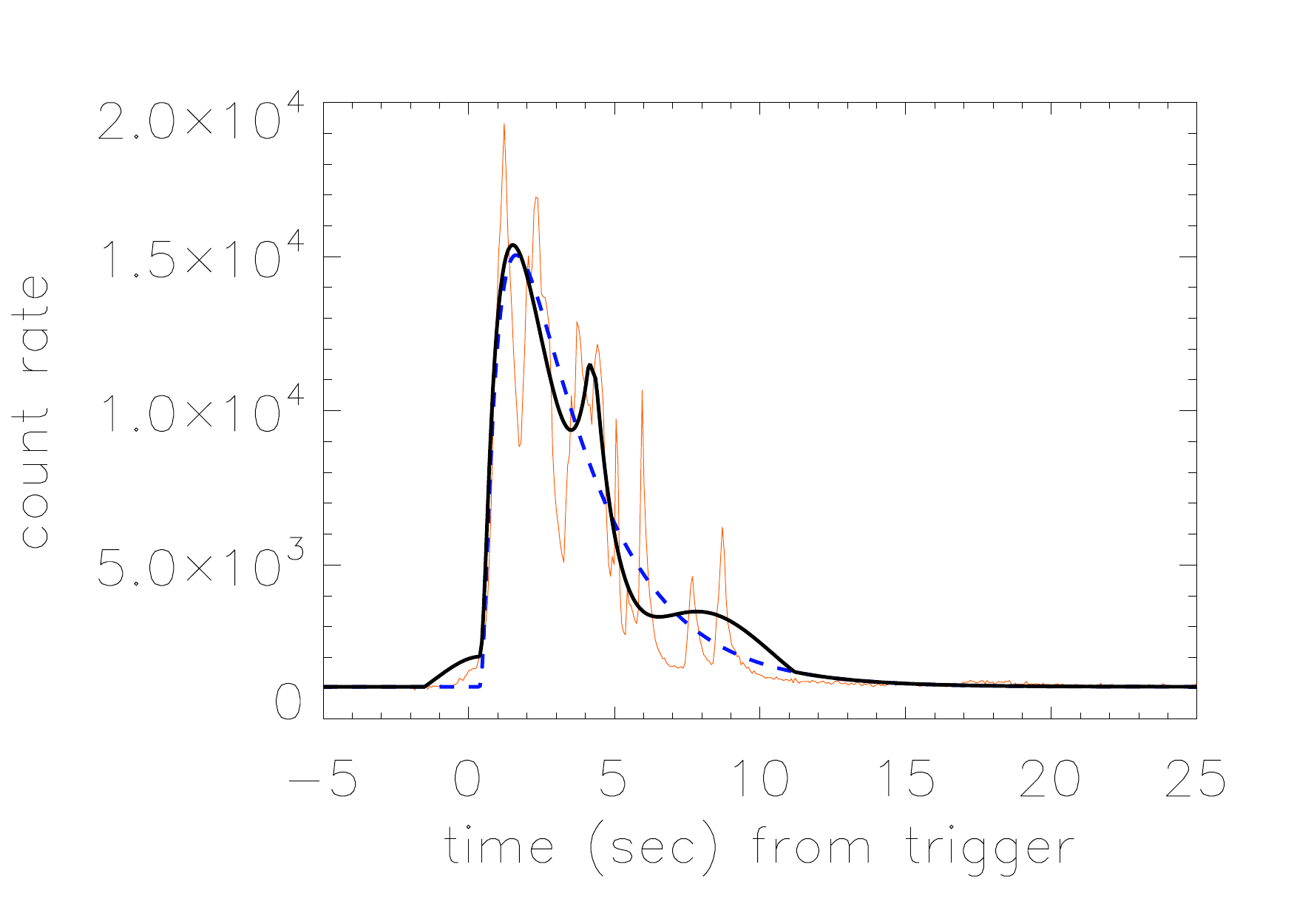}
\caption{BATSE trigger 143p1 observed at $S/N = 131.5$. The left panel shows the \cite{hak14} residual structure, while the right panel shows both the \cite{nor05} pulse fit (dotted line) and the combined \cite{nor05} pulse plus \cite{hak14} residual fit (solid line). In this and all subsequent plots, `residual count rate' and `count rate' refer to the number of counts per 64 ms bin.\label{fig:143}}\end{figure}

\begin{figure}
\plottwo{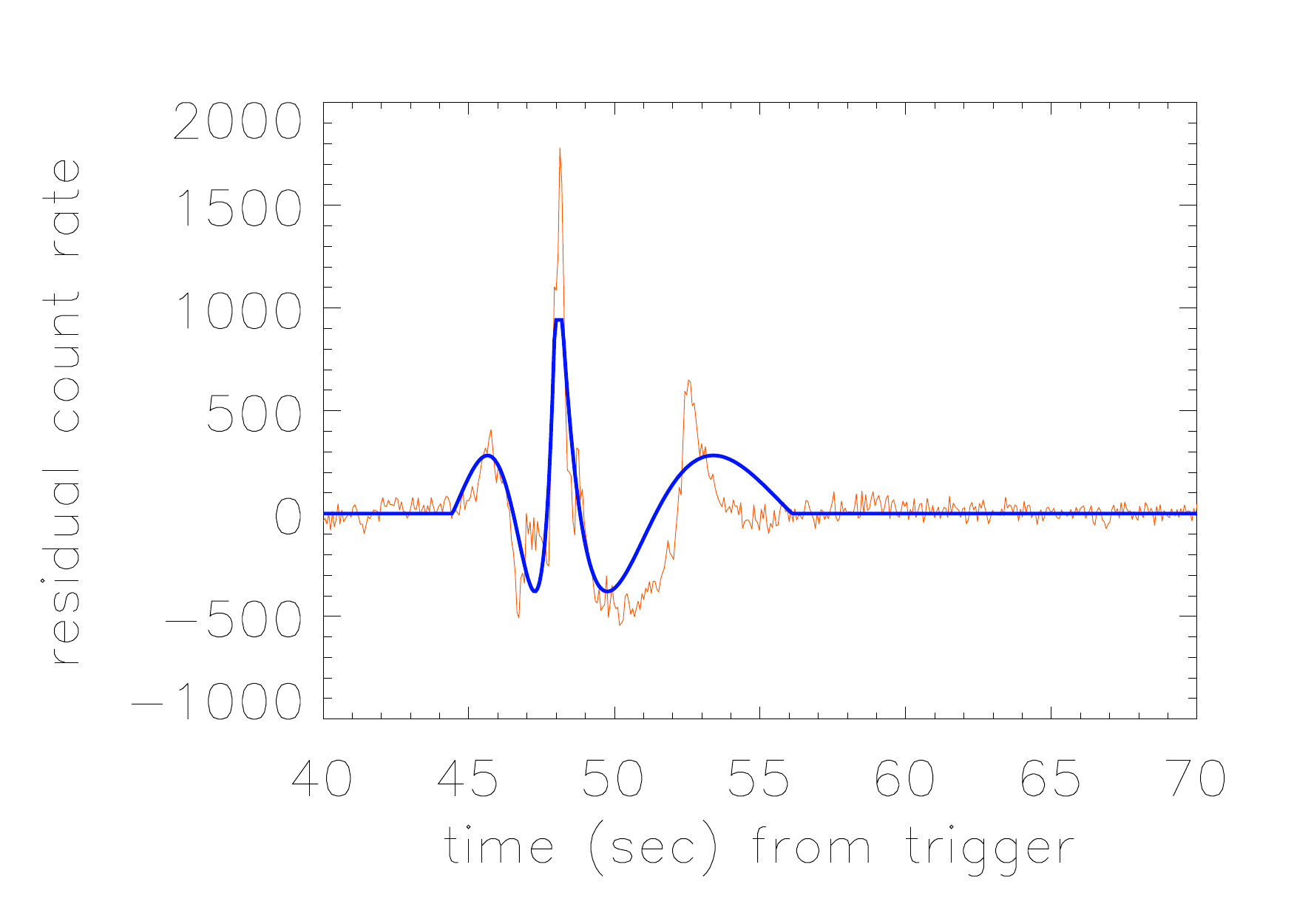}{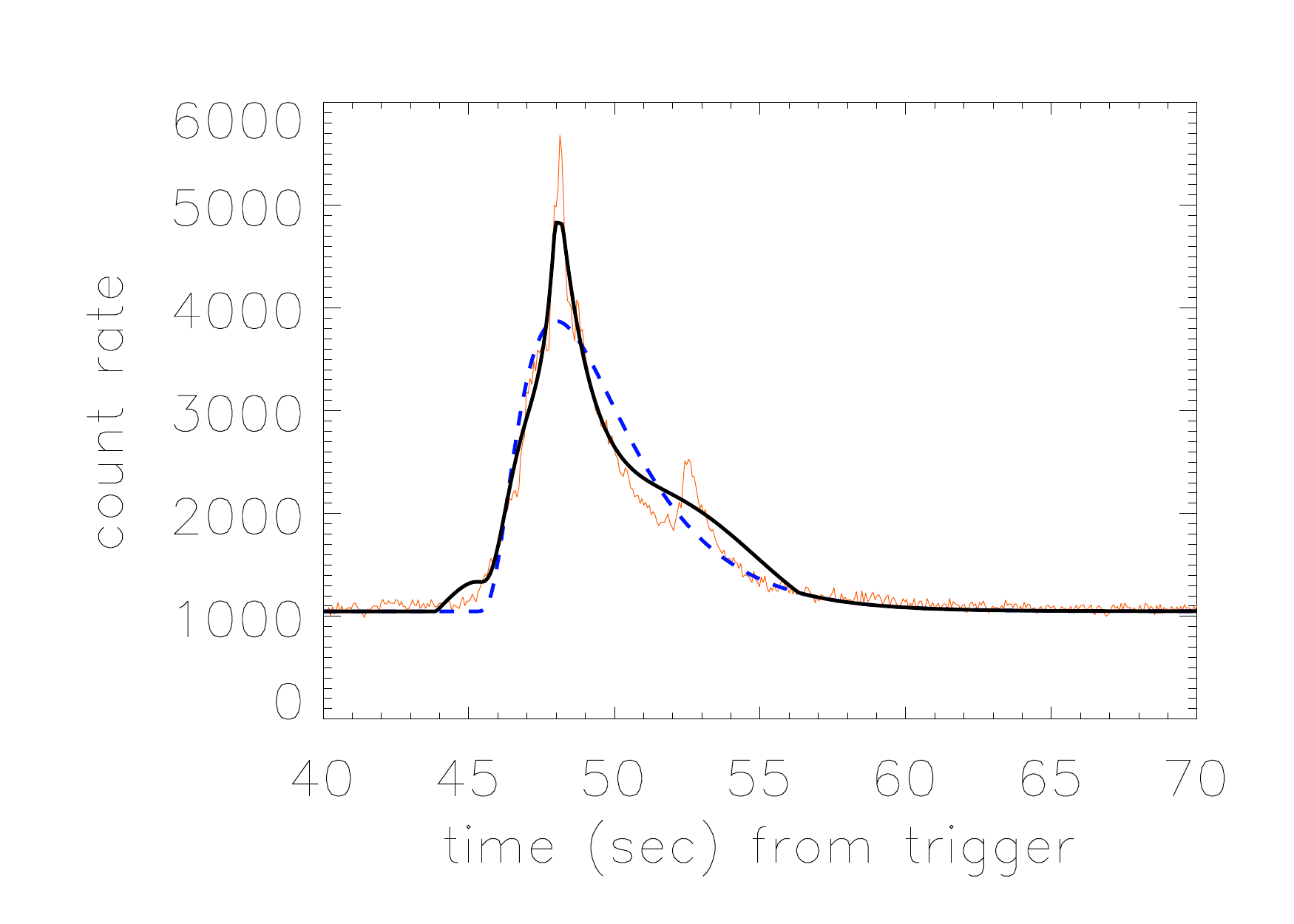}
\caption{BATSE trigger 143p2 observed at $S/N = 55.9$.  The left panel shows the \cite{hak14} residual structure, while the right panel shows both the \cite{nor05} pulse fit (dotted line) and the combined \cite{nor05} pulse plus \cite{hak14} residual fit (solid line). \label{fig:143p2}}\end{figure}

\begin{figure}
\plottwo{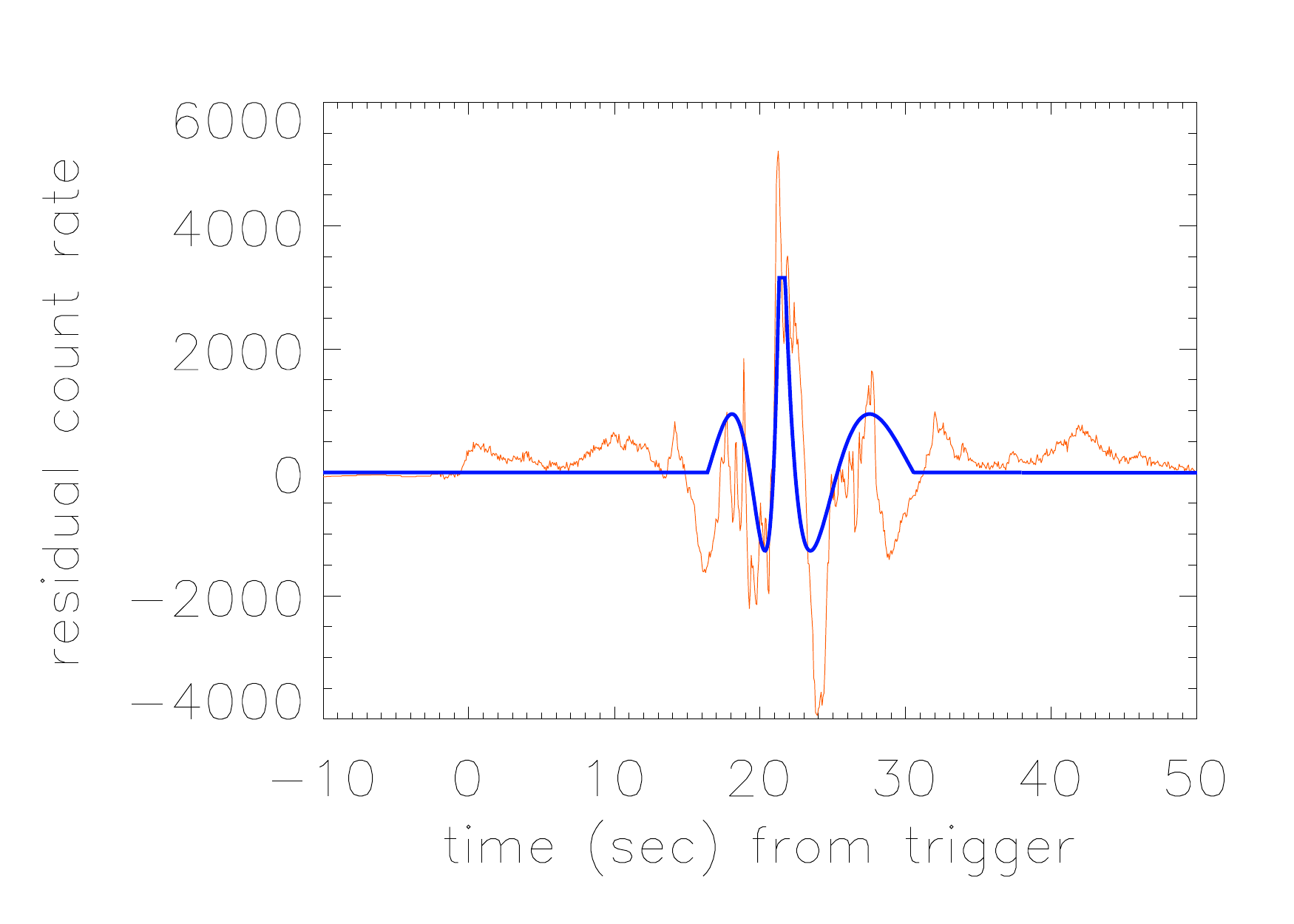}{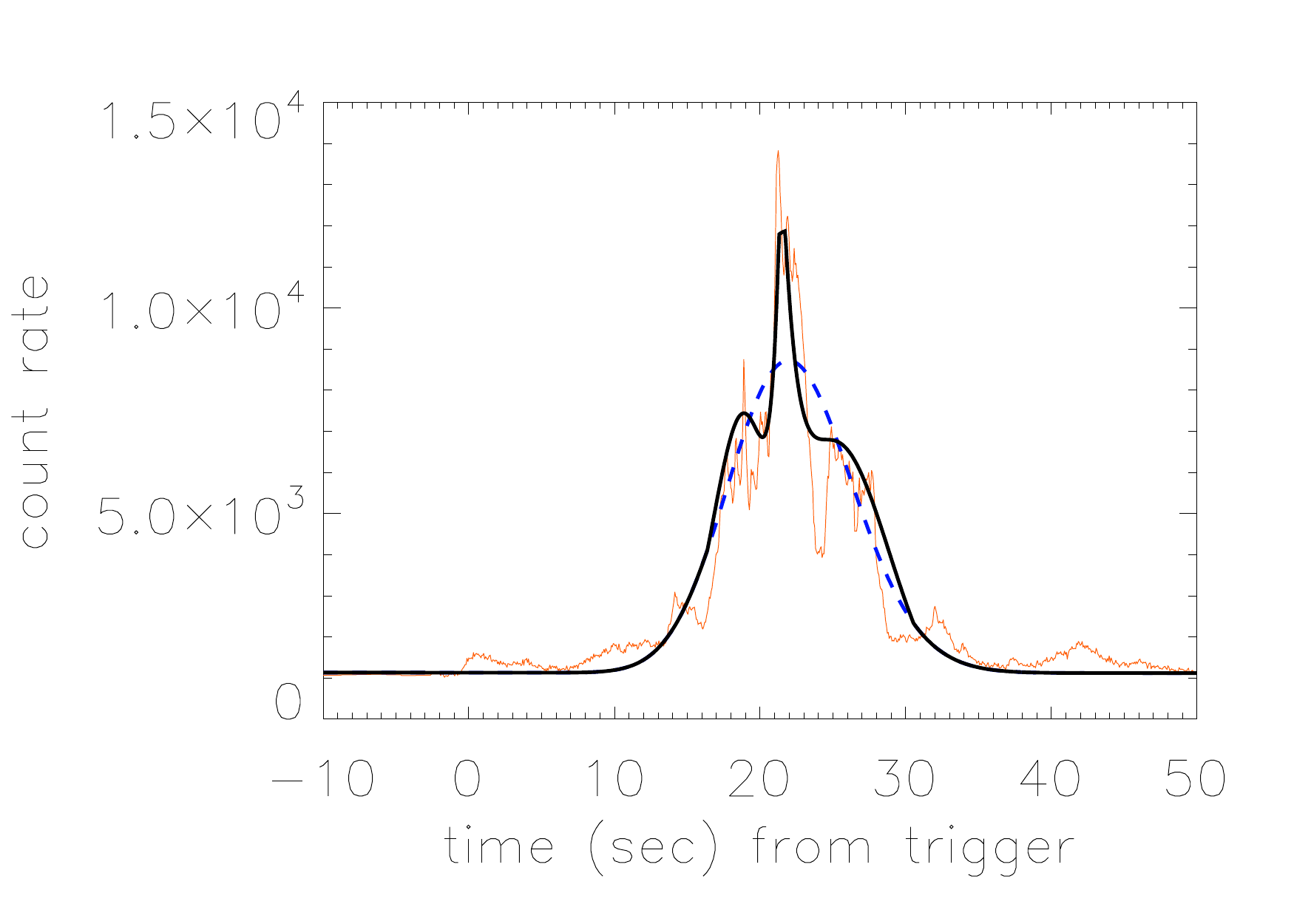}
\caption{BATSE trigger 249 observed at $S/N = 108.0$.  The left panel shows the \cite{hak14} residual structure, while the right panel shows both the \cite{nor05} pulse fit (dotted line) and the combined \cite{nor05} pulse plus \cite{hak14} residual fit (solid line).\label{fig:249}}\end{figure}

\begin{figure}
\plottwo{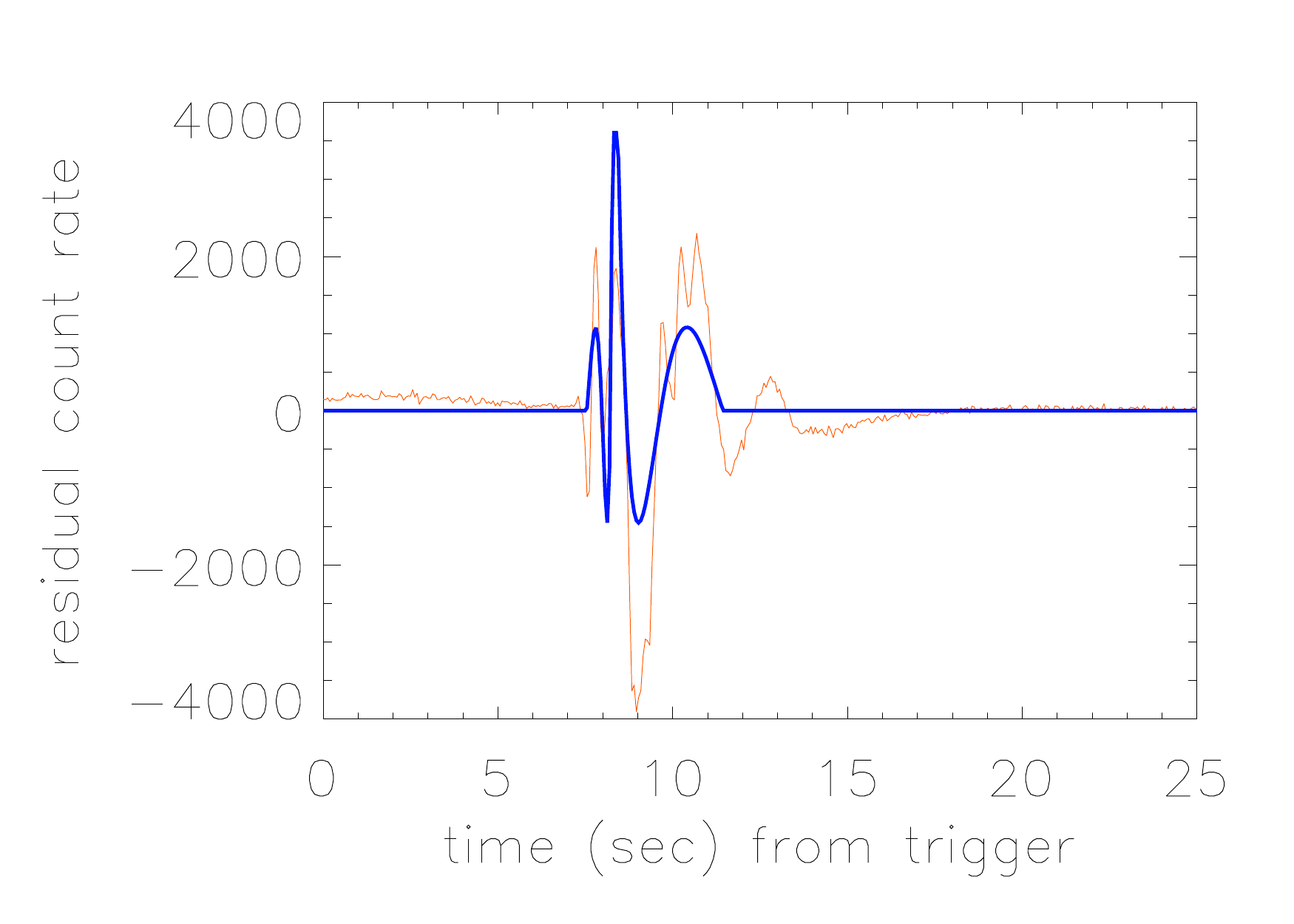}{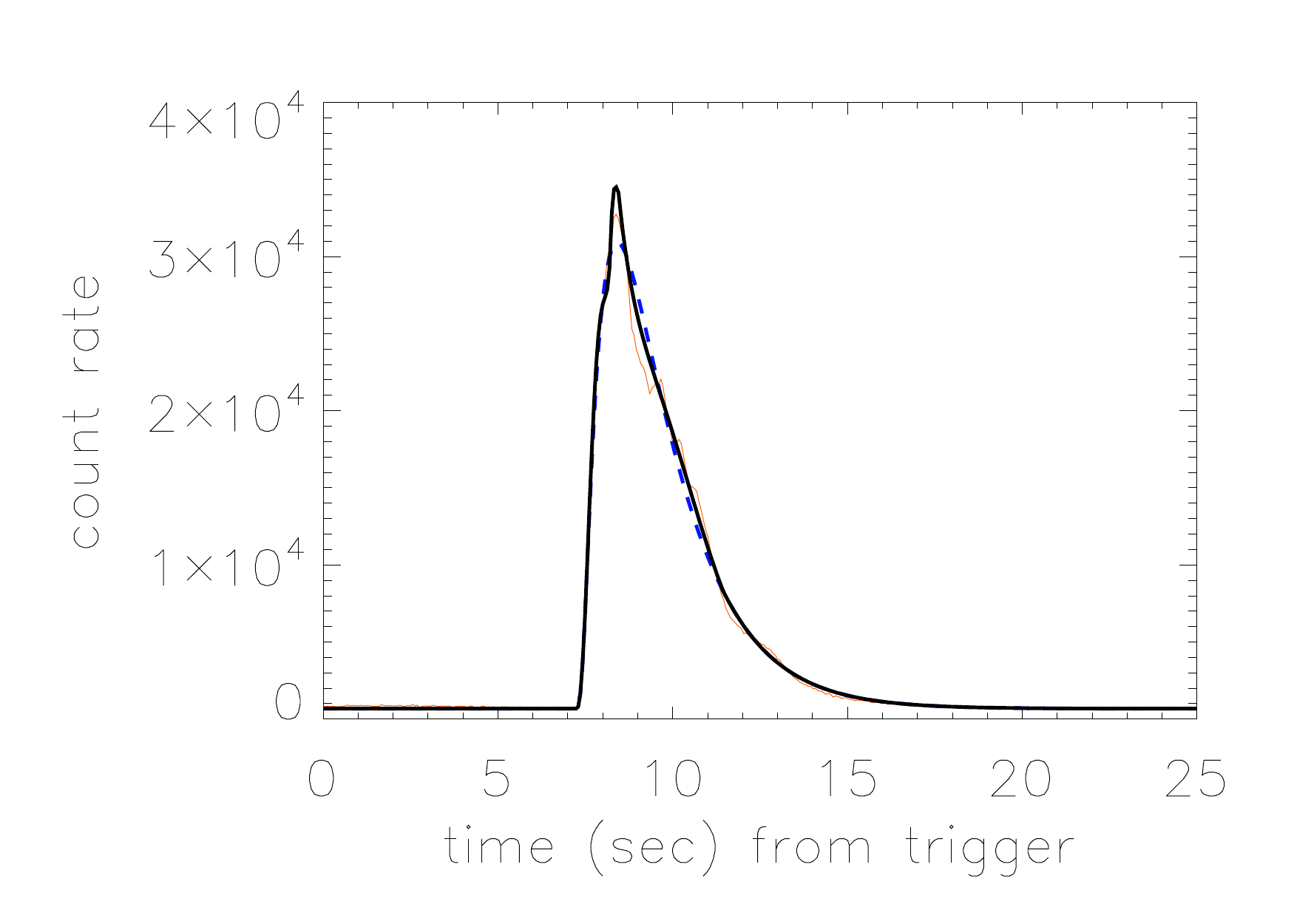}
\caption{BATSE trigger 5614 observed at $S/N = 177.2$. The left panel shows the \cite{hak14} residual structure, while the right panel shows both the \cite{nor05} pulse fit (dotted line) and the combined \cite{nor05} pulse plus \cite{hak14} residual fit (solid line).\label{fig:5614}}\end{figure}

\begin{figure}
\plottwo{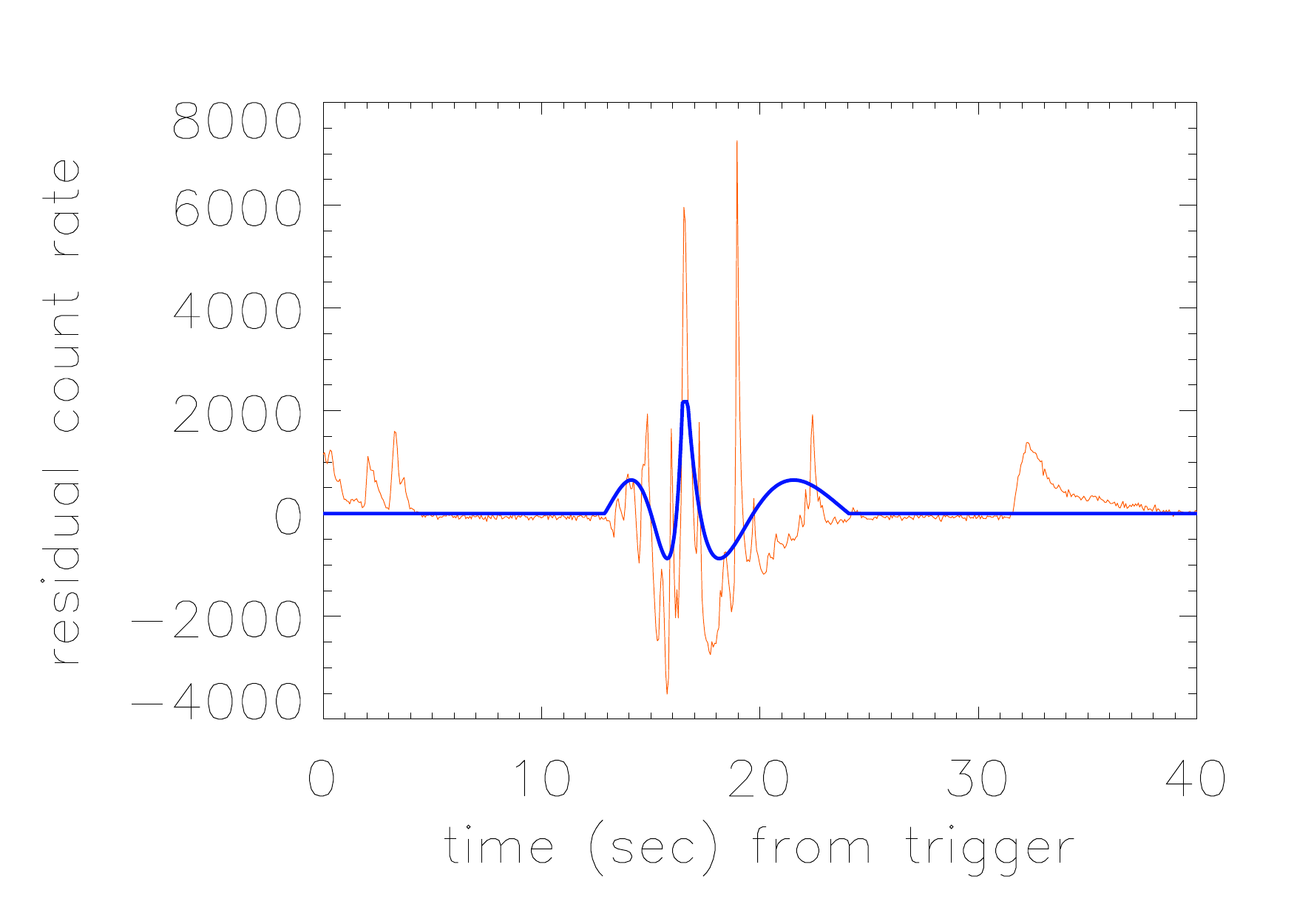}{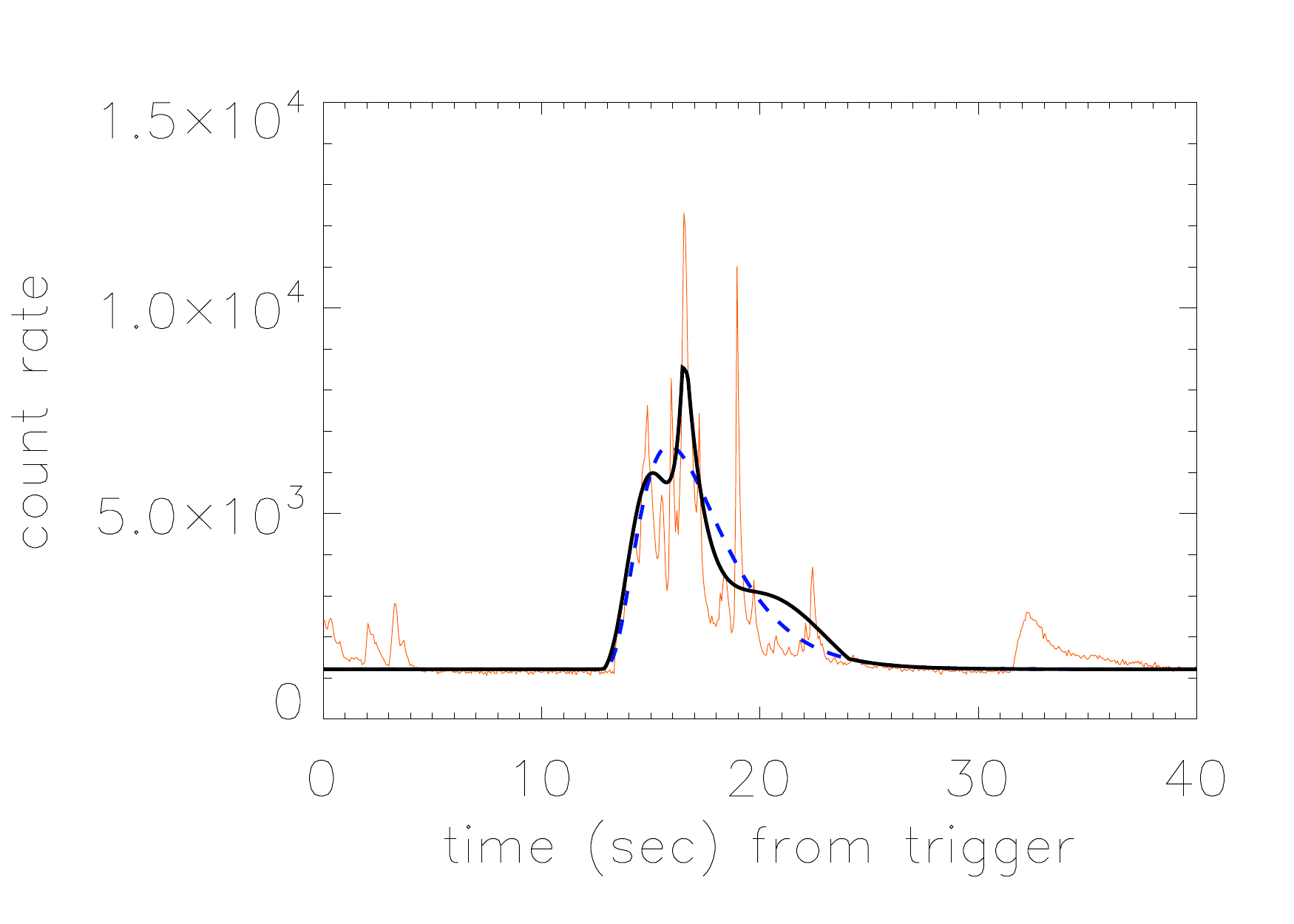}
\caption{BATSE trigger 7301p1 observed at $S/N = 100.9$.  The left panel shows the \cite{hak14} residual structure, while the right panel shows both the \cite{nor05} pulse fit (dotted line) and the combined \cite{nor05} pulse plus \cite{hak14} residual fit (solid line).\label{fig:7301}}\end{figure}

\begin{figure}
\plottwo{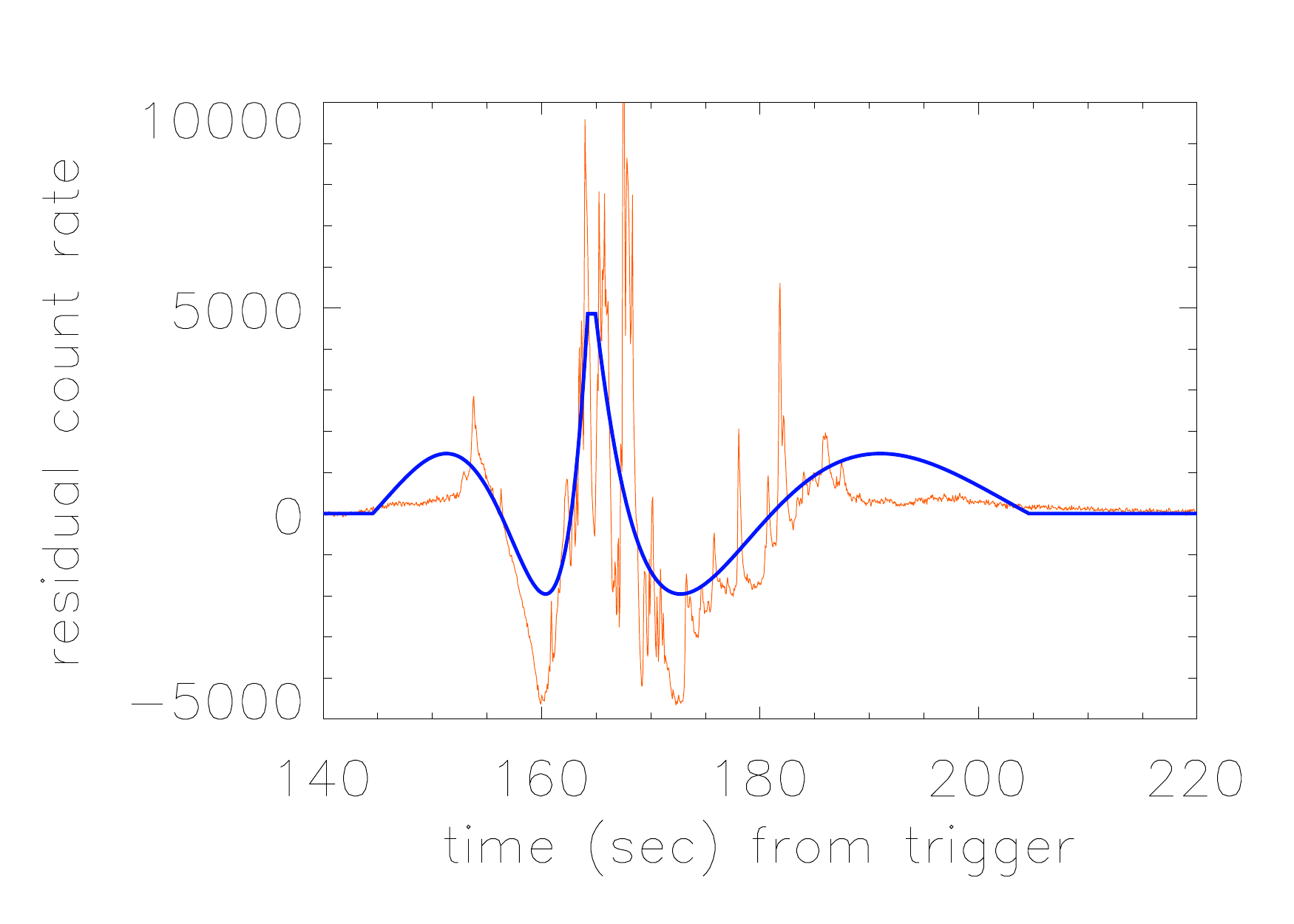}{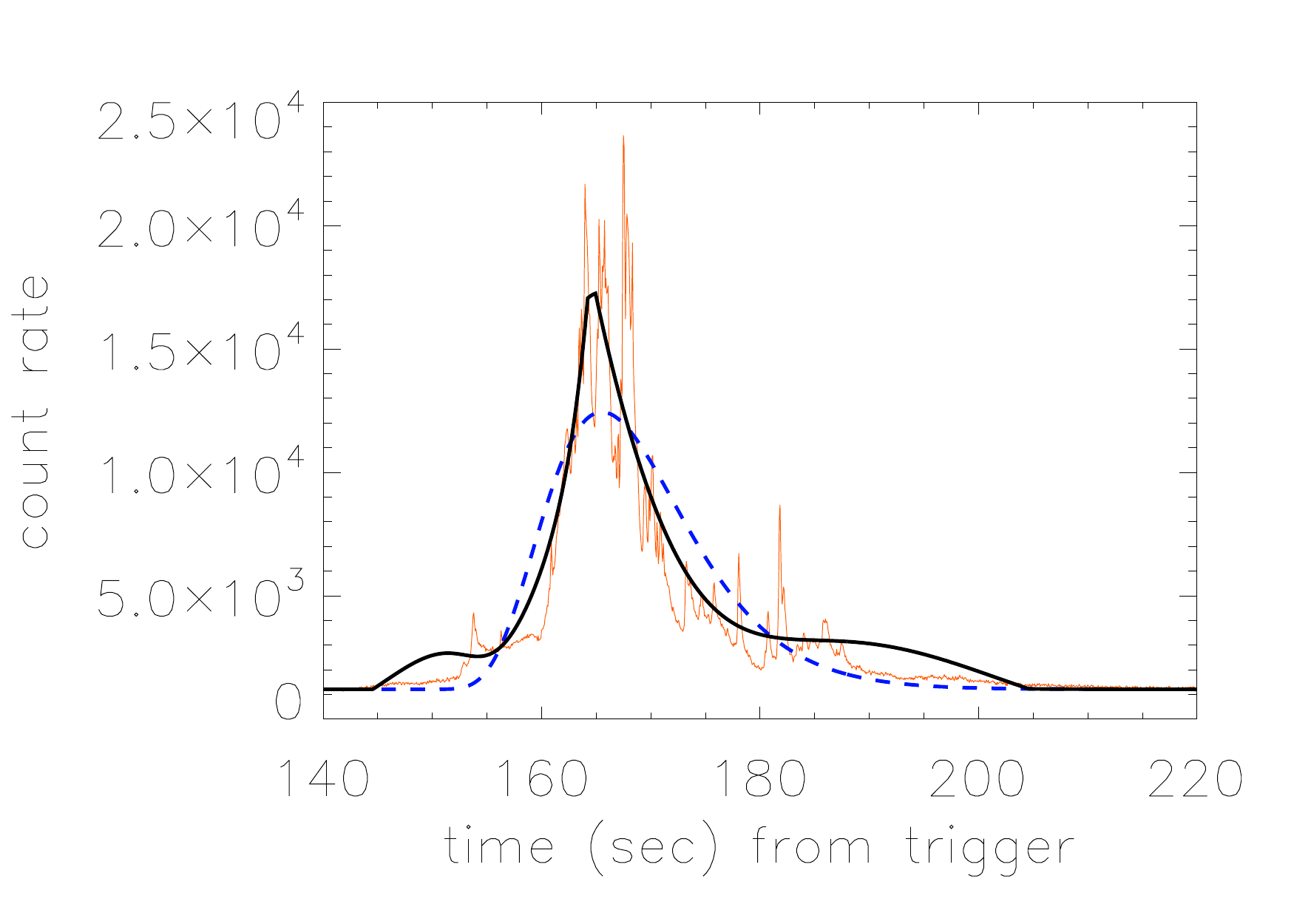}
\caption{BATSE trigger 7301p2 observed at $S/N = 144.3$.  The left panel shows the \cite{hak14} residual structure, while the right panel shows both the \cite{nor05} pulse fit (dotted line) and the combined \cite{nor05} pulse plus \cite{hak14} residual fit (solid line).\label{fig:7301p2}}\end{figure}

Fits to the six pulses using the monotonic model described by Equation \ref{eqn:function} and the residual model described by Equation \ref{eqn:residual} are shown in Figures \ref{fig:143} $-$ \ref{fig:7301p2}.  The left panels of Figures \ref{fig:143} $-$ \ref{fig:7301p2} show the residual structure while the right panels show both the monotonic fit (dotted line) and the combined fit (solid line). Each pulse is classified as complex using the formal criteria described in Section \ref{sec:resmodel}, indicating that the combination of the monotonic pulse model and the residual model do not adequately describe the pulse light curves. However, in each case it can be seen that the models contribute to {\bf only} an approximate first-order fit by creating envelopes that contain the most noticeable structures, and that the residual fits are only rough approximations to the actual residual structure.  

\subsection{Smoke: Pulse Properties as Functions of Signal-to-Noise}

We estimate the properties of these six BATSE pulses at lower $S/N$ by reducing their assumed ``intrinsic" counts distributions and adding stochastic instrumental noise. The pulse plus residual fits described thus far have been based on the {\em summed} counts in BATSE's four energy channels $1+2+3+4$. Since some pertinent pulse properties are energy-dependent ({\em e.g.,} spectral lags), we use a process that produces reductions in each individual energy channel rather than only in the summed four-channel light curve.

For each of the six BATSE pulses listed, we estimate their properties at lower
$S/N$ values by removing the background from each energy channel and assuming that the remaining photon flux distribution is a fair representation of the ``intrinsic'' flux distribution. When summed together, these channel-dependent counts allow us to reproduce the ``intrinsic" summed four-channel counts observed by the instrument. We then define roughly ten equal logarithmic four-channel count intervals between the observed peak count rate and a measure close to background (ostensibly near $S/N \approx 3$). For each energy channel we iteratively decrease the peak counts proportionally through these logarithmic decrements, add energy-dependent Poisson noise to the results, and recombine the energy channel-dependent light curves to obtain a reduced-intensity four-channel light curve.

This process assumes that the reduced pulse would have been detected with a similar background rate if it had been detected at a lower $S/N$, an assumption that is only partially valid because the background rate varied slowly and regularly throughout the Compton Gamma-Ray Observatory's mission in response to the satellite's eccentric orbit. The ``noisification" process also assumes that there is no systematic coupling between detector energy channels, and that the detector response does not depend significantly on count rate. There are potential problems with both of these assumptions: the non-diagonal energy response of BATSE's Large Area Detectors (LADs) created a coupling between energy channels, and low $S/N$ observations were affected by spacecraft orientation and relative direction between the Earth and a GRB ({\em e.g.,} see \cite{pen96}). However, we believe that both assumptions are valid as first-order approximations.

\subsubsection{Pulse Structure}\label{sec:structure}

As $S/N$ decreases, pulse structure becomes less noticeable. At high $S/N$, pulse structure is easily distinguishable and separable from Poisson noise. At low $S/N$, structure becomes indistinguishable from noise, while the underlying smooth monotonic pulse remains identifiable and measurable. This becomes an issue when determining the number of pulses potentially present in a GRB light curve: at low $S/N$ the monotonic pulse dominates over structure, and the light curve appears to be that of a single pulse. At moderate and high $S/N$ these structures are clearly not the result of noise, and can even appear to be so pronounced that they can be mistaken for separate pulses. However, confusion in the number of pulses can be reduced by recognizing that the structures are linked in phase with the main body of emission, either sitting atop the monotonic pulse ({\em e.g.,} 143p1; see Figure \ref{fig:143}) or a connected to it as a continuum of episodes like cutout paper doll chains ({\em e.g.,} 249; see Figure \ref{fig:249}). The importance of this relationship will be discussed further in Section \ref{sec:hr_evol}.

\begin{figure}
\plottwo{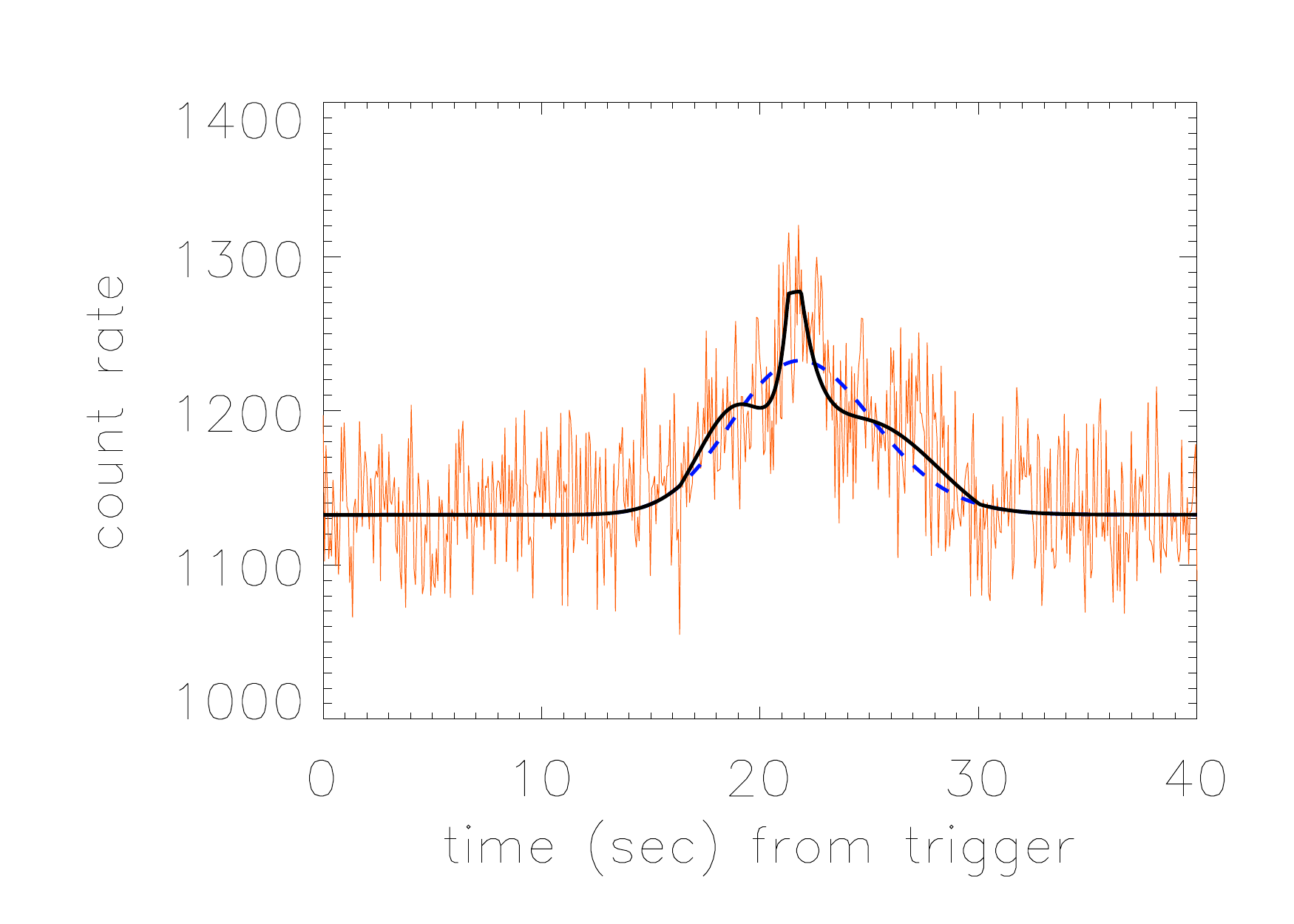}{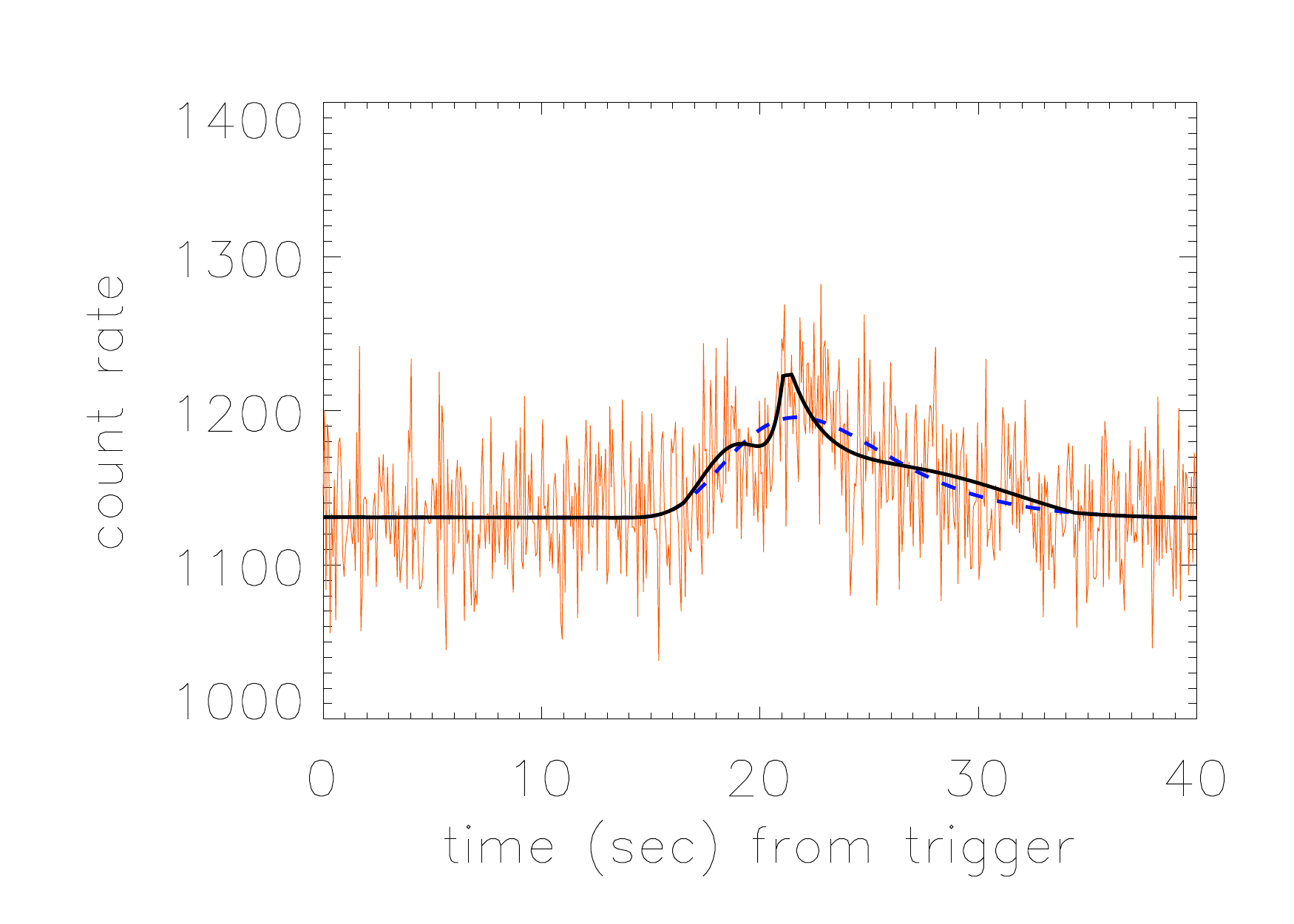}
\caption{Two stages in the intensity reduction of 249. The left panel shows what the pulse can look like at $S/N=5.2$, where the pulse and residual fits combine to place the pulse in the blended class. The right panel shows what the pulse can look like at $S/N=4.3$, where the residuals do not statistically improve the fit and the pulse is classified as a simple pulse. Although the addition of the residual function does not significantly improve the fit at $S/N=4.3$, the combined pulse and residual fits are shown for visual comparison. \label{fig:red249}}\end{figure}

Figure \ref{fig:red249} demonstrates two stages in the intensity reduction of BATSE trigger 249. If this pulse had been detected at $S/N=5.2$ (left panel), then it would be classified as blended because both monotonic and residual pulse models are needed to completely account for the pulse light curve variations. If the pulse had been detected at $S/N=4.3$ (right panel), then a fit to the residual function would not be statistically significant (as measured by $p_{\Delta}$) and the pulse would be classified as simple.

\begin{figure}
\epsscale{0.50}
\plotone{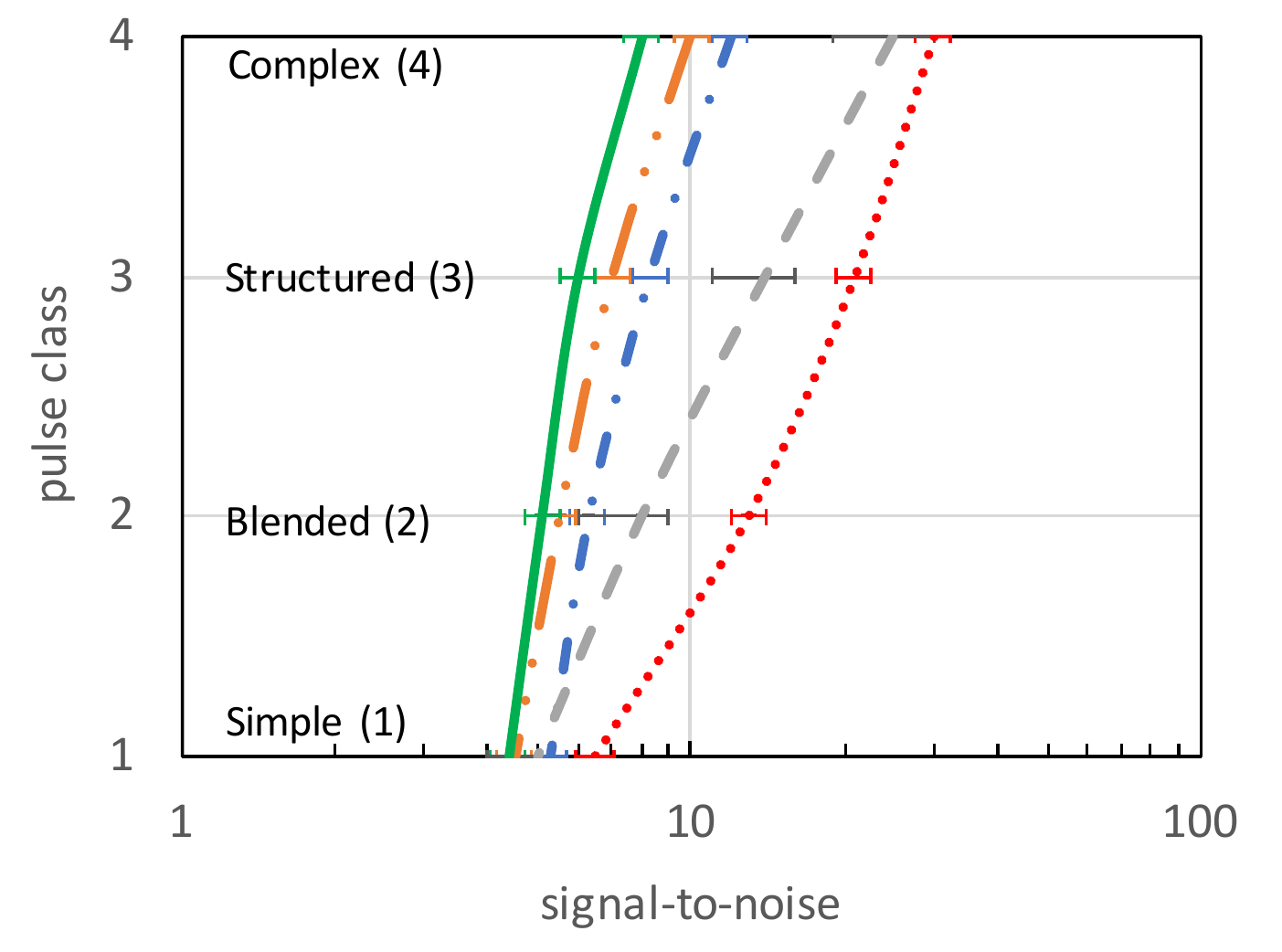}
\caption{GRB pulse classification as $S/N$ is reduced. Indicated are the general paths taken during $S/N$ reduction by each of the five pulses shown: 5614 (red dotted line), 143p1 (blue dashed-dotted line), 143p2 (gray dashed line), 249 (orange dashed-double-dotted line), and 7301p1 (green solid line). The reduced classification of 7301p2 is not plotted, but is similar to that of 249. Simple pulses (class 1) are best characterized by monotonic pulses, blended pulses (class 2) are best characterized by a monotonic pulse overlaid by a smooth triple-peaked residual function, structured pulses (class 3) are improved but not adequately characterized by a pulse plus the residual function, and complex pulses (class 4) are too structured to be described by these simple pulse models.}\label{fig:class}\end{figure}

Figure \ref{fig:class} summarizes the effect of $S/N$ on pulse classification for the pulses in this study. All pulses are classified as complex at high $S/N$, and all pulses are classified as simple at low $S/N$. The $S/N$ values at which these transitions occur (complex to structured, structured to blended, and blended to simple) vary from pulse to pulse depending on the intrinsic pulse structure, the background at the time the burst was detected, and the way in which the random background fluctuations have interacted with the reduced pulse.

It is useful to compare the $S/N$ distributions of pulses populating each class
to the overall pulse $S/N$ distribution.
To do this we must {\em estimate} the numbers of pulses in each class
because each pulse must be fitted
before it can be classified, and this is a time-consuming procedure.  
Even estimating the overall pulse $S/N$ distribution must be performed with care because
measuring the mean background for each burst
first requires us to separate the background from the pulse (and from multiple
pulses when the burst contains more than one); this again requires us to fit
each pulse before we can characterize its $S/N$. 

\begin{figure}
\epsscale{0.50}
\plotone{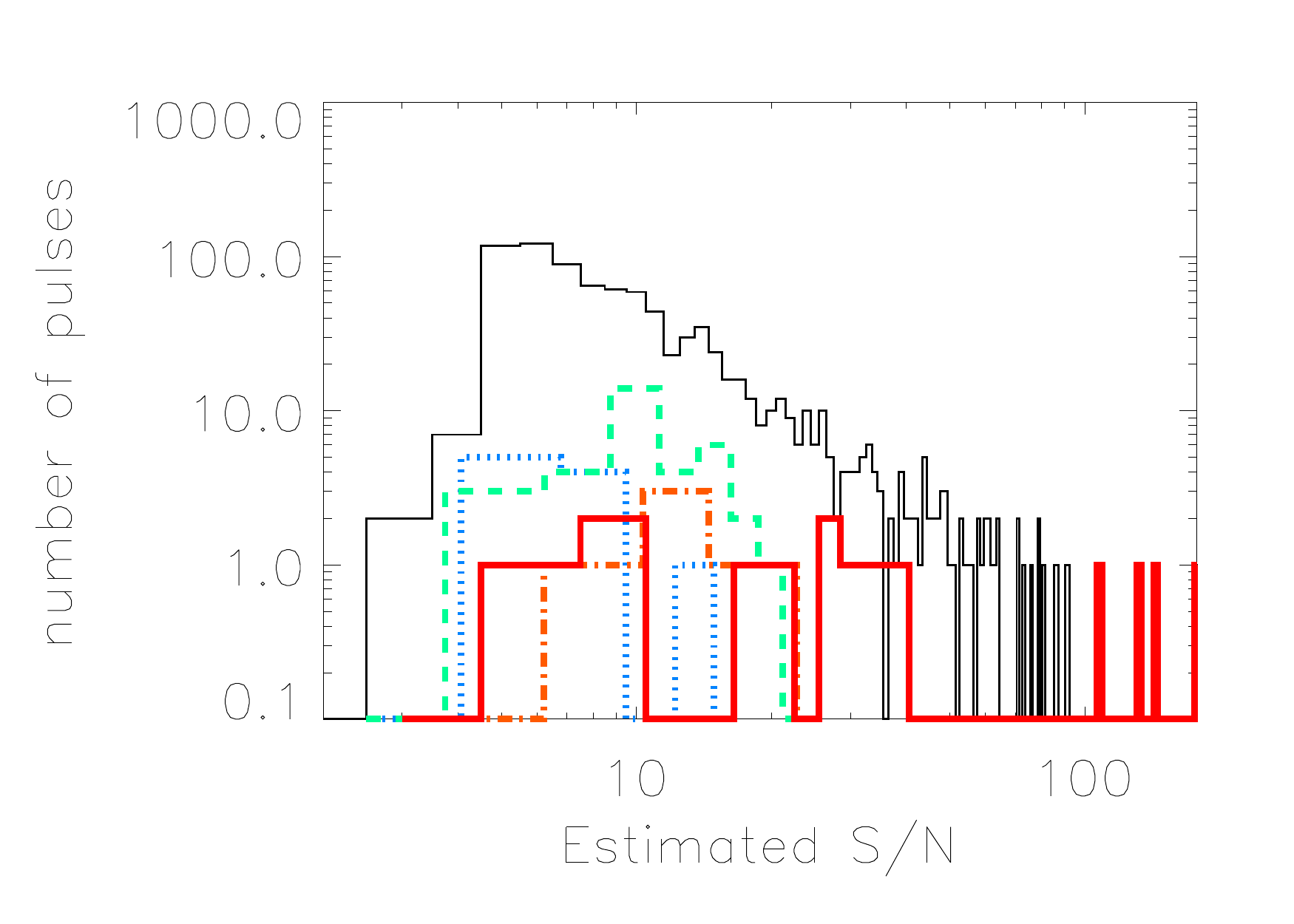}
\caption{The estimated $S/N$ distribution of the brightest BATSE pulse in each burst (solid black line). The plot is overlaid with the measured $S/N$ values of classified BATSE pulses (primarily from \cite{hak14,hak15}) Simple pulses (blue dotted line) are found at the lowest $S/N$ ($\langle \log(S/N) \rangle = 0.83$ or $S/N=6.7$), blended pulses (dashed green line) have ($\langle \log(S/N) \rangle = 1.03$ or $S/N=10.7$), structured pulses (dotted-dashed orange line) have ($\langle \log(S/N) \rangle = 1.15$ or $S/N=14.1$), and complex pulses (solid red line) have ($\langle \log(S/N) \rangle = 1.51$ or $S/N=32.7$). The large $S/N$ of the pulses in this sample ($S/N > 100$) can be contrasted with the rest of the distribution. Note also that all pulses with $S/N \ge 25$ have been classified as complex.} \label{fig:SNhist}\end{figure}

In order to carry out our estimate we first note that there is a strong correlation between $S/N$ values and published BATSE 1024 ms $C_{\rm max}/C_{\rm min}$ values (the count rate at which a GRB triggers relative to the minimum count rate for a trigger to occur):
\begin{equation}
S/N=4.606 (C_{\rm max}/C_{\rm min})^{0.669},
\end{equation}
with a correlation coefficient of $R^2=0.933$. From this correlation we can estimate the $S/N$ distribution of the brightest pulses in each BATSE burst using $C_{\rm max}/C_{\rm min}$ (again, we cannot identify the $S/N$ distribution of all pulses without going through the time-consuming process of fitting all pulses). Second, we obtain the distribution of BATSE GRB pulses fitted in previous papers ({\em e.g.,} \cite{hak14,hak15}). We make sure to exclude GRBs with $T_{90}$ durations less than 2 seconds ($T_{90}$ is the duration spanned by 90\% of the flux) because 64 ms binning distorts any structure present; \cite{hak18}). We overlay the estimated pulse distribution with the observed distributions of simple, blended, structured, and complex pulses in Figure \ref{fig:SNhist}. It is apparent that the current fitted pulse distribution favors pulses exhibiting the triple-peaked structure ({\em e.g.,} blended and structured) because of our bias towards looking for evidence of it.

Based on the classification criteria of these pulses and on the distribution of BATSE pulse $S/N$ values, we estimate that at least $50\%$ of pulses from long BATSE GRBs are found at $S/N \le 8$. Most of these are likely to be classified as either simple or blended, although pulses containing more structure are also found at these $S/N$ values. This result explains both why \cite{hak14} were able to establish a simple residual model using low-$S/N$ pulses and why the residual model did not suffice when applied to bright pulses. {\em It is entirely likely that all GRB pulses exhibit complex structures, but only a small fraction of pulses exhibit these structures without having them smoothed by instrumental effects related to $S/N$}. 

However, caution must be exerted in generalizing and interpreting why GRB pulse structure exists at low $S/N$. Not all faint pulses are simple; some have been classified as blended, structured, and even complex. Smeared-out, complex, intrinsic pulse structures can appear indistinguishable from overlapping pulses at low $S/N$.


\subsubsection{Pulse Duration and Asymmetry}

\begin{figure}
\plottwo{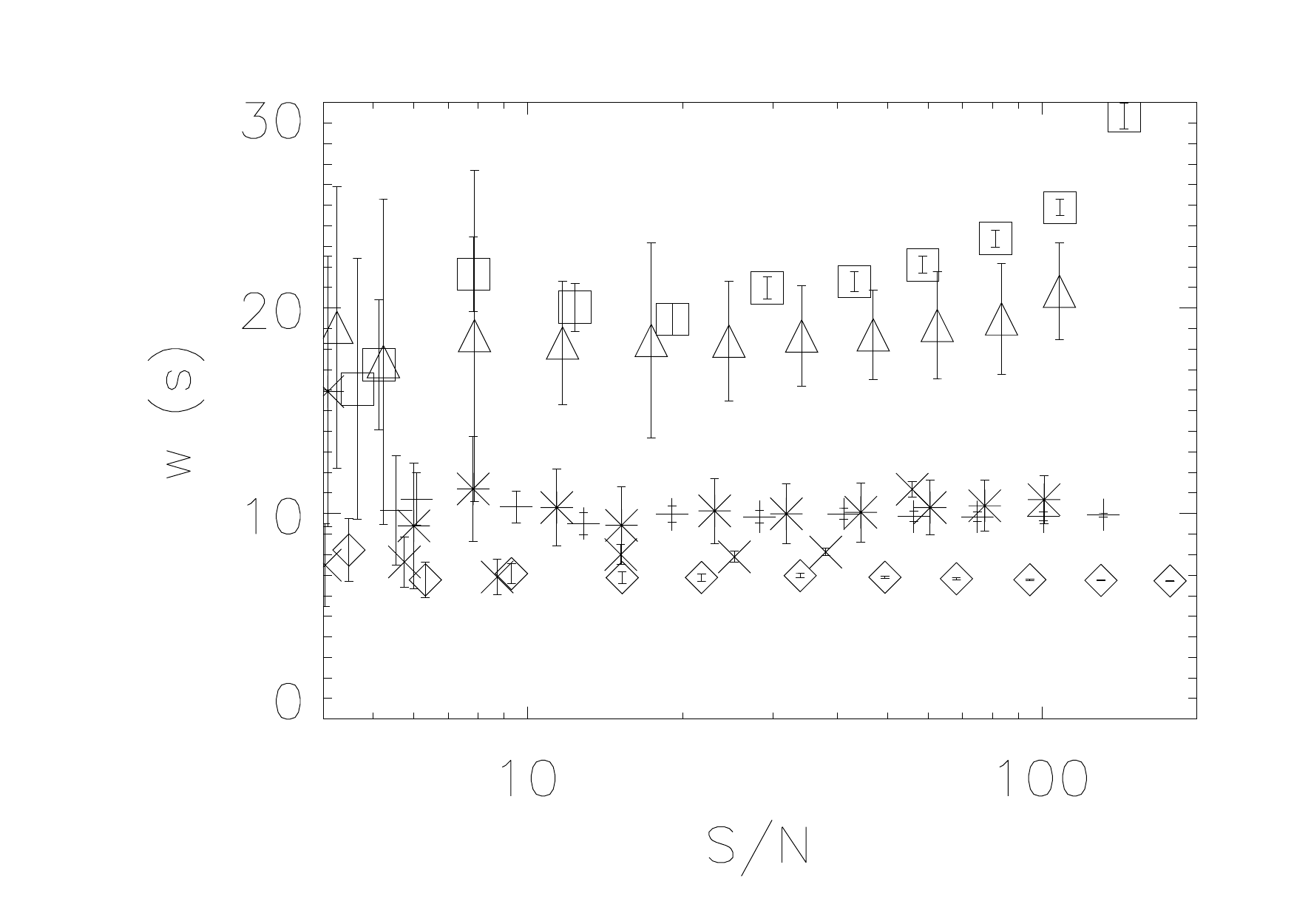}{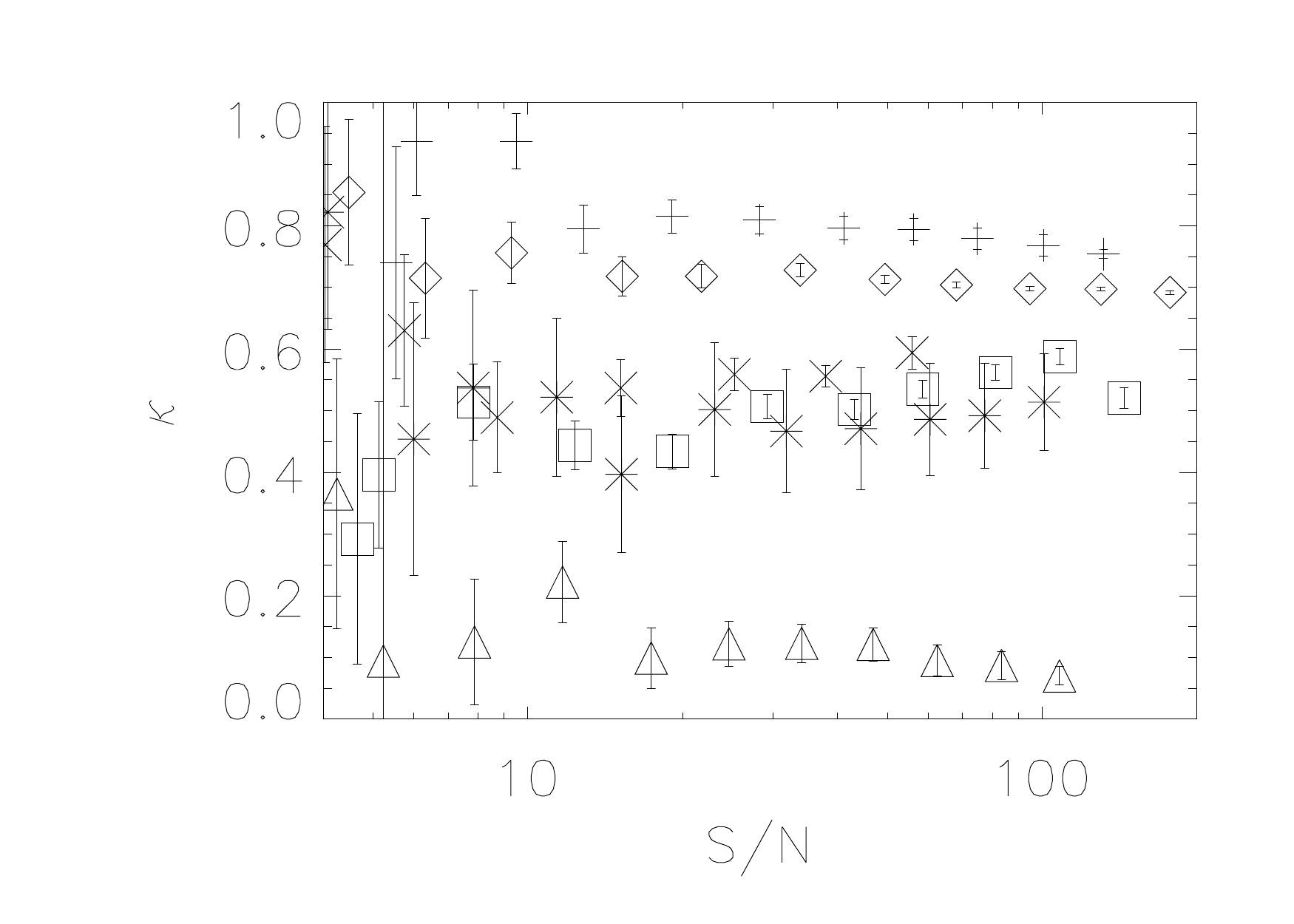}
\caption{Fitted durations (left panel) and asymmetries (right panel) of pulses 143p1 (plusses), 143p2 (crosses), 249 (triangles), 5614 (diamonds), 7301p1 (asterisks), and 7301p2 (squares) as their $S/N$ values are reduced. Fitted pulse durations and asymmetries do not change systematically with decreased $S/N$, with the exception of 7301p2 (due to insufficient contributions to the fit from the loss of emission in the pulse's broad ``wings"). Some measurements appear to undergo discontinuous shifts as $S/N$ drops; this happens when pulse structure abruptly disappears into the noise and the fit parameters have to fit a more simply-structured pulse. Uncertainties increase for all pulse measurements when $S/N < 6$.\label{fig:prop}}\end{figure}

Although observing a GRB pulse at reduced $S/N$ smears out its pulse structure, the pulse's duration ($w$) and asymmetry ($\kappa$) extracted from the monotonic model shown in Equation \ref{eqn:function} do not appear to be systematically affected. Figure \ref{fig:prop} demonstrates this for the bright pulses in this sample as $S/N$ is reduced.  Measured $w$ and $\kappa$ values are, however, affected statistically by large uncertainties at low $S/N$ ($S/N < 6$). We also note that the pulses with the largest amounts of structure preceding and following the main monotonic component (249, 7301p1, and 7301p2) have the largest duration and asymmetry uncertainties at all $S/N$ values. Pulses in which structure overlays the monotonic component (143p1, 143p2, and 5614) have smaller uncertainties that actually serve to guide the monotonic pulse fit and decrease uncertainty in measuring the pulse fit parameters ($\tau_1$, $\tau_2$, $t_s$, and $A$). 

\subsubsection{Pulse Spectral Lags}

Spectral lags indicate the time difference between the optimal alignment of pulse's high-energy and low-energy fluxes. Lags are measured using the cross-correlation function (CCF). The CCF is either applied to the data (data lag); {\em e.g.,} \cite{che95,nor96,ban97}) or to the fitted pulses (fit lag; \cite{nor05,hak11}). Spectral lags correlate with GRB luminosity \citep{nor96} and provide a simple measurement of spectral properties. Although there appears to be a general relationship between both types of lag and pulse evolution ({\em e.g.,} \cite{koc03,sch04,ryd05,hak11,hak15,hak18}), some pulses have negative data lags \citep{ukw10,roy14,cha18} and data lags are often inconsistent with fit lags ({\em e.g.,} \cite{ukw10}). 


We predict that data lags obtained from structured pulse data will be different than fit lags of the same pulses. Short-duration pulses have shorter lags (of both types) than long-duration pulses ({\em e.g.,} \cite{nor05,pen07,hak08}). This is not terribly surprising since a pulse's lag must be shorter than the pulse's duration since the flux distributions generally overlap \citep{hak11}. Similarly, structured pulses must have short data lags, since structure often occurs on short timescales, and since structure is generally bright enough to contribute significantly to the CCF peak. Fit lags measured by applying the CCF to multi-channel \cite{nor05} pulse fits, however, are not constrained to be short, because the fits effectively smear out any energy-dependent structures by redistributing them.
Since the broad CCF distributions made from these smeared light curves smoothly span the entire pulse duration, the extracted fit lags can exceed the timescales of the structures. Additionally, pronounced structure in one energy channel but not another can shift the CCF, and therefore the lag obtained from it. 

We calculate the data lags from the CCF for several energy channels for each pulse, and compare these to the lags of the smooth fitted monotonic pulse components (fit lags), the results of which can be found in Table \ref{tab:lags}. We have chosen to demonstrate lags between channels 3 and 1, channels 3 and 2, and channels 4 and 1 because these are commonly cited energy channels. As discussed above, we find that the greatest similarities between data lags and \cite{nor05} fit lags (obtained by cross-correlating the energy-dependent pulse fits) occur in the smoothest, least-structured pulses (143p2 and 5614), while the greatest differences are found in the most-structured pulses (143p1, 249, 7301p1, and 7301p2). The data lags and \cite{nor05} fit lags of 143p1 are relatively similar, even though this pulse shows pronounced structure. The reason for this appears to be that the rapidly-varying structure is primarily embedded within the pulse decay and does not markedly change the fit parameters of any of the energy-dependent pulse fits.

\begin{deluxetable*}{crrrrrr}
\tablenum{1}
\tablecaption{Pulse Spectral Lags\label{tab:lags}}
\tablewidth{0pt}
\tablehead{
\colhead{Pulse} & \colhead{31 data lag} & \colhead{31 fit lag} & \colhead{32 data lag} & \colhead{32 fit lag} & \colhead{41 data lag} & \colhead{41 fit lag} \\
}
\startdata
143p1 & $0.04 \pm 0.03$ & $0.11 \pm 0.03$ & $0.03 \pm 0.03 $ & $0.07 \pm 0.03$ & $0.10 \pm 0.03 $ & $0.21 \pm 0.03$  \\
143p2 & $0.02 \pm 0.03$ & $-0.07 \pm 0.03$ & $0.01 \pm 0.03 $ & $-0.02 \pm 0.03$ & $0.10 \pm 0.04 $ & $0.03 \pm 0.03$ \\
249 & $0.07 \pm 0.03$ & $0.37 \pm 0.03$ & $0.02 \pm 0.03 $ & $0.14 \pm 0.03$ & $0.18 \pm 0.04 $ & $0.38 \pm 0.03$  \\
5614 & $-0.22 \pm 0.03$ & $-0.15 \pm 0.03$ & $-0.06\pm 0.03 $ & $-0.06 \pm 0.03$ & $-0.31 \pm 0.03 $ & $-0.13 \pm 0.03$  \\
7301p1 & $0.03 \pm 0.03$ & $-0.56 \pm 0.05$ & $0.01 \pm 0.03 $ & $-0.35 \pm 0.04$ & $0.03 \pm 0.03 $ & $-1.13 \pm 0.09$  \\
7301p2 & $0.03 \pm 0.03$ & $1.84 \pm 0.03$ & $0.01 \pm 0.03 $ & $1.075 \pm 0.03$ & $0.03 \pm 0.0 $ & $2.33 \pm 0.05$  \\
\enddata
\tablecomments{Data lag is the time delay between emission in a high-energy channel and a low-energy channel, measured using the CCF from the time at which the maximum value of the cross-correlation function occurs. Fit lag refers to a lag obtained by cross-correlating the \cite{nor05} pulse fits in each energy channel, and is thus a smoothed version of the standard CCF lag.}
\end{deluxetable*}

\begin{figure}
\plottwo{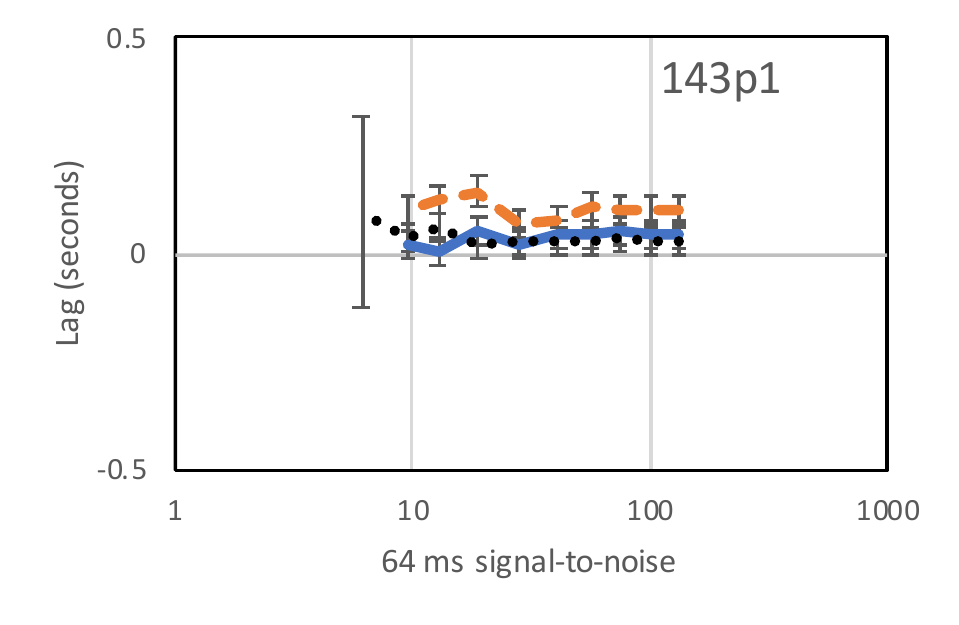}{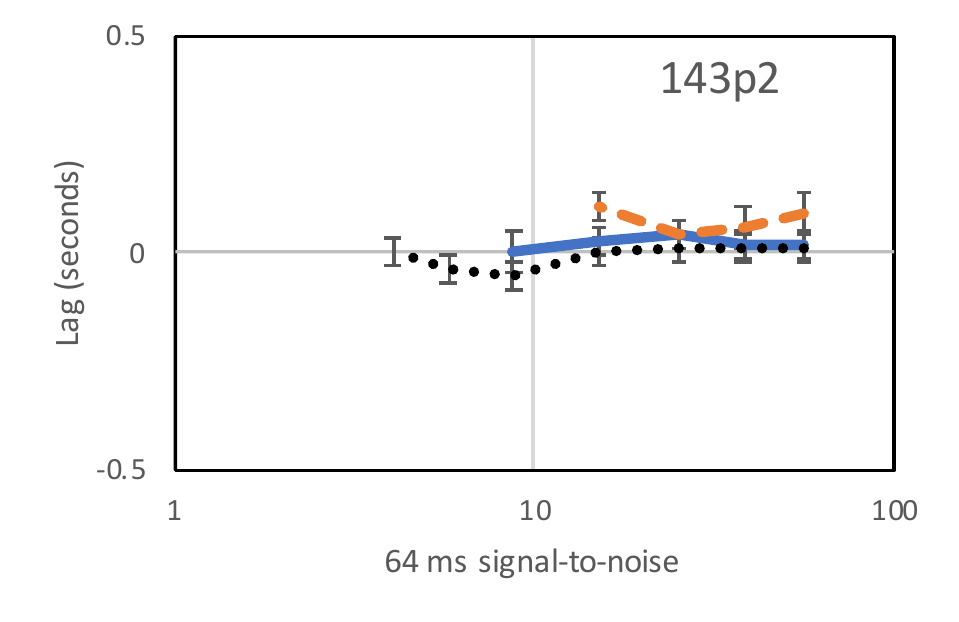}
\caption{Data lags of pulses 143p1 (left panel) and 143p2 (right panel). Black dotted lines indicate lags between BATSE channels 3 and 2, blue solid lines indicate lags between BATSE channels 3 and 1, and orange dashed lines indicate lags between BATSE channels 4 and 1. Lags in both 143p1 and 143p2 decrease systematically with decreasing $S/N$.\label{fig:lag1}}\end{figure}
\begin{figure}
\plottwo{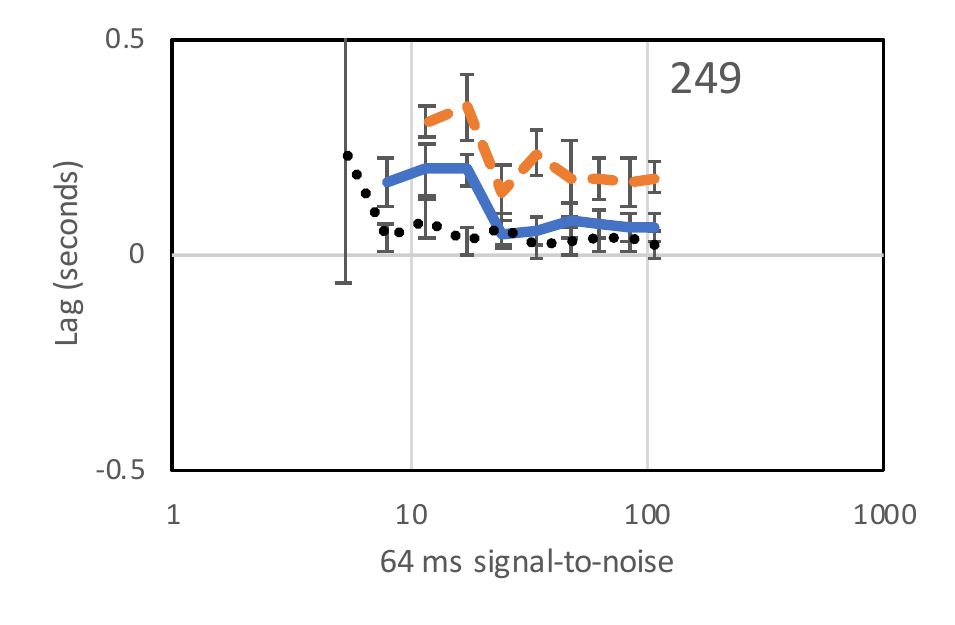}{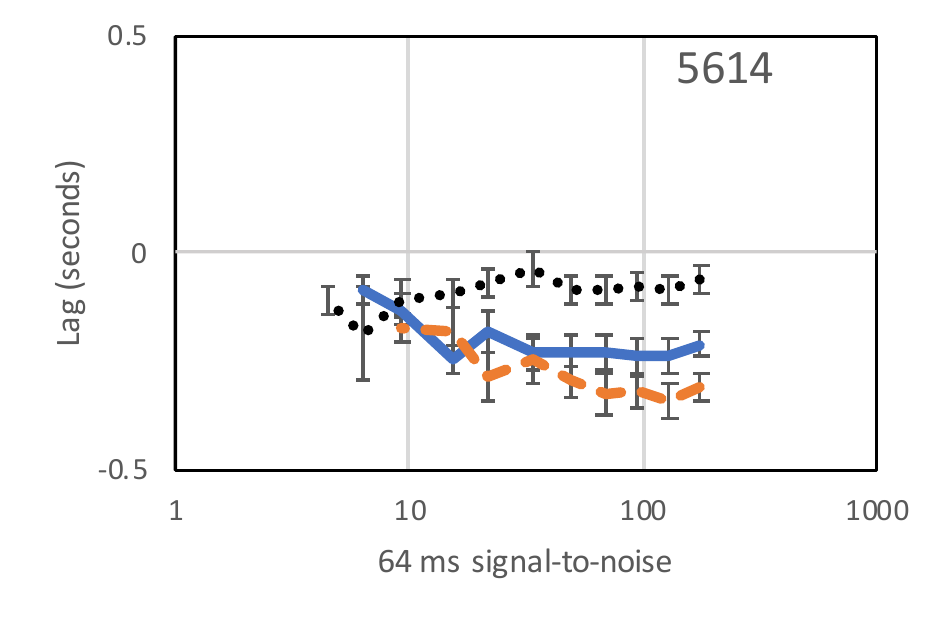}
\caption{Data lags of pulses 249 (left panel) and 5614 (right panel). Black dotted lines indicate lags between BATSE channels 3 and 2, blue solid lines indicate lags between BATSE channels 3 and 1, and orange dashed lines indicate lags between BATSE channels 4 and 1. Lags in 249 remain constant with decreasing $S/N$ while those of 5614 increase systematically with decreasing $S/N$.\label{fig:lag2}}\end{figure}
\begin{figure}
\plottwo{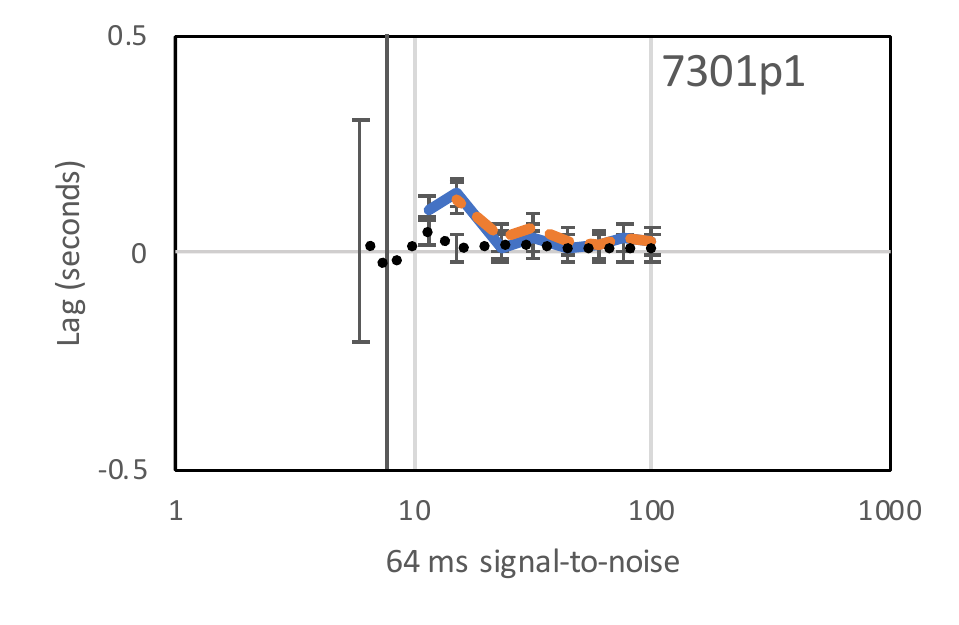}{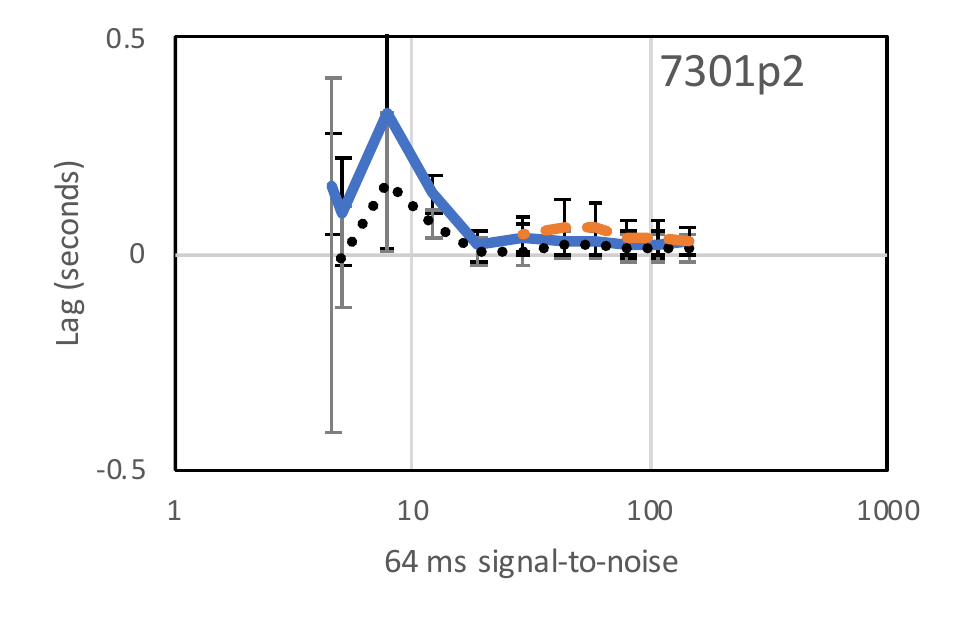}
\caption{Data lags of pulses 7301p1 (left panel) and 7301p2 (right panel). Black dotted lines indicate lags between BATSE channels 3 and 2, blue solid lines indicate lags between BATSE channels 3 and 1, and orange dashed lines indicate lags between BATSE channels 4 and 1. Lags in 7301p1 do not undergo systematic changes, while lags in 7301p2 increase systematically by large amounts with decreasing $S/N$. \label{fig:lag3}}\end{figure}

We can also see if the spectral lags of bright pulses change as they are observed closer to the instrumental noise limit. Pulse data lag measurements across a variety of $S/N$ values are shown in Figures \ref{fig:lag1} $-$ \ref{fig:lag3}. Figure \ref{fig:lag1} shows the data lags of pulses 143p1 (left panel) and 143p2 (right panel), Figure \ref{fig:lag2} shows the data lags of pulses 249 (left panel) and 5614 (right panel), and Figure \ref{fig:lag3} shows the data lags of pulses 7301p1 (left panel) and 7301p2 (right panel). The data lags of most pulses in our sample do not change systematically when viewed at lower $S/N$, although the uncertainties in measuring these values do increase and some lags become unmeasurable due to the loss of signal in a particular energy channel (usually channels 1 or 4).  Possible exceptions are the channel 3 vs.~1 and channel 4 vs.~1 lags of pulses 249 and 5614; these exhibit slight systematic increases not measured in their channel 3 vs.~2 lags. These increases are caused by the loss of pulse structure in noisy channel 1 but not in the other channels; without this structure the lag measures a cross-correlation between structured channel 3 or 4 emission and a now smoother channel 1 light curve. The resulting data lags are larger than the data lags of the original unreduced pulses (which are sensitive to structure) and smaller than the fit lags of the unreduced pulses (which are insensitive to structure). With these caveats, the loss of pulse structure at low $S/N$ does not appear to introduce systematic changes in spectral lag measurements.

Pulse lag and duration are both luminosity indicators \citep{hak08}, yet neither seems to be systematically affected by decreased $S/N$, although lag results can be marginally affected when one lag channel is significantly noisier than the other. {\em Thus, measurements of these characteristics at low $S/N$ are generally reliable, with uncertainties mostly produced by small number photon counting statistics.}

\subsection{Mirrors: Temporally Reflected Residuals}

Because instrumental noise has been shown to smear out GRB pulse structure, we are forced to confront the fact that our best chance to learn about GRB pulse structure is to study the spectro-temporal evolutionary properties of bright, complex GRBs. The study of complex GRB pulse characteristics has generally been avoided primarily because authors have disagreed on the number of pulses represented in bursts containing structure if the structure cannot be attributed to noise. However, as a result of our analysis in Section \ref{sec:structure}, we are able to approach each bright episode under the assumption that it represents a single structured pulse.

The six pulses in our sample suggest that bright pulse residuals exhibit a wavelike function similar to that described by the residual model, but with a larger number of complete waves than the three used to develop Equation \ref{eqn:residual}. The complex residual structures of these pulses can be seen in the left panels of Figures \ref{fig:143} $-$ \ref{fig:7301p2}. \cite{hak14} assumed that the wave used in the empirical fitting function did not extend outside the Bessel function's third zeros at $J_0 (x = \pm 8.654)$ because the low- to moderate-$S/N$ pulse residuals showed no evidence of this. That assumption is not true at the higher $S/N$ values of pulses in this sample, where 143p1 appears to exhibit nine sets of up-and-down residual fluctuations, 143p2 has five, 249 has at least seven, 5614 has six, 7301p1 has roughly six, and 7391p2 has three. However, since the residual function successfully characterized these structures at lower $S/N$, we suspect that the high-$S/N$ structures are extensions of this phenomenon, and that they have similar origins even though each pulse shows structures that occur on different timescales with differing amplitudes. 

We have unsuccessfully attempted to extend the residual model to these pulses beyond the Bessel function's third zeros. The additional wave features present at high $S/N$ cannot be adequately fitted by simply extending the Bessel function limits and changing the $t_0$, $a$, $\Omega$, and $s$ values. The function incorrectly models these additional waves because the residual model was fine-tuned to work with moderate $S/N$ GRB pulse structures; it cannot be extrapolated to the unexpected structures we encounter at high $S/N$.

The wavelike structures at the beginning of each pulse are observed to have similar characteristics to those at the end of the pulse, except that they appear to be temporally inverted.  The residual function described in Equation \ref{eqn:residual} partially explains this, as it was developed with an implicit recognition that GRB pulse residuals were best fit by temporally-inverted structures. The reason why this model worked was not understood. However, closer examination shows that the many maxima and minima of the pulse residual structures occur in a temporally-mirrored order, with near-reversible shapes. {\em The structures at the beginning of each pulse are decent reversed facsimiles of the structures at the end of the pulse. Furthermore, the wavelike features at the beginning of pulses are compressed relative to those at the ends of pulses, apparently coordinated with the pulse's asymmetry.} When an odd number of waves is found, this indicates that the time of reflection occurs near a wave maximum. When an even number of waves is found, this indicates that the time of reflection lies at a wave minimum. 

Since the residual function is an incomplete representation of the residual structure, we choose to more accurately characterize the mirrored residual structure in our sample. We do this by folding the residuals over at a pulse's temporal symmetry point (the {\em time of reflection}, which we call $t_{\rm 0;mirror}$), then stretching the folded reversed residuals by an amount $s_{\rm mirror}$ until the folded wave prior to the reflection time (generally on the rising part of the pulse light curve) matches the wave following the reflection time (generally on the decaying part of the light curve). We find these values iteratively, by cycling $t_{\rm 0;mirror}$ through a range of possible values, folding and reversing the residuals at each possible time of reflection, then cycling $s_{\rm mirror}$ through a range of possible values so as to stretch the reversed-residuals to match the forward- and reversed stretched residual light curves. The CCF calculated at each step is used to compare the quality of the matches, with the optimal CCF value identifying the best characterization of $t_{\rm 0;mirror}$ and $s_{\rm mirror}$. We note that $t_{\rm 0;mirror}$ is similar to $t_0$ (defined in Equation \ref{eqn:residual}) except that $t_{\rm 0;mirror}$ is no longer necessarily measured from the time when the residual function is a maximum ($t_0$ is), and that $s_{\rm mirror}$ is similar to $s$ except that it is measured from $t_{\rm 0;mirror}$ rather than from $t_0$. Our folding and stretching approach also uses all residuals rather than just those fitted by Equation \ref{eqn:residual}. In other words, $t_{\rm 0;mirror}$ and $s_{\rm mirror}$ are less model-dependent definitions than $t_0$ and $s$, since the former require only a pulse fit to be made prior to their measurement whereas the latter require both a pulse fit and a residual fit.

\begin{figure}
\plottwo{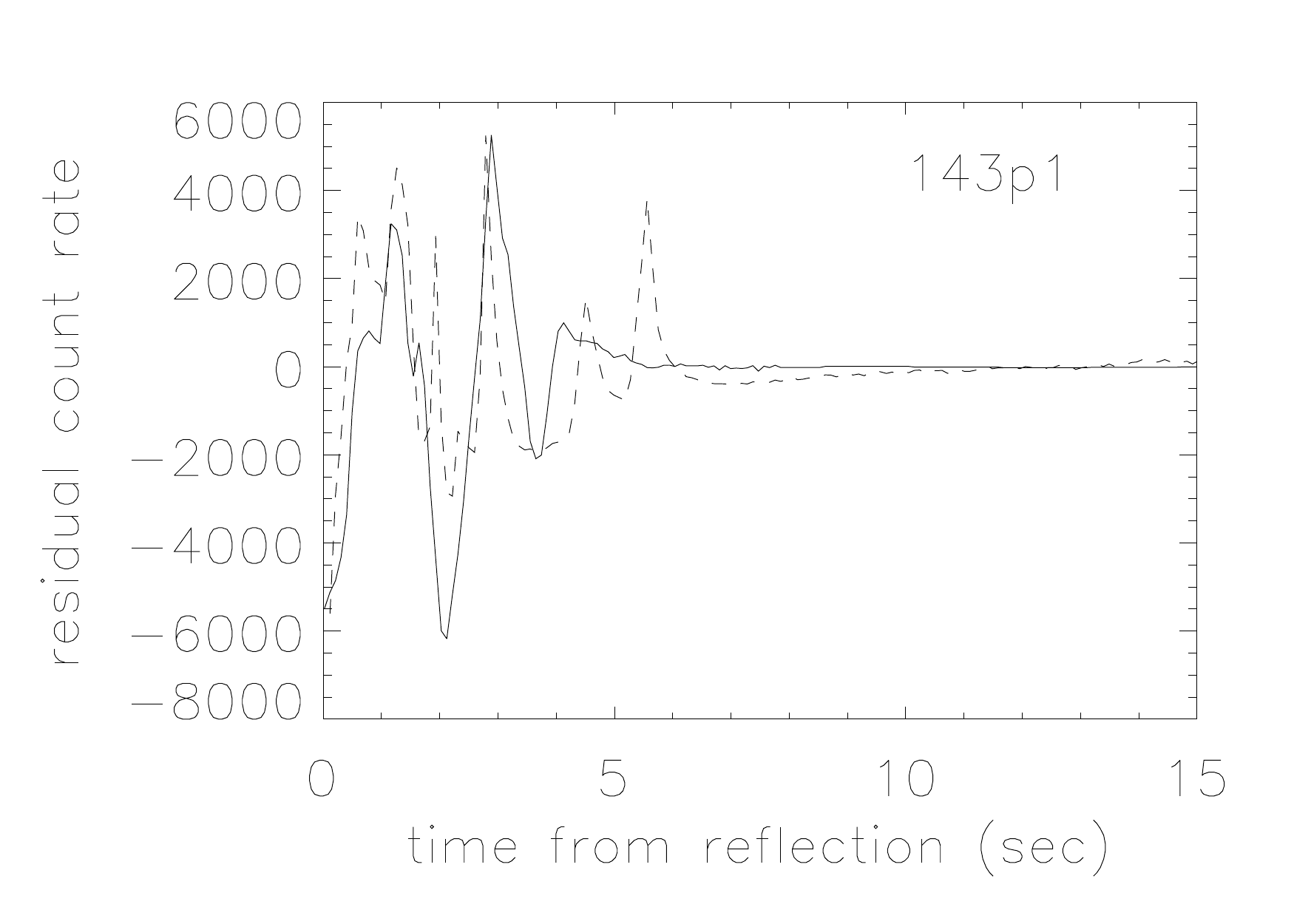}{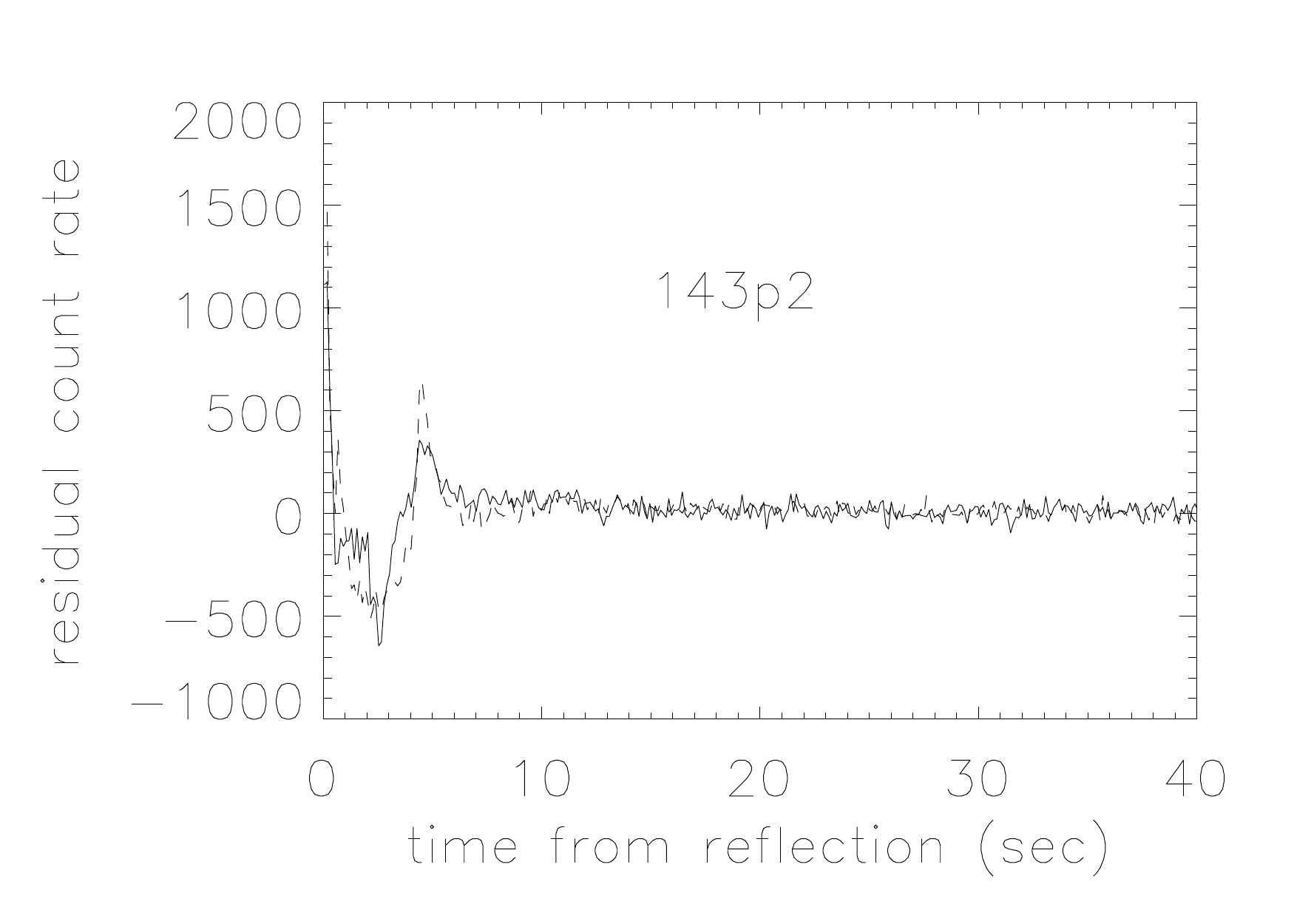}
\caption{The mirrored and stretched residuals of BATSE triggers 143p1 and 143p2 (after removing the \cite{nor05} fit). The cross-correlation function yields values of 
0.60 (143p1) and 0.84 (143p2) when the residuals from the pulse rise (solid line) are inverted, stretched, and overlaid with the residuals from the decay side (dotted line).  \label{fig:mirror1}}\end{figure}

\begin{figure}
\plottwo{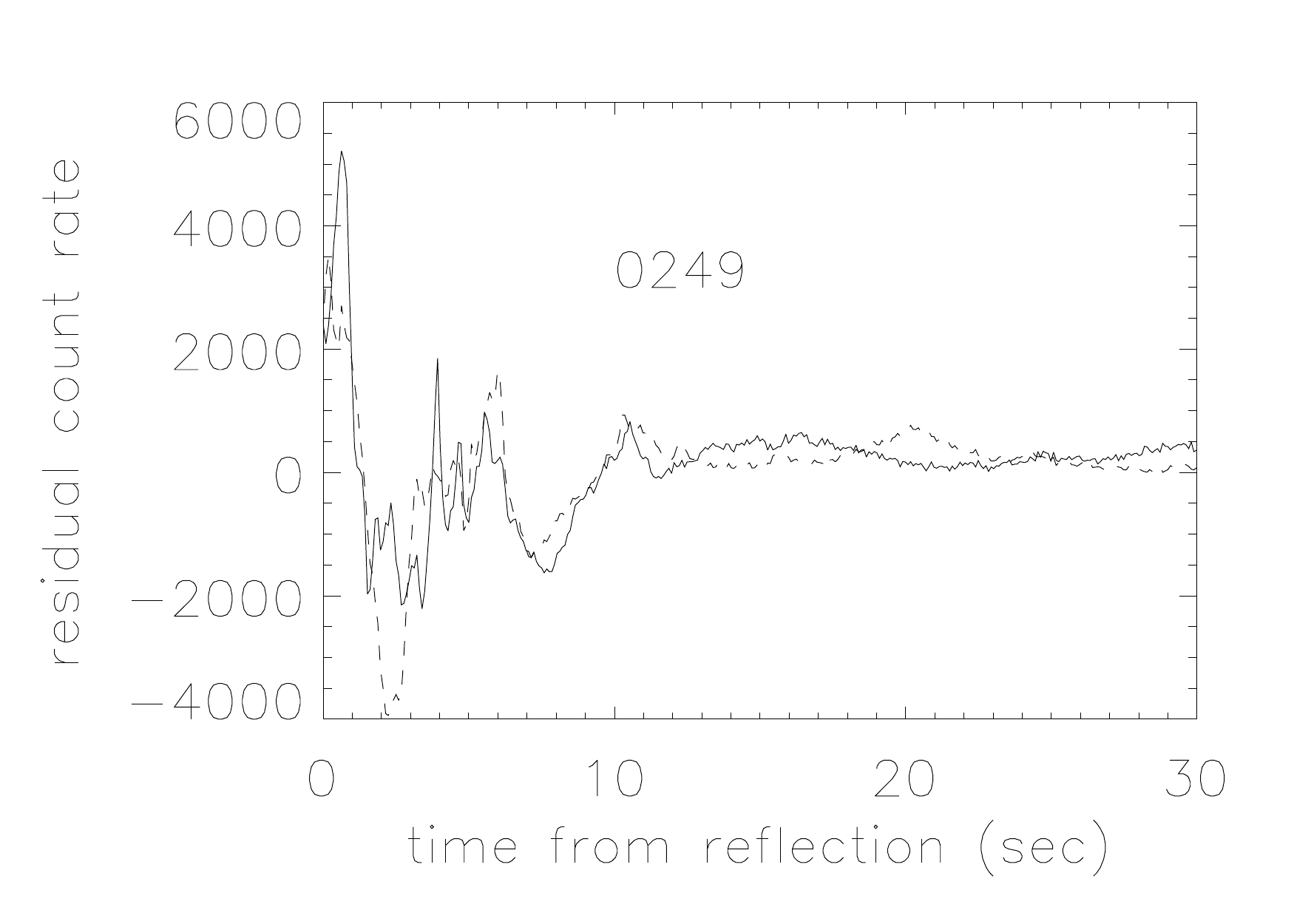}{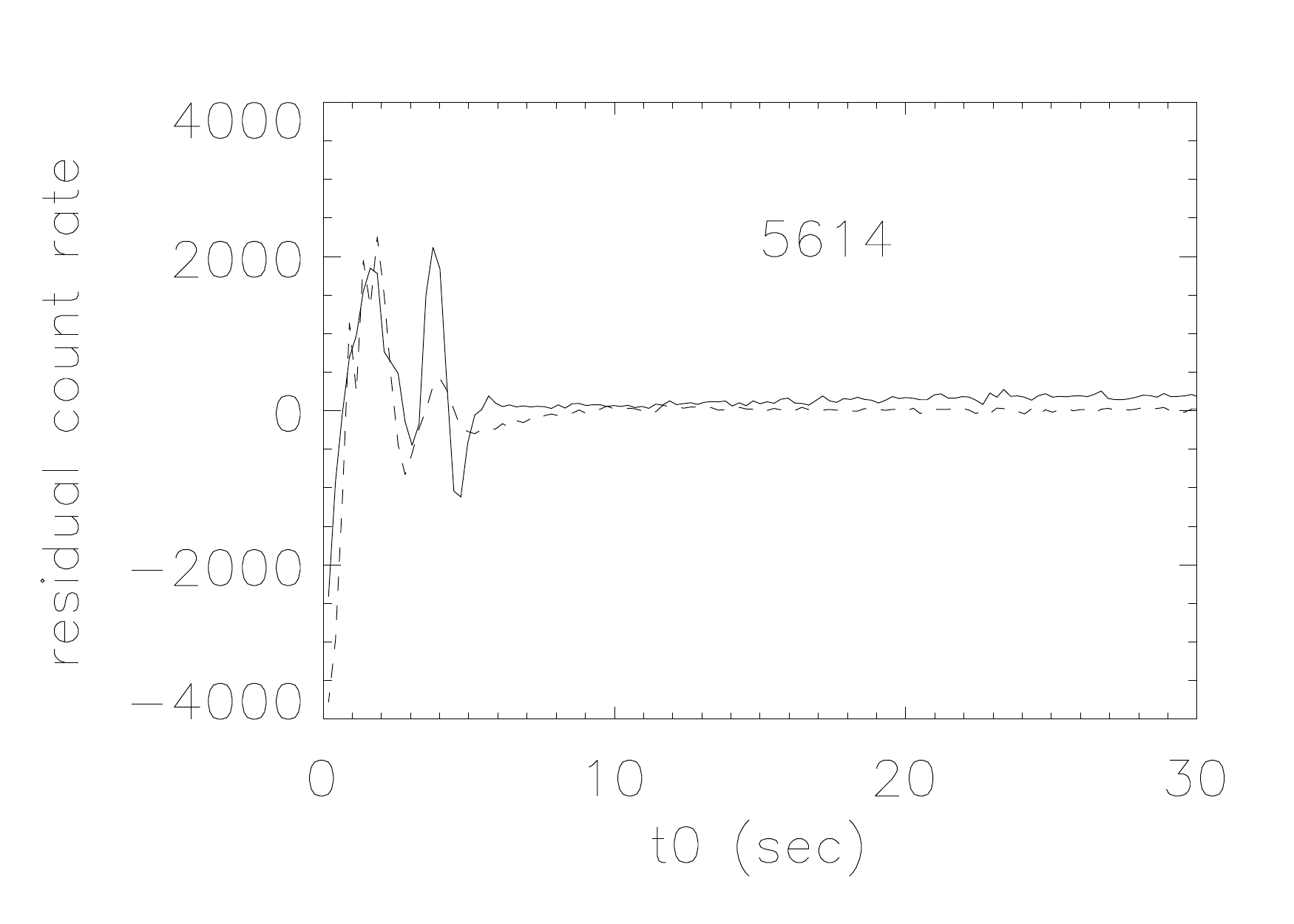}
\caption{The mirrored and stretched residuals of BATSE triggers 249 and 5614 (after removing the \cite{nor05} fit). The cross-correlation function yields values of 0.72 (249) and 0.81 (5614) when the residuals from the pulse rise (solid line) are inverted, stretched, and overlaid with the residuals from the decay side (dotted line).\label{fig:mirror2}}\end{figure}

\begin{figure}
\plottwo{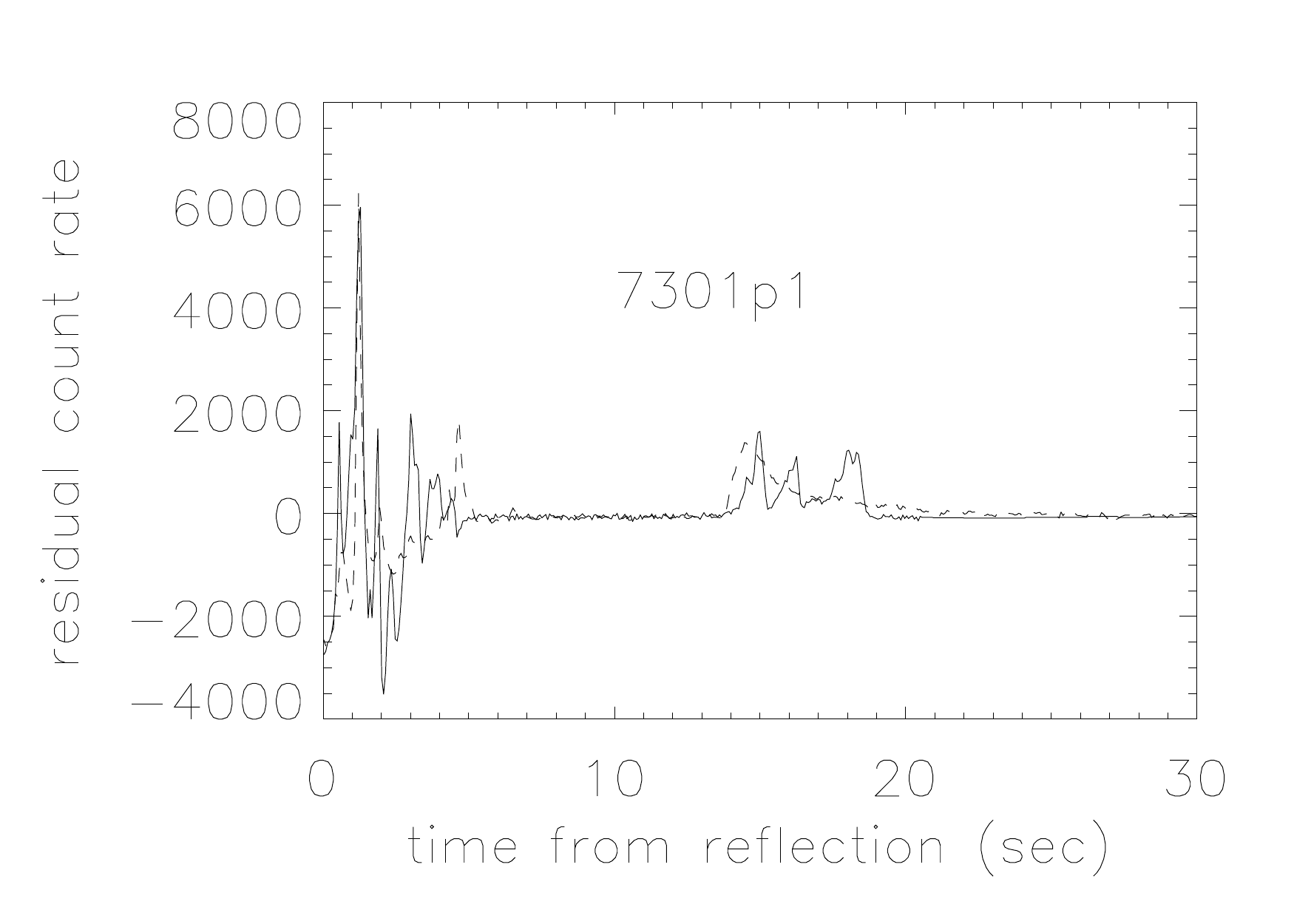}{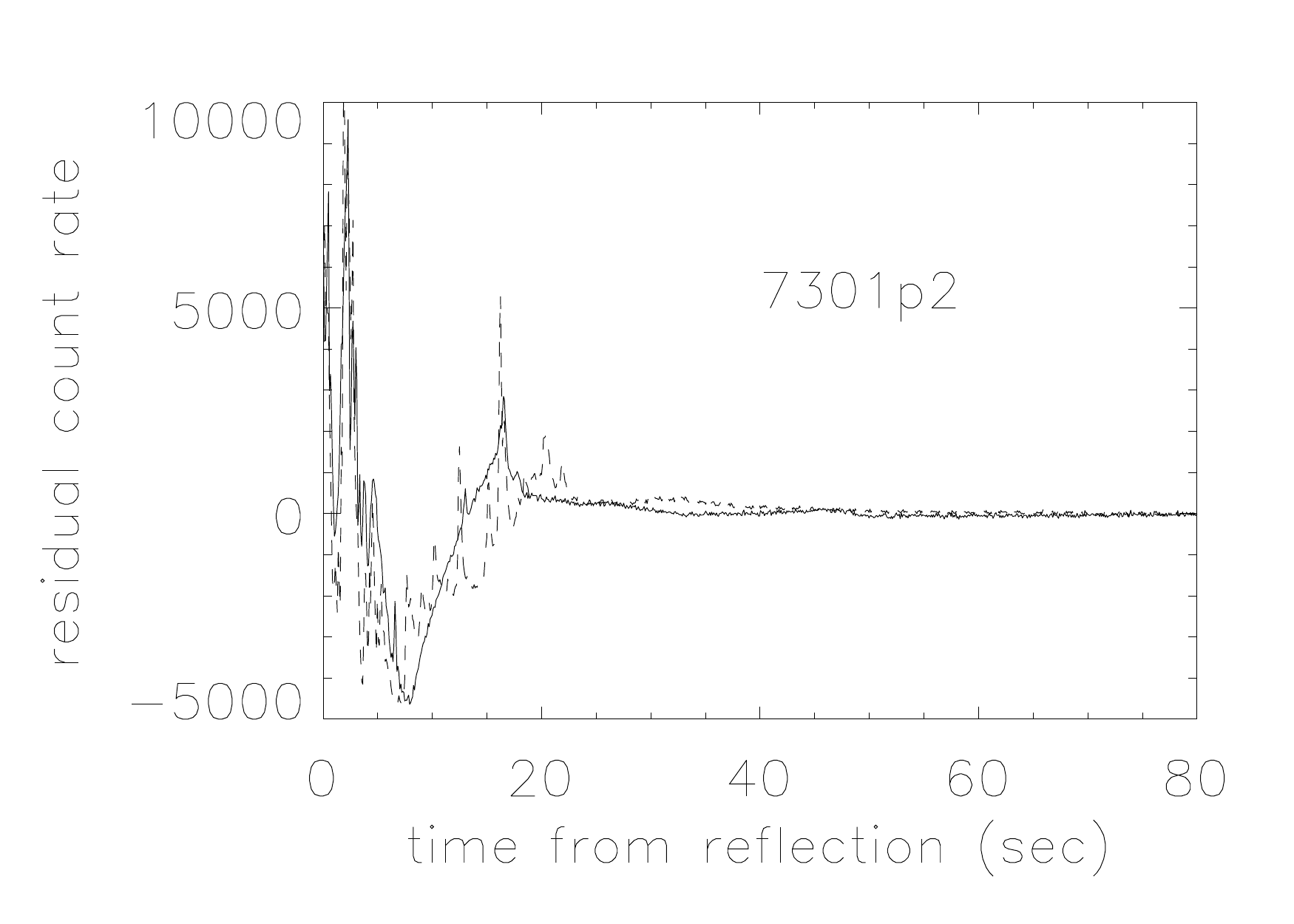}
\caption{The mirrored and stretched residuals of BATSE triggers 7301p1 and 7301p2 (after removing the \cite{nor05} fit). The cross-correlation function yields values of  0.66 (7301p1) and 0.81 (7301p2) when the residuals from the pulse rise (solid line) are inverted, stretched, and overlaid with the residuals from the decay side (dotted line).\label{fig:mirror3}}\end{figure}

The results of mirroring and stretching the residuals of our six pulses are shown in Figures \ref{fig:mirror1} $-$ \ref{fig:mirror3}.
The times given in these figures are referenced to $t_{\rm 0;mirror}$ (seventh column of Table \ref{tab:res}) rather than to the trigger time ($t=0$ seconds), and these times only describe the decay portion of the residuals (dotted line), because the rise times (solid line) have been time-inverted and stretched. Despite the bright, non-randomly distributed residual structures, the fits are good. The optimum values of the CCF exceed $0.6$ for these residuals, indicating a strong correlation between the structures preceding the time of reflection and those following it. The CCFs of three of the pulses (143p2, 5614, and 7301p2) exceed 0.8. We note that the fit we chose for 143p1, with a CCF of 0.60, was slightly worse than the optimal fit of 0.64 $-$ we chose this fit because it pairs all but one of the residual peaks whereas the optimal fit misses three peaks in order to pair the two deepest troughs.

The fitted characteristics of the mirrored residual parameters $s_{\rm mirror}$
and $t_{\rm 0;mirror}$ are compared to their counterparts $s$ and $t_0$ in Table \ref{tab:res}.
The pulse asymmetry $\kappa$ is included for comparison with the $s$ value for 
each pulse, as these parameters have previously been shown to anti-correlate;
this anti-correlation is only loosely observed from these six pulses.
Also indicated are the best-fit CCF values used to obtain the time-reversible characteristics.
The values are generally in good agreement, and differ primarily
for the reasons described above.

We estimate the significance of the mirroring by comparing the time-inverted, stretched residuals preceding the time of reflection to those following it. We use the $p-$values obtained from a Spearman Rank-Order test to compare the two distributions, after first interpolating the time-reversed and stretched residuals so that they contain the same number of data points as the better-sampled post-$t_{\rm 0;mirror}$ residuals. This approach gives us a likelihood that the residuals are related, without providing information about how closely the intensities match one another. {\em The $p-$ values found for all pulses are significant ($p < 10^{-8}$), indicating that residuals following $t_{\rm 0;mirror}$ are time-reversed and stretched versions of those preceding it.}

To ensure that our fitting procedure has not inadvertently biased the results, we repeat the analysis after excluding all intensities less than $3\sigma$ away from the mean background rate (this data truncation eliminates any temporal correlations that might result from incompletely-fitted monotonic pulse components and/or backgrounds). We define {\em $p-$bright} to be the Spearman Rank-Order correlation $p-$value obtained from comparing these truncated data. The results of the Spearman Rank-Order test are shown in the final column of Table \ref{tab:res}; all six pulses have $p-$bright $ < 10^{-4}$. The strongest matches are found for 7301p2, 249, and 143p2 (small $p-$bright values indicate support for the stretched temporal inversion hypothesis). The least significant match is found for 5614 due to the small number of bright data points this pulse contains. {\em We conclude that the time-reversed and stretched residuals are not an artifact of our data analysis procedure.}  Furthermore, we will show in Section \ref{sec:hr_evol} that the increased significance of $p-$ relative to $p-$bright, when coupled with hardness evolution, indicates additional evidence for faint time-reversed and stretched structures extending farther from the time of reflection as opposed to inadequate monotonic pulse-fitting. 

We note that half the measured $t_{\rm 0;mirror}$ values (143p1, 5614, and 7301p1) occur at times where the residuals are at a minimum rather than at a maximum. Since $t_0$ by definition measures reflection from a maximum value of the residuals, $t_{\rm 0;mirror}$ and $t_0$ can differ significantly (see  Table \ref{tab:res}), as can $s$ and $s_{\rm mirror}$. This result demonstrates that Equation \ref{eqn:residual} is not an accurate characterization of complex residual structures. It also explains why some pulse duration ($w$) and asymmetry ($\kappa$) measurements shown in Figure \ref{fig:prop} undergo discontinuous changes as $S/N$ is reduced: residual fit parameters change as structure is incrementally washed out by noise.


\begin{deluxetable*}{cccccccc}
\tablenum{2}
\tablecaption{Fitted Pulse vs. Time-Reversible Residual Properties\label{tab:res}}
\tablewidth{0pt}
\tablehead{
\colhead{Pulse} & \colhead{$\kappa$} & \colhead{CCF} & \colhead{$s$} & \colhead{$s_{\rm mirror}$} & \colhead{$t_0$} & \colhead{$t_{\rm 0;mirror}$} & \colhead{$p-$bright} \\
}
\startdata
143p1 & $0.75 \pm 0.01$ & $0.60$ & $0.45 \pm 0.04$ & $0.67$ & $4.07 \pm 0.05 $ & $3.15$ & $2\times 10^{-6}$ \\
143p2 & $0.59 \pm 0.02$ & $0.84$ & $0.50 \pm 0.01$ & $0.51$ & $48.09 \pm 0.01 $ & $48.03$ & $2\times 10^{-14}$  \\
249 & $0.07 \pm 0.02$ & $0.72$ & $0.56 \pm 0.02$ & $0.74$ & $21.51 \pm 0.03 $ & $21.70$ & $3\times 10^{-22}$ \\
5614 & $0.69 \pm 0.01$ & $0.81$ & $0.19 \pm 0.02$ & $0.31$ & $8.30 \pm 0.01 $ & $8.92$ & $2\times 10^{-4}$  \\
7301p1 & $0.51 \pm 0.08$ & $0.66$ & $0.82 \pm 0.06$ & $0.96$ & $16.62 \pm 0.03 $ & $17.8$ & $2\times 10^{-7}$  \\
7301p2 & $0.52 \pm 0.02$ & $0.81$ & $0.48 \pm 0.02$ & $0.72$ & $165.40 \pm 0.07 $ & $165.60$ & $3\times 10^{-45}$  \\
\enddata
\end{deluxetable*}

\section{Discussion}\label{sec:hr_evol}

GRB pulse temporal and spectral behaviors are closely linked, with pulse structure playing an important role in characterizing spectral evolutionary behavior. Pulse spectro-evolution is often characterized in terms of ``hard-to-soft'' and ``intensity tracking'' behaviors ({\em e.g.,} \cite{whe73, gol83, nor86, pac92}); these terms unfortunately suggest two distinct and bimodal characterizations when in fact a large continuous range of behaviors exists ({\em e.g.,} \cite{kar94, bha94, for95}). We shall demonstrate that much of this variation may not be real, but may instead result from the lack of a consistently-applied pulse definition. We have previously demonstrated that essentially all pulses undergo hard-to-soft spectral evolution in that they are harder at the pulse onset than they are late in the decay phase \citep{hak15,hak18}. Despite this gradual softening, spectral re-hardening of moderately bright GRB pulses typically occurs around the times that pulse structure occurs, as found from the three residual peaks. Spectral evolution is therefore subject to the relative hardnesses of the residual peaks. Pulses appear to evolve ``hard-to-soft'' if their central peaks are softer than their precursor peaks, whereas they follow ``intensity tracking'' behaviors if their central peaks are harder than their precursor peaks. Depending on the relative hardnesses of the peaks, there exist a wide range of intermediary behaviors \citep{hak15} that authors often either exclude from their analyses or force-classify into either the ``intensity tracking'' or ``hard-to-soft'' categories. 

Asymmetry also plays a role in pulse hardness evolution: asymmetric pulses are harder overall and have
pronounced hard-to-soft evolution; these contrast with symmetric pulses 
that are softer and have weak hard-to-soft evolution \cite{hak15}. 
This weak evolution can result in softer precursor peaks
than central peaks, producing intensity tracking behaviors.

Using BATSE's four energy channels, we define a counts hardness ($hr$) in each
time bin $i$ as:
\begin{equation}
hr_i= \frac{C_{3i}+C_{4i}}{C_{1i}+C_{2i}}
\end{equation}
where $C_{1i}, C_{2i}, C_{3i}$, and $C_{4i}$ are the counts/bin in
channels 1 ($20 - 50$ keV), 2 ($50 - 100$ keV), 3 ($100 - 300$ keV), 
and 4 (300 keV $- 1$ MeV), respectively. This counts hardness has been
chosen instead of a standard ratio like $hr_{32}=C_3/C_2$ in order to
increase the number of photon counts in each bin, making the hardness more reliable 
for use on short timescales and thus in probing spectral evolution.

The pulse hardness evolution of 
moderately bright BATSE pulses (64 ms resolution), Swift pulses (64 ms
resolution), and short BATSE pulses (4 ms resolution)
have been studied previously in aggregate samples of pulses \citep{hak15,hak18}.
The technique used was to combine the energy-dependent light curves
of many pulses by 1) scaling
the durations of the pulse sample to the unitless fiducial timescale, 2) subtracting out the monotonic pulse
model of each pulse to obtain residuals, 3) aligning the residuals to the 
zeros, minima, and maxima of the residual function to account for asymmetry, 
4) summing the energy-dependent residual counts from different
pulses to obtain composite time-dependent counts in each energy channel,
and 5) using these composite counts to obtain time-dependent hardnesses
to describe aggregate pulse evolutionary behavior. 
The fiducial time interval used in these analyses (Equation \ref{eqn:res}) 
had the additional benefit of recognizing that residual structures 
often begin before and end after the
monotonic pulse component does.

The six pulses used in this analysis are bright enough that
the counts hardness evolution 
of each can be studied individually,
rather than in aggregate.
However, the fiducial timescale no longer completely encapsulates
the structured components of all sampled pulses
because some of the complex residual wave structures extend beyond the 
three known residual peaks, and therefore beyond the boundaries of
the fiducial timescale.
It thus makes more sense to 
bound each pulse's hardness evolution by its mirrored residual timespan 
rather than by its fiducial timescale.

\subsection{BATSE Trigger 143}

Both of BATSE trigger 143's asymmetric pulses
exhibit hard-to-soft spectral evolution. However, 143p1 is characterized by a 
complex residual structure whereas 143p2 is triple-peaked; the added structure
in 143p1's light curve causes it to undergo more frequent hardness fluctuations.

\begin{figure}
\plottwo{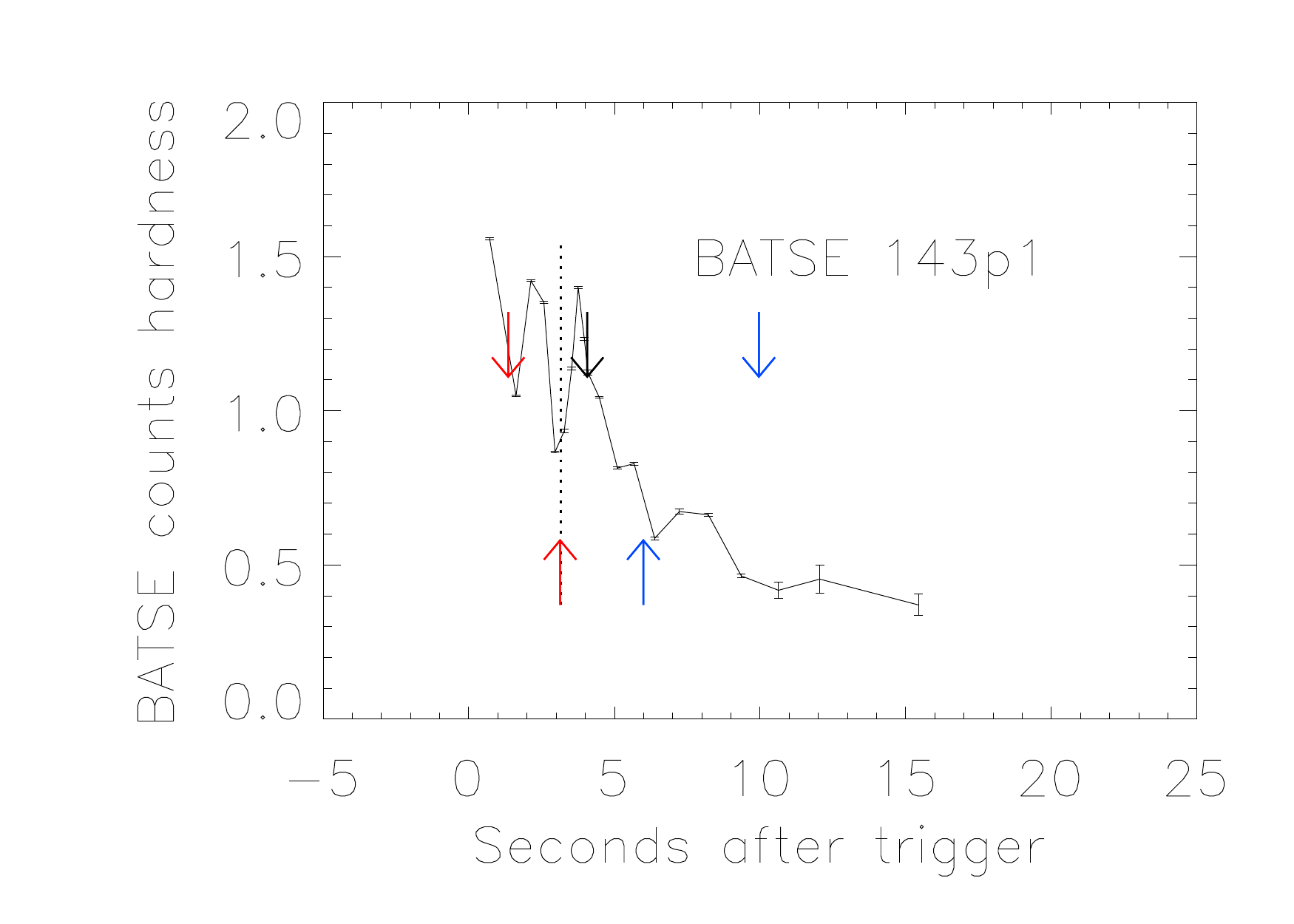}{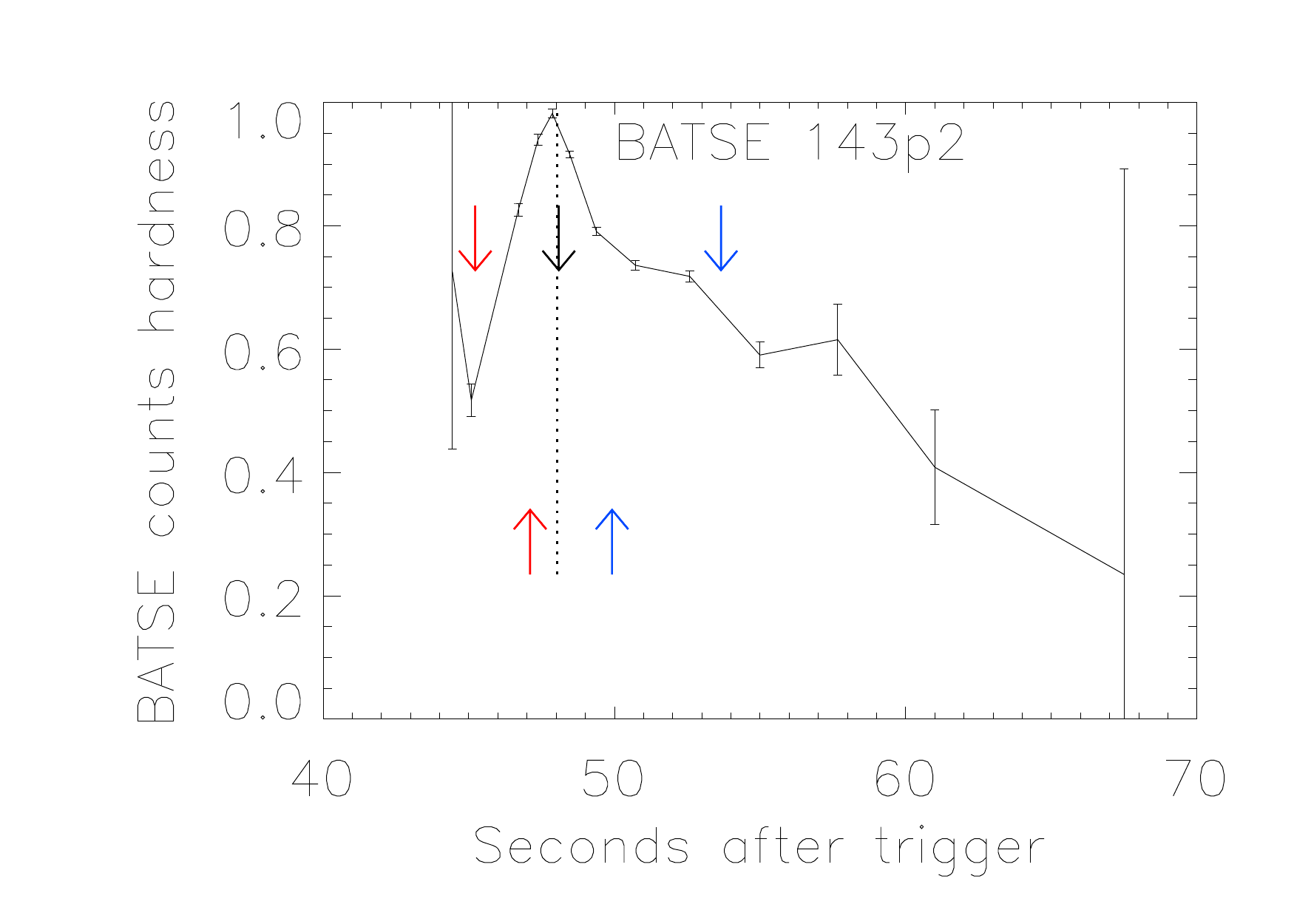}
\caption{Counts hardness ($hr$) evolution of BATSE pulses 143p1 (left panel) and 143p2 (right panel). In this and subsequent figures, downward facing arrows indicate the approximate times of the precursor peak (red), central peak (black), and decay peak (blue). Upward facing arrows indicate the time of the valley separating the precursor peak (red) and decay peak (blue) from the central peak.\label{fig:hr1}}\end{figure}

Pulse 143p1's rapid intensity variations are accompanied by spectral re-hardening 
(Figure \ref{fig:hr1}, left panel); these are difficult to temporally resolve
given the small numbers of counts available per bin during 
the short variable intervals.
We demonstrate these variations using a combination of the hardness 
evolution plot (Figure \ref{fig:hr1}, left panel),
the fitted pulse and residual light curves (Figure \ref{fig:143}), and the time-reversible
and stretched structures (Figure \ref{fig:mirror1}; left panel).

Links between the triple-peaked pulse structure
and hardness evolution are shown using the 
approach of \cite{hak15, hak18}.
Figure \ref{fig:hr1} (left panel) shows 
the hardness evolution relative to the times of
the precursor peak (downward-facing red arrow),
the valley between the precursor and central peaks (upward-facing
red arrow), the central peak (downward-facing black arrow), 
the valley between the central and decay peaks (upward-facing
blue arrow), and the decay peak (downward-facing blue arrow).
The precursor peak and the decay peak provide
rough estimates of the fiducial timescale, and therefore of
the monotonic pulse duration.

Pulse 143p1 is odd in that the peak of the monotonic pulse ($\tau_{\rm peak} = 1.7$ seconds)
does not align well with the peak of the residuals ($t_0 = 4.1$ seconds; see Table \ref{tab:res}).
The central residuals peak of 143p1 roughly aligns with a series of fluctuations 
occurring on the downslope of the light curve. The time of
reflection found from the time-reversed and stretched residuals 
is identified in Figure \ref{fig:hr1} (left panel) by a vertical dotted line.

The three peaks of the residual function are smoothed representations
of a more complex residual structure that
can be seen by comparing 143 p1's hardness evolution to both 
the fitted residual structure (Figure \ref{fig:143}, left panel)
and the time-reversible residual structure (Figure \ref{fig:mirror1}; left panel).
The time of reflection at $t_{\rm 0;mirror} = 3.15$ takes place almost 
a full second before the fitted residual peak; 
this corresponds to a deep residual valley
(left panel of Figure \ref{fig:mirror1}). The pulse undergoes
hardening immediately prior to and following $t_{\rm 0;mirror}$;
this hardness bump encompasses a set of unresolved peaks occurring at around $t = 2$ 
seconds that are mirrored with three closely-spaced peaks occurring 
around $t = 6$ seconds. These mirrored structures are seen in
the left panel of Figure \ref{fig:mirror1} occurring between 1 and 2 seconds after
the time of reflection.
Two bright peaks at $t = 1$ and $t = 6$ seconds are also mirrored,
as is the hard bump at the trigger ($t = 0$ seconds) with two peaks at $t = 8$ and $t = 9$ seconds.
The spectrally hardest part of the pulse occurs at this trigger bump. The bump shows up on the
time-reversed light curve (Figure \ref{fig:mirror1}; left panel)
roughly 4 seconds after $t_{\rm 0;mirror}$, where it coincides
with two spikes at $t = 8$ and $t = 9$ seconds (Figure \ref{fig:143}; left panel).

The residuals of 143p2 exhibit three sharp peaks (Figure \ref{fig:143p2})
that align well with the residual function; these are
the precursor peak at 45 seconds, the central peak at 48 seconds,
and the decay peak at 52 seconds. Two other faint bumps
(at 43 seconds and 59 seconds) are also visible beyond the fiducial time interval. 
The stretched and mirrored residuals (right panel of Figure \ref{fig:mirror1})
are centered at $t_{\rm 0;mirror} = 48.03$ seconds, which coincides with the
position of the central peak (identified by the dotted vertical line
in the right panel of Figure \ref{fig:hr1}). The precursor and decay peaks
are found 4 seconds after $t_{\rm 0;mirror}$, and the two bumps
can be seen about 12 seconds after $t_{\rm 0;mirror}$. 
Pulse 143p2 is softer and undergoes more gradual
hard-to-soft evolution than 143p1 (right panel of Figure \ref{fig:hr1}),
re-hardening at each peak and bump, and demonstrating that the bumps
represent faint pulse components. 

\subsection{BATSE Trigger 249}

BATSE trigger 249 exhibits multiple residual peaks (Figure \ref{fig:249})
that build in intensity from the trigger to the
central peak and lessen thereafter.
The residual model places the precursor
peak at 17 seconds, the central peak at 22 seconds, and the decay
peak at 27 seconds. However, two other sets of paired peaks
are also present (the first pair at 10 seconds and 42 seconds, and 
the second pair at 15 seconds and 32 seconds).
The central peak aligns closely with $t_{\rm 0;mirror}$, and
the pairing of the various sets of peaks can be seen in the left panel of Figure \ref{fig:mirror2}.

\begin{figure}
\plottwo{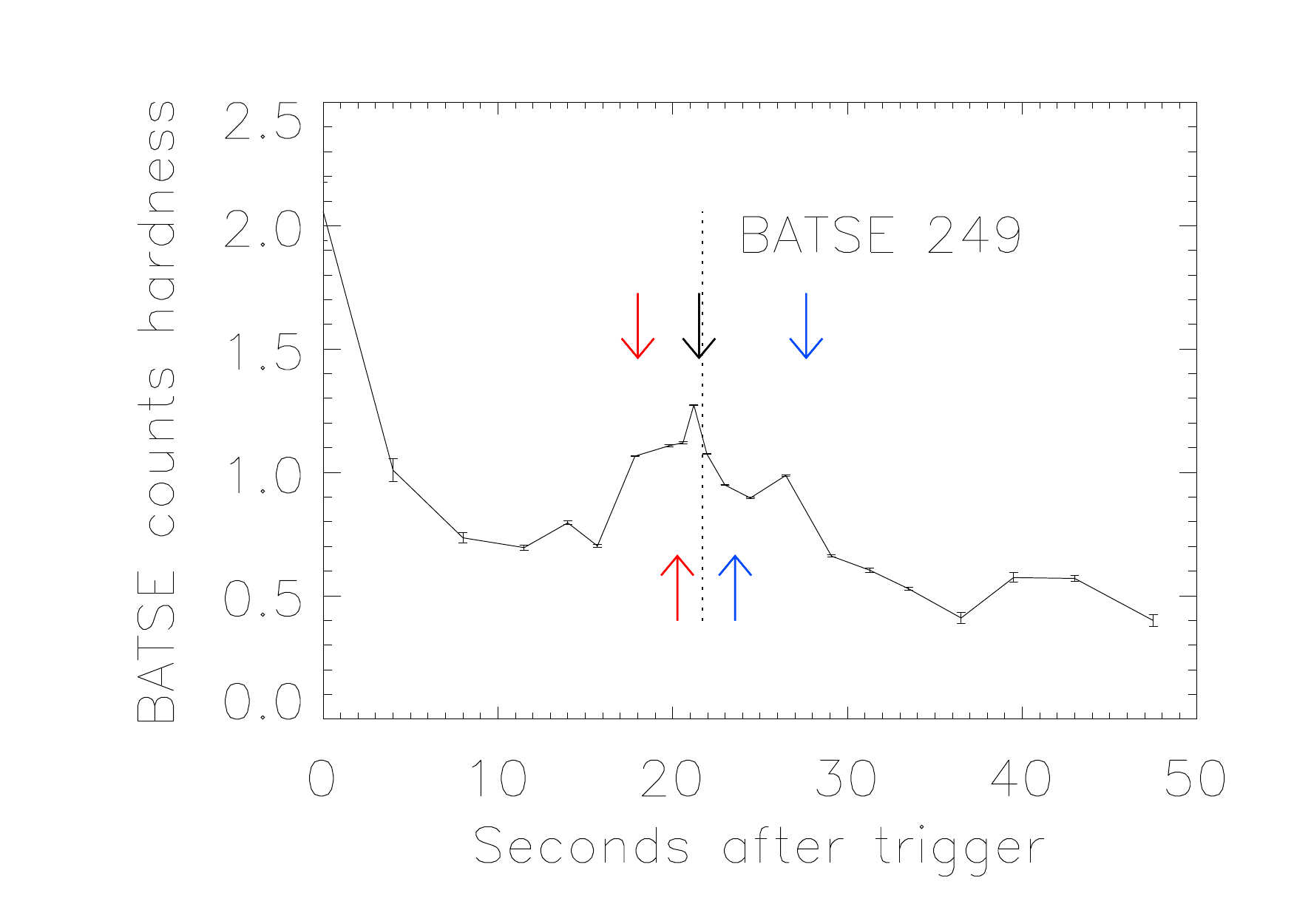}{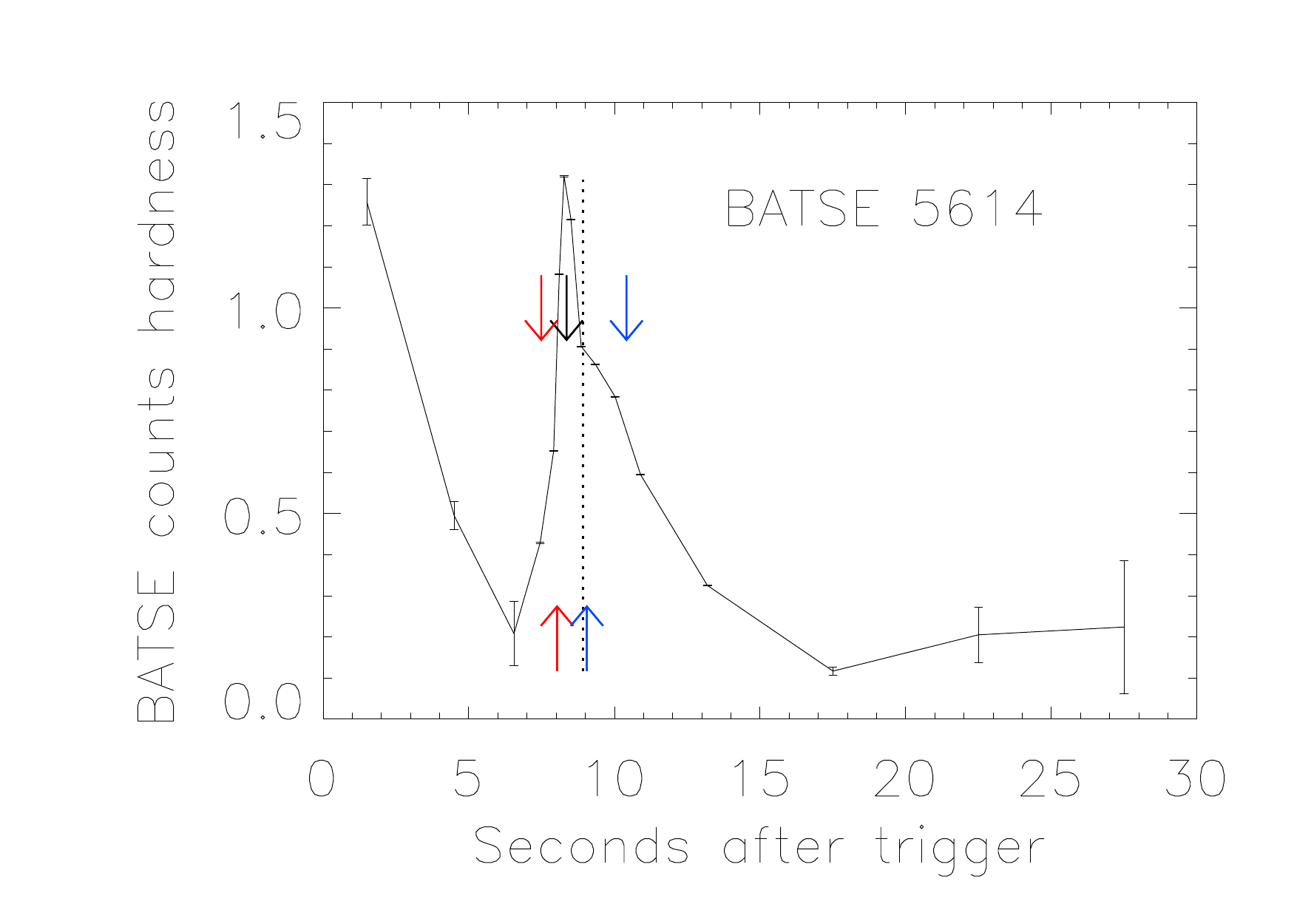}
\caption{Counts hardness ($hr$) evolution of BATSE pulses 249 (left panel) and 5614 (right panel).\label{fig:hr2}}\end{figure}

Trigger 249 demonstrates the inadequacy of characterizing a pulse's duration 
based solely on monotonicity assumptions.
The central {\bf monotonic} part of the pulse, extending from roughly 15 to 28 seconds,
contains most of the flux, but it is clear from mirrored residuals in 
Figure \ref{fig:mirror2} and the hard-to-soft evolution in
Figure \ref{fig:hr2} (left panel) that this pulse starts at the trigger and
continues for roughly 50 seconds (see also Figure \ref{fig:249}).
The pulse evolves consistently from hard-to-soft over the timescale
spanned by the mirrored residuals, while
the central portion of the pulse undergoes a re-hardening at the precursor
peak that is generally maintained until the end of the decay peak; this hardness
plateau representing the monotonic part of the pulse 
(left panel of Figure \ref{fig:hr2}) seems to stand apart from the 
pulse's overall hardness evolution.

\subsection{BATSE Trigger 5614}

Trigger 5614's smooth light curve (Figure \ref{fig:5614}) is nearly monotonic
and has only low-amplitude residuals. These can almost be fitted by 
the triple-peaked residual model.
However, the residual structure has too many lobes, 
and these (Figure \ref{fig:mirror2}, right panel) are consistent 
with stretched and mirrored structures similar to those found in the other pulses.
The time of reflection for 5614 occurs at $t_{\rm 0;mirror} = 8.92$
seconds. This occurs, similar to 143p1, at a minimum in the residuals,
indicating that a pulse's peak intensity is not always augmented 
by the peak intensity of the residuals. The pulse has two distinct
sets of mirrored and stretched residuals: the central peak located at $t = 8.3$ seconds
correlates with a three-peaked structure at $t = 10.5$ seconds (this is the first
peak found 1.5 seconds after the time of reflection in Figure \ref{fig:mirror2}, right panel), and 
a peak at $t = 7.8$ seconds matches a bump at $t = 13$ seconds 
(this is the second peak found 4 seconds after the time of reflection in Figure \ref{fig:mirror2}, right panel).
As in the case of 249, 5614 undergoes significant re-hardening at the
time of the precursor peak and maintains this throughout the central
and decay peaks (right panel of Figure \ref{fig:hr2}). 
In fact, the hardening is so pronounced that the pulse appears
to follow intensity-tracking behavior rather than the expected hard-to-soft evolution.
Supportive evidence for this interpretation comes from the pulse's
spectral lags, which are negative. 
GRB pulses canonically evolve from hard-to-soft,
with the amount of evolution primarily guided by a pulse's
overall hardness and asymmetry.
How can we reconcile 
the apparent existence of an intensity tracking pulse
with our interpretation of a hard-to-soft evolution?
How can we interpret negative spectral lags 
in terms of hard-to-soft evolution? The answers to these questions lie in
information supplied by 5614's mirrored residual function.

Equation \ref{eqn:residual} only partially captures 5614's pulse structure. 
This can be seen in the pulse and residual light curves (Figure \ref{fig:5614}), and it is also indicated 
by the seven-second delay between the trigger time and the rise of the monotonic
pulse intensity. The episode that triggers 5614 is clearly part of
the burst, and the presence of extended time-reversed and stretched residuals
leads us to believe that this low intensity triggering bump must also be part of the pulse.
When we extend the pulse hardness evolution to include this episode 
we find that it is spectrally very hard.
{\em With the addition of this early peak, the pulse evolves hard-to-soft
as expected instead of following its apparent intensity tracking behavior.}
Furthermore, Figure \ref{fig:hr2} (right panel) shows
weak evidence of an extended mirrored tail at about $t = 25$ seconds. 
Although the other residuals clearly match up in the 
mirrored residuals plot (Figure \ref{fig:mirror2}, right panel), the faint trigger bump
around 25 seconds after $t_{\rm 0;mirror}$ is only barely visible.

\begin{figure}
\plottwo{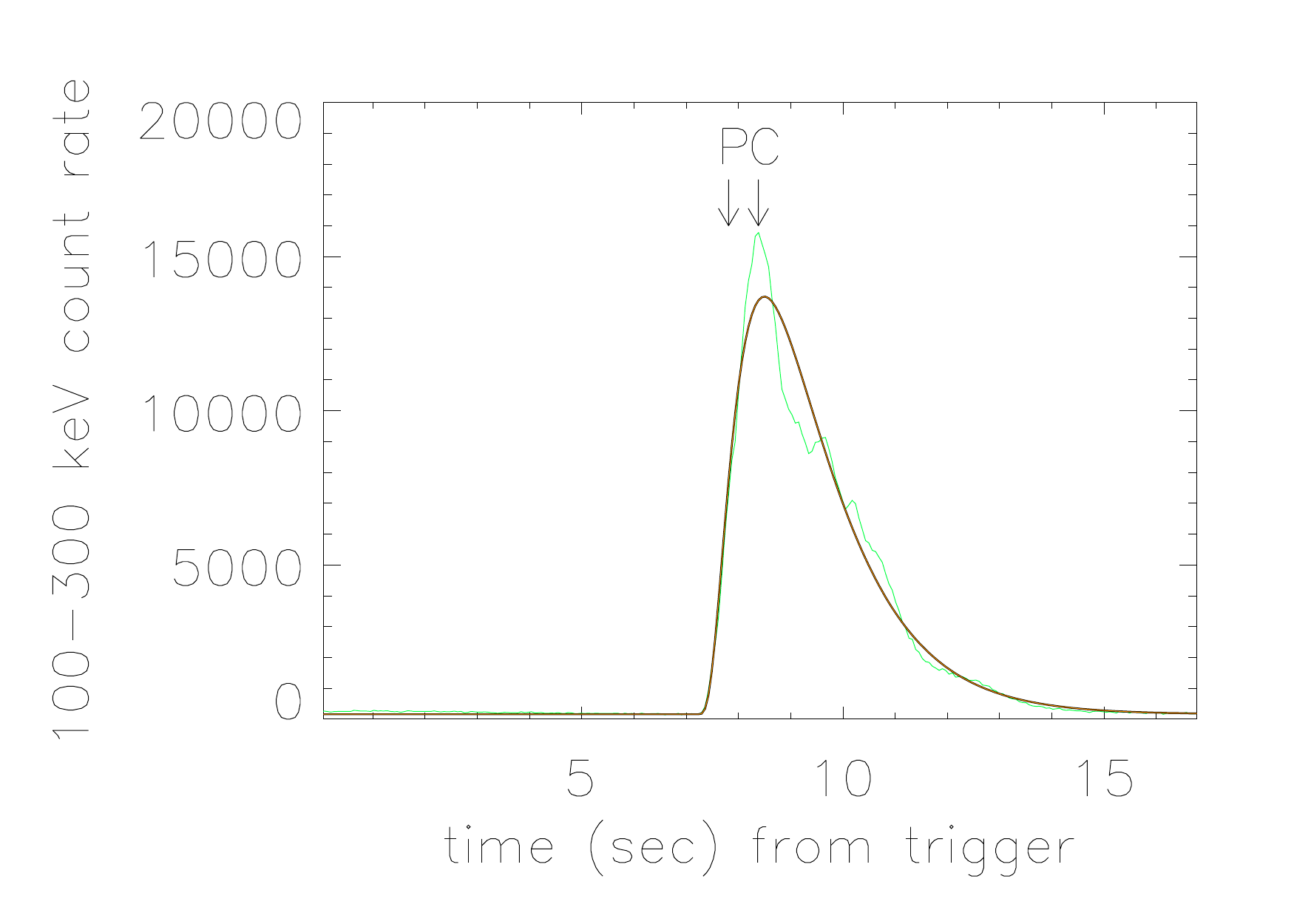}{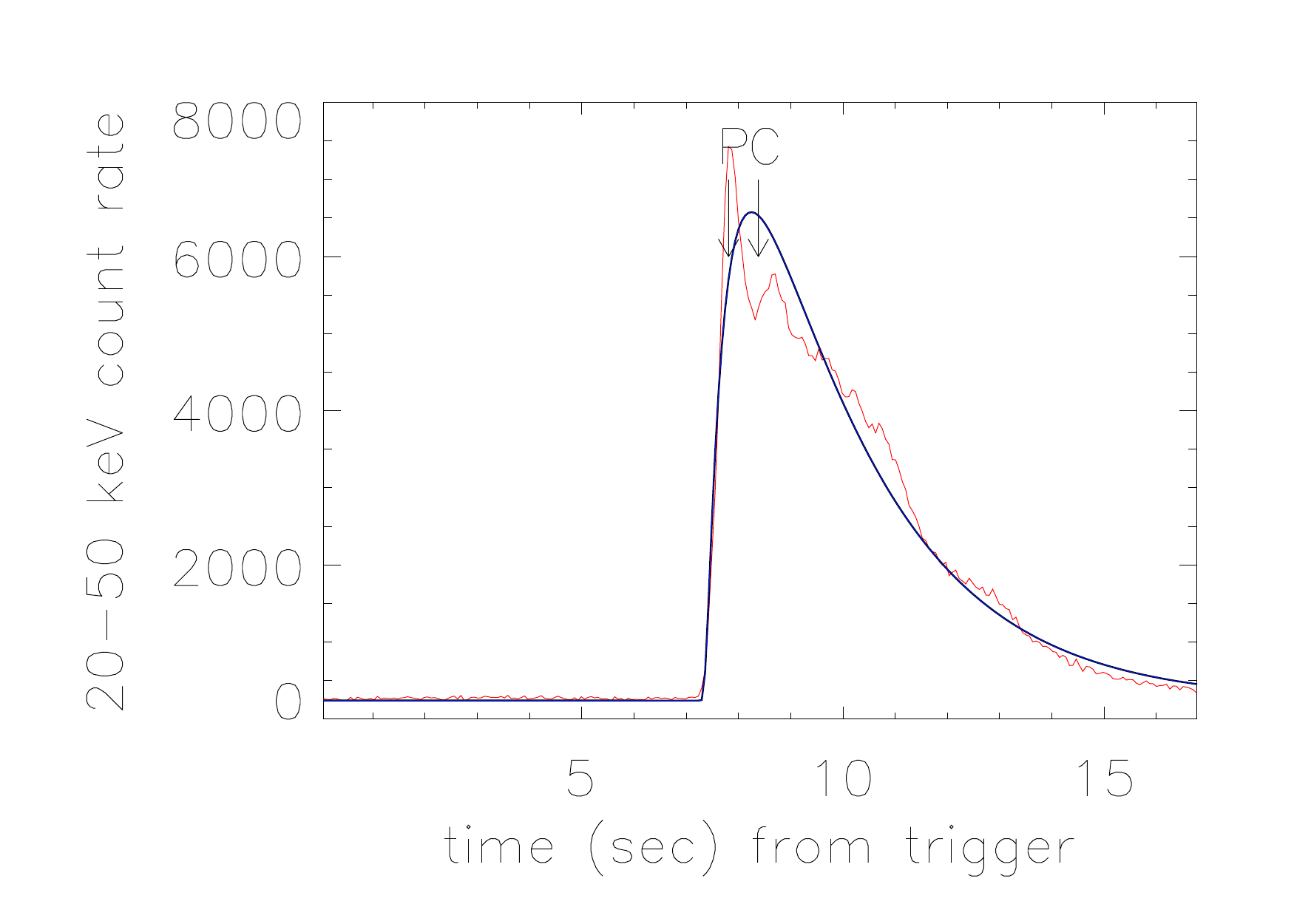}
\caption{Pulse structure of 5614 in channel 3 (100 - 300 keV; left panel) and in channel 1 (20 - 50 keV; right panel). Arrows indicate the times of the brightest channel 1 emission (P, corresponding closely with the time of the precursor peak found from the residuals) and brightest channel 3 emission (C, corresponding closely with the time of the central peak found from the residuals). The precursor peak has a smaller channel 3/channel 1 counts ratio than the central peak, indicating that it is the softer of the two peaks. The precursor peak's unexpected brightness causes the cross-correlation function to measure a negative lag when it aligns the brightest channel 1 emission with the brightest channel 3 emission, even though the right panel of Figure \ref{fig:hr2} shows that 5614 evolves from hard-to-soft. \label{fig:5614nlag1}}\end{figure}

The pulse undergoes a significant re-hardening 
at the time of the fitted precursor peak which it roughly maintains throughout 
the pulse's central and decay peaks. This is similar to what is observed
in pulses 143p2 and 249. However, the
precursor peak of 5614 is not as hard as the central peak.
Even more interesting, the precursor peak 
is brighter in channel 1 (20 - 50 keV) than the
central peak is in channel 3 (100 - 300 keV), which can be seen in Figure \ref{fig:5614nlag1}.
As a result of this peculiarity, the spectral lag calculated between channels
3 and 1 is negative, even though the pulse still undergoes hard-to-soft
evolution with re-hardening at each of the peaks.
In other words, the negative lag is an artifact created when the CCF mistakenly correlates 
the enhanced soft emission of the precursor peak for the pulse's central peak intensity to find a lag.
{\em This type of spectro-evolutionary
behavior, coupled as it is with GRB pulse residual structure, may
explain the existence of many negative GRB spectral pulse lags.}

\subsection{BATSE Trigger 7301}

\begin{figure}
\epsscale{0.50}
\plotone{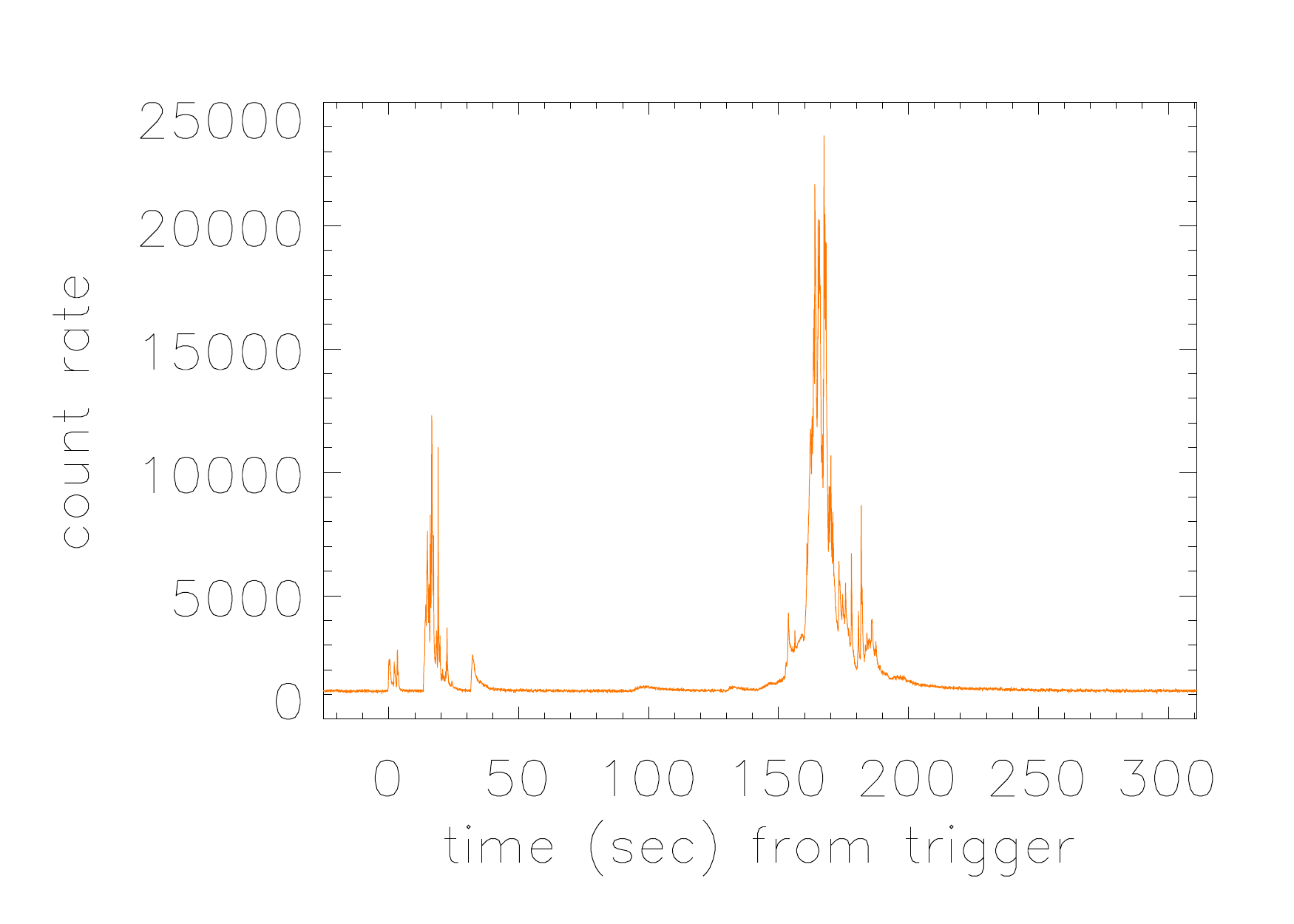}
\caption{Full light curve of 7301.\label{fig:7301lc}}\end{figure}

BATSE 7301's behavior is complex, even when compared to the other complex bursts being analyzed here. 
The complete light curve of 7301 is shown in Figure \ref{fig:7301lc}. In addition to 7301p1 and 7301p2,
the burst contains two faint emission episodes starting $t = 0$ seconds and at $t = 32$ seconds,
low intensity bumps at around $t = 100$ seconds and $t = 135$ seconds, and a long emission tail
following 7301p2. This emission complicates simple explanations of 7301's spectro-temporal evolution.

The duration of 7301p1 depends on whether or not the pulse encompasses two faint emission episodes at $t = 0$ seconds (the three-peaked `comb') and $t = 32$ seconds (the `shoe').  The optimum CCF value of 0.66 occurs for the mirrored residuals if these two faint emission episodes are treated as part of the 7301p1 pulse structure (shown in Figure \ref{fig:mirror3}). This inclusion improves the CCF, which would be 0.58 otherwise.  The value of $s$ measured from the mirrored residuals is $s_{\rm mirror; with}=0.99$ (essentially unstretched) compared to $s_{\rm mirror; without}=0.72$ when the comb and the shoe have been excluded. Each of these measures are in moderate agreement with the value of $s=0.82$ measured from the residual function.

\begin{figure}
\plottwo{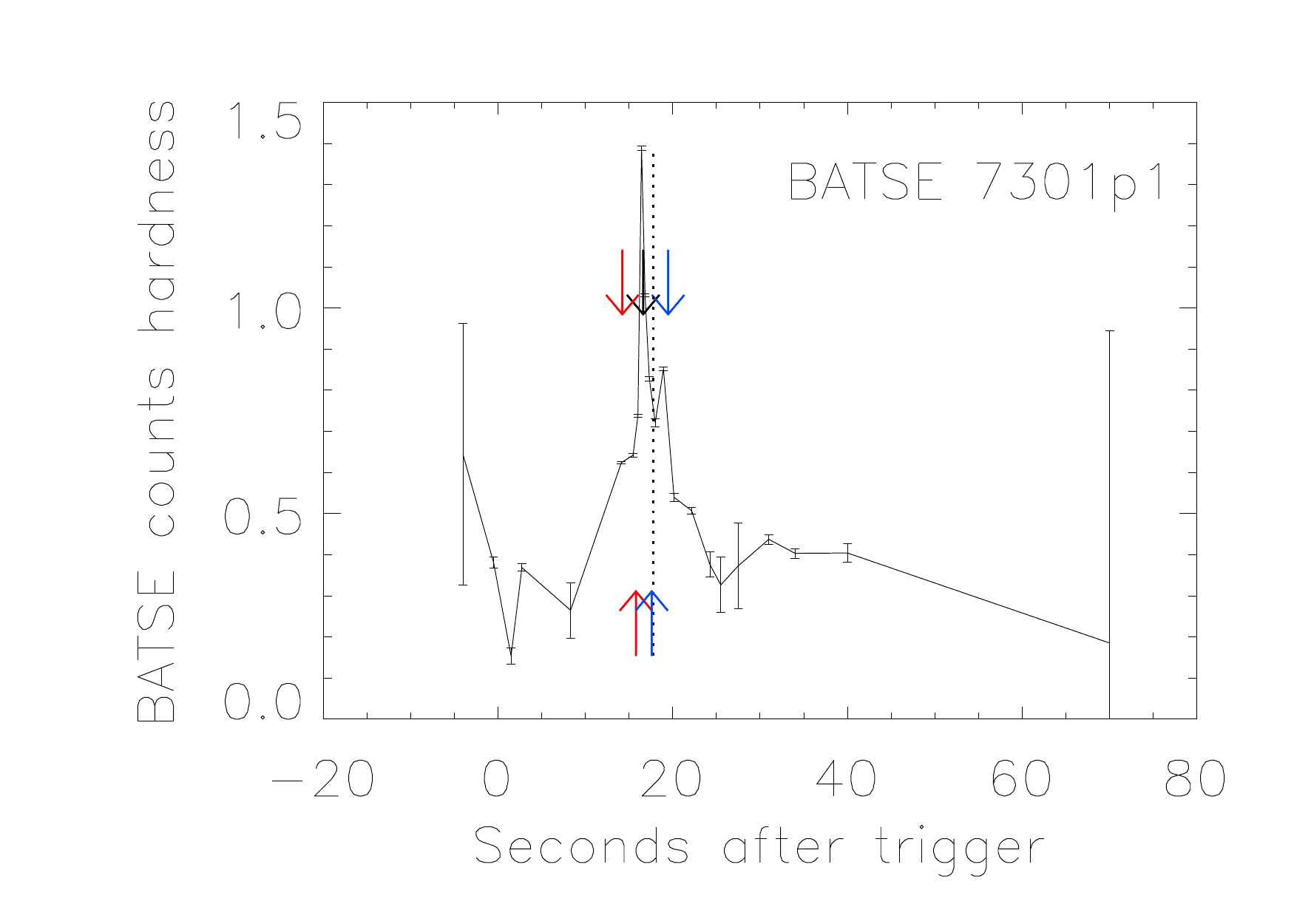}{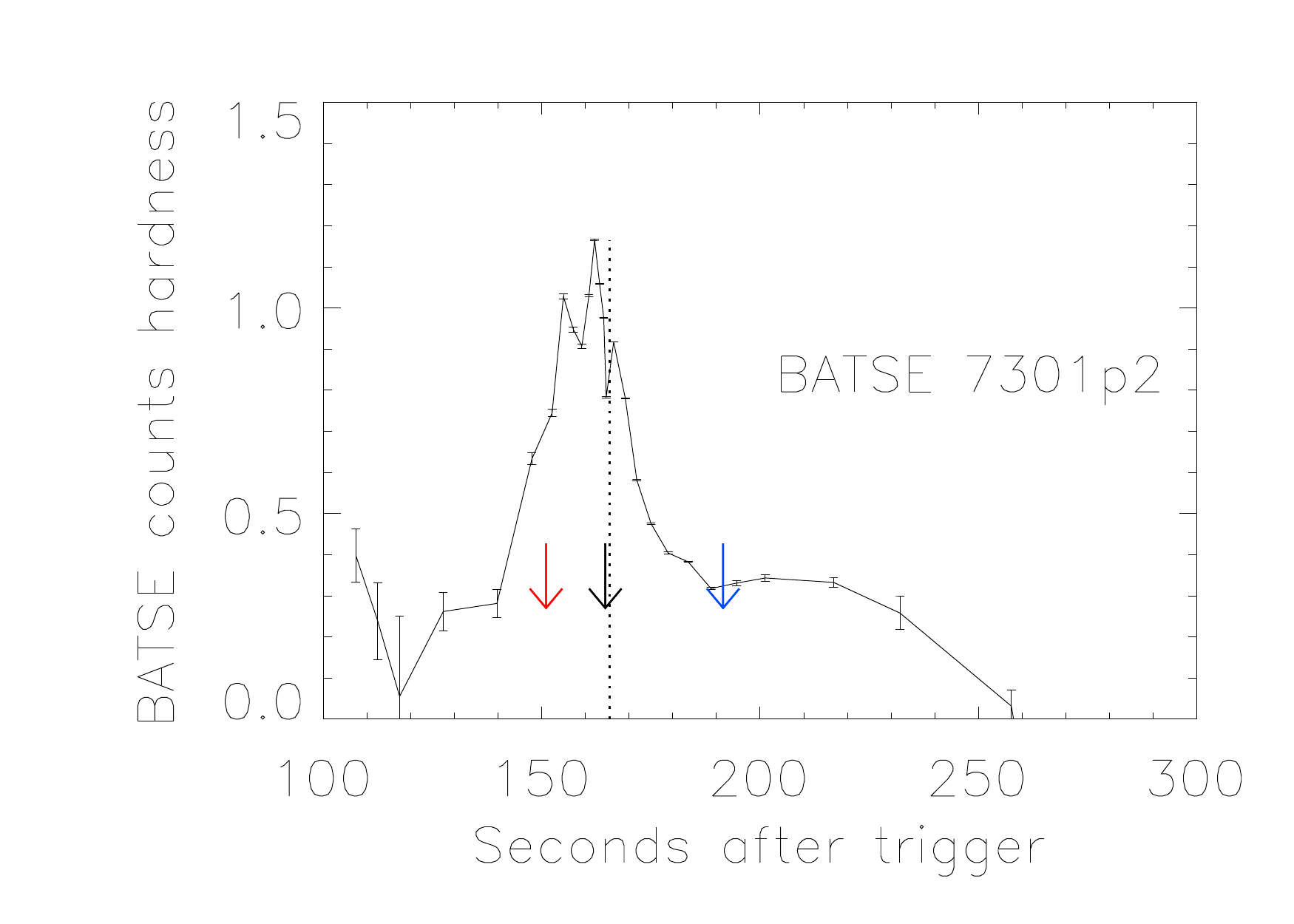}
\caption{Counts hardness ($hr$) evolution of BATSE pulses 7301p1 (left panel) and 7301p2 (right panel).\label{fig:hr3}}\end{figure}

We favor the hypothesis that the comb and the shoe belong to 7301p1. If they do, then the two brightest peaks in the main pulse (at $t = 16.6$ seconds and $t = 19$ seconds in Figure \ref{fig:7301} and occurring 1.2 seconds after $t_{\rm 0;mirror}$ in Figure \ref{fig:mirror3}, left panel) are found to align, as do the two valleys immediately preceding and following them (the deep valley at $t = 15.5$ seconds and the shallower valley at $t = 20.5$ seconds  seen in the left panel of Figure \ref{fig:7301}); these valleys can also be seen 3 seconds after $t_{\rm 0;mirror}$ in the left panel of Figure \ref{fig:mirror3}. Furthermore, it is no longer disturbing that this alignment causes $t_{\rm 0;mirror}$ to occur at a residual function minimum rather than at a maximum, since similar behavior was found in 143p1 and 5614. Finally, although the hardness evolution (shown in the left panel of Figure \ref{fig:hr3}) is consistent with the hard-to-soft spectral evolution of three separate pulses, it is also consistent with the gradual hard-to-soft evolution of a single pulse that re-hardens at each emission peak. Occam's Razor leads us to think that these behaviors represent a single pulse rather than multiple pulses.



BATSE pulse 7301p2 exhibits very strong evidence of temporally-mirrored and stretched residuals. The time of reflection ($t_{\rm 0;mirror} = 165.6$ seconds) occurs at a narrow residual peak which is closely bounded by a pair of mirrored and stretched peaks at $t = 163$ seconds and $t = 167$ seconds (seen 2 seconds after $t_{\rm 0;mirror}$ in the right panel of Figure \ref{fig:mirror3}), and by another broader pair of peaks at $t = 153$ seconds and $t = 186$ seconds (16 seconds after $t_{\rm 0;mirror}$). Additionally, the faint emission at $t = 100$ seconds aligns with the long 7301p2 pulse tail extending out past $t = 200$ seconds; this can be observed from 20 to 50 seconds after $t_{\rm 0;mirror}$.

The hardness evolution of 7301p2 is similar to what is observed in pulses 249, 5614, and 7301p1. This behavior is characterized by weak hard-to-soft spectral evolution punctuated by sustained hard emission during the central `monotonic' pulse. Spectral re-hardening occurs at each intensity peak. Of all the pulses studied in this sample, 7301p2 has the most gradual hard-to-soft evolution, and is thus most comparable to an intensity-tracking pulse.

Occam's Razor leads us to believe that BATSE 7301 contains only two distinct pulses characterized by large and rapid intensity fluctuations. The durations of these two pulses, as measured from their stretched and mirrored residuals, span over 250 seconds. This interpretation is considerably different than what would be made by assuming that a pulse constitutes any emission spike not attributed to instrumental noise. 

Incorporating the comb and shoe features into 7301p1 increases the pulse's duration and makes it similar to that of 7301p2. Pulses 143p1 and 143p2 also have similar durations. Although it is hard to draw conclusions about relative pulse durations from only two GRBs, we note that this suspected relationship might warrant further study if other bright, structured, multi-pulsed GRBs can be found and analyzed.


\subsection{Are these Pulse Characteristics Ubiquitous?}

The six bright GRB pulses analyzed here exhibit measurable spectro-temporal features that are not likely to have been washed out by instrumental effects. If these properties are representative of properties that would be found in fainter pulses, then these are probably defining GRB pulse characteristics that can be used to constrain GRB physics. These characteristics suggest that GRB pulses exhibit smooth monotonic components that evolve from hard-to-soft and structured components that undergo time-reversed and stretched variations. Monotonic component asymmetry occurs in tandem with compression and stretching of the time-reversible structured component.

We do not know if the time-reversed and stretched residuals found here apply to all GRB pulses. The greatest
difficulty in ascertaining how generic these characteristics are relates to our inability to
thoroughly characterize GRBs at low $S/N$. At low $S/N$ it is difficult
to delineate complex single-pulsed GRBs from multi-pulsed ones,
and measurement of the characteristics that might otherwise prove helpful
are buried in statistical uncertainties. We also note that many
faint GRBs have different properties than those found in bright GRBs.
For example, \cite{nor00} show that some faint GRBs have long CCF data lags and soft spectra, but they 
also show that there are no bright versions of these bursts. Our analysis suggests that
these characteristics cannot be created simply by observing a bright, hard,
short-lag burst at lower $S/N$. 

We have performed a cursory exploration of a few complex BATSE GRB pulses, and
have found more of these types of events. For example, again using the cross-correlation function
to describe fit quality, we find that triggers 1008 ($S/N = 22.0$, $CCF = 0.64$),
1114 ($S/N = 29.2$, $CCF = 0.54$), 3954 ($S/N = 32.4$, $CCF = 0.50$),
5417 ($S/N = 39.0$, $CCF = 0.40$), and 6422 ($S/N = 36.5$, $CCF = 0.43$)
all show evidence of time-reversible and stretched structures even though they
are observed at lower $S/N$ than the six pulses in our sample. That pulses with these
$S/N$ values show complex and possibly time-reversible and stretched structure is consistent with the 
results shown in Figure \ref{fig:SNhist}; if time-reversed and stretched residuals
are characteristic of all GRB pulses, then as many as 100
BATSE GRBs should have been observed at large enough $S/N$ to exhibit these characteristics. 

\cite{hak18} have recently used 4-ms resolution BATSE TTE (time-tagged event) data 
to demonstrate that short GRBs also exhibit pulse structure consistent with this paradigm.
Roughly 40 GRB pulses in the TTE Pulse Catalog have complex structures
that might exhibit time-reversed and stretched residuals. However, few of these
pulses have $S/N > 15$. The $S/N$ of these pulses can be increased by using coarser time binning
(say, 64 ms). However, using coarser temporal binning also washes out pulse structure
by decreasing temporal resolution (see also \cite{bur14}).

Time-reversible and stretched residual structures are likely to be measurable in GRB pulses observed by other
satellite experiments, such as Swift and Fermi's GBM experiment. Swift and GBM have better temporal resolution
than BATSE. However, Swift operates at lower energies where the sky is noisier, and 
GBM has a smaller surface area than BATSE. Both of these
characteristics reduce $S/N$. A preliminary analysis of Swift GRBs
has yielded few with complex structures among hundreds with simple structures. However, simple single-pulsed GRBs 
were favored in the selection of that sample, so work on Swift data is ongoing.
We have begun studying GBM data in conjunction with members of the GBM team; we do not yet have estimates
of how many GRB pulses might exhibit time-reversible and stretched residual structures, but we are encouraged by GBM's low systematic noise.

\begin{figure}
\epsscale{0.50}
\plotone{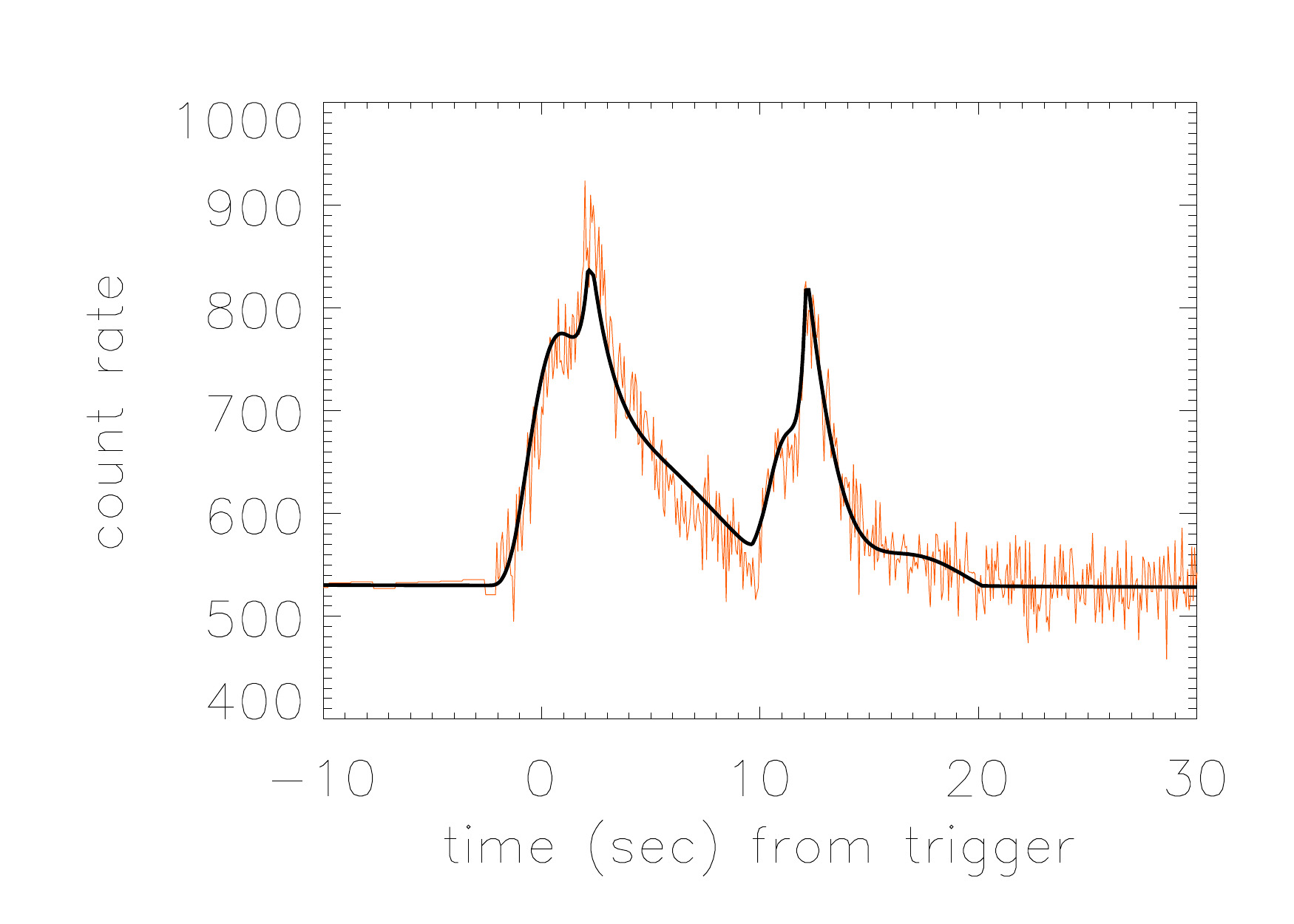}
\caption{Fitted light curve of BATSE trigger 148, showing the delineation of two overlapping pulses.\label{fig:148}}\end{figure}

Some GRBs characterized by overlapping emission episodes
are found to contain distinguishable yet overlapping pulses.
Knowledge of the residual function can
help us in these cases. As an example we show fits for BATSE trigger 
148 (Figure \ref{fig:148}). This burst can be
successfully fitted by two overlapping pulses; 148p1 is a blended pulse
with $S/N = 17.1$ while 148p2 is a blended pulse with $S/N = 12.9$.
We note that this technique only works for
a small subset of overlapping BATSE GRBs that are bright with
bright residual structures, and thus cannot generally help
resolve confusion over the number of pulses in a faint GRB.

\section{Smoke and Mirrors: Defining GRB Pulses}

``Smoke and Mirrors'' refers to the obscuration of a truth
using misleading or irrelevant information. 
Our own study has inadvertently obscured the truth by stating that GRB
pulse durations are not affected by $S/N$.
Although we demonstrated that {\em fitted} pulse durations can be accurately 
recovered over a range of $S/N$, we subsequently 
showed that these monotonic components
often have shorter durations than their associated residual structures. Unlike
the monotonic pulse components, these faint residual structures can disappear into the background
at low- to moderate-$S/N$ when monotonic pulse components do not.
Furthermore, faint residual ``bumps'' observed in some GRBs can be
misinterpreted as being separate monotonic pulses.
Treating these structures as separate pulses
misleads us to think that pulses are shorter, more common, and 
more widely separated than they actually are.

Should time-reversed and stretched residuals be used to redefine GRB pulses?
Revising the pulse definition would certainly be an improvement over
the current situation, where investigators explain GRB physics in terms of 
a wide range of different and often incompatible pulse definitions. This
incongruity makes it nearly impossible to tell whether or not two models are
consistent with one another, or even if they are both consistent with the data.
We suspect that it is too early to ask the GRB community to accept a new 
pulse definition. First of all, such a definition would only be viable when applied
to bright (high $S/N$) bursts; bursts of intermediate intensity only exhibit
the triple-peaked residual structure while faint bursts have the appearance of monotonic bumps.
Second, the definition might be instrument-dependent; more work needs to be
done to clarify selection biases introduced by satellites with different sensitivities
working in different energy ranges. Third, accepting a new definition 
prematurely might prevent us from recognizing variations of this behavior or perhaps 
other behaviors belonging to different classes of objects. Without formally recommending
a new GRB pulse definition, we ask that other researchers be aware of 
the non-monotonic nature of most GRB pulses, and of the pulse
characteristics generally associated with non-monotonicity, so that
results of various analyses can be more readily compared.

\section{Interpretation}

\subsection{Physical Explanations for the Time-Reversed Residual Structures}

Despite the results of our analysis, we find it hard
to imagine a physical mechanism capable of producing
structured, time-reversible, stretched residuals.
Since we think it {\bf is} unlikely that the temporal mirroring and stretching found in GRB pulse structure is 
due to a time-reversed process, we suspect that it must instead indicate a {\em kinematic} process. 

Natural objects rarely exhibit time-symmetric light curves,
but kinematic mechanisms explain the ones that do. 
Simple time-reversed light curves are produced by {\em gravitational 
microlensing} when a distant source passes behind a compact foreground object 
\citep{pac86}. When the foreground object is a binary, microlensed light curves can
exhibit any of a wide range of complex symmetric temporal structures ({\em e.g.,} see \cite{lie15}).
Although we do not suspect that time-reversed and stretched GRB light curve structures are produced by
gravitational microlensing, we note that the black holes formed during the GRB
emission might be capable of providing the necessary environmental conditions.

We suspect that
kinematic motions causing mirrored and stretched
residuals are directly related to jets and the expansion accompanying GRB explosions.
Furthermore, since the mirroring is often coupled to both
pulse asymmetry and evolving spectral 
hardness, we suspect that it is an indicator of how and when energy is 
fed into a GRB system. 

Current GRB models do not naturally explain the existence of time-reversed and stretched
residuals. Although time-reversed properties and stretching went into the development of the \cite{hak14} residual function, 
these characteristics have largely gone unmentioned in the literature other than by the authors, who
attributed them to forward and reverse shocks. This interpretation now seems 
unlikely given that the residual features extend well beyond three simple peaks.
To our knowledge no standard GRB model has specifically predicted the 
existence of time-reversed GRB pulse structures.

The mechanism by which GRB prompt emission is produced
is still largely unknown. 
The {\em standard model} states that it
originates from material ejected relativistically during stellar
black hole formation \citep{ree94}. The rapid variability
of GRB prompt emission indicates that
it arises internally from within the expanding material 
\citep{sar97}. Following the recent \cite{dai17}) review,
we list three general mechanisms by which 
GRBs might produce pulsed emission:
\begin{enumerate}
\item Synchrotron shocks \citep{mes93,tav95}. Mildly relativistic shocks
are thought to occur when faster jet ejecta catches up with slower ejecta
\citep{ree94,kob97,dai98}; synchrotron shocks 
occur in optically thin regions and models predict smoothly-varying
light curves for which pulse duration
reflects variations in the intensity of the central engine
\citep{kob97,dai98,pan99}. 
\item Photospheric emission ({\em e.g.,} \cite{pac86a,goo86,pee08,bel11}). 
A quasi-thermal spectrum is produced
when the ejecta becomes transparent to its own radiation;
these spectra can be made non-thermal with the addition of a low-energy tail produced
by synchrotron emission and a high-energy tail produced by comptonization 
({\em e.g.,} \cite{tho94,ree05,pee06,bel10,ryd11}). 
\item Magnetic reconnection. If magnetization is large and 
emission occurs in an optically thin region where shocks
cannot develop, then accelerated electrons should be able
to radiate nonthermal or quasi-thermal spectra
\citep{spr01,dre02,gia05,zha11}. 
\end {enumerate}
Very little is known about the detailed microphysics capable of producing
emission in any of these models.

Pulse durations and
separation times can provide valuable clues to understanding GRB kinematics. Among
other things they 
can be used to determine whether the progenitor
undergoes a long-lasting or an impulsive ejection process. 
The relative rarity of GRB pulses per burst, pulse durations as defined by
residuals that extend beyond the monotonic component, 
the existence of time-reversed and stretched residuals, 
and hard-to-soft pulse spectral evolution all place surprising and potentially 
strict constraints on kinematic characteristics of
GRB models. 

Based on our findings, we propose four general kinematic models
to demonstrate how these observations impose constraints. \newcounter{bean} We
recognize that not all kinematic models are compatible with the
three variations of the standard model described above.

\begin{description}
\item[Model 1] Radiation is emitted when an impulsive impactor
encounters obstructions in a medium along its path, followed by an abrupt course 
reversal that causes it to pass through the medium in reverse order. In agreement with
the standard model, the impactor-cloud interactions must be mildly relativistic because both the
obstructions and mirror must move toward the observer faster than the impactor.
\item[Model 2] An impulsive impactor moves through a 
symmetrically-structured medium, causing radiation occurring
at the beginning of the interaction to be in reverse order to that occurring 
at the end of the interaction. The structure on the central engine-facing
side is probably compressed relative to the structure on the other side.
\item[Model 3] Sustained radiation is produced as a 
symmetrically-structured impactor moves through a simple medium.
Despite general structural symmetry, the impactor is probably compressed
on its leading face as it moves away from the central engine.
\item[Model 4] Waves are generated as material moves through a
symmetrically-distributed non-uniform medium. Through some unknown 
process the waves generated at the beginning of the event are similar but 
compressed relative to those generated at the end of the event.
\end{description}

These models all provide convenient explanations not only for the residual
structure but also for hard-to-soft GRB pulse evolution. 
The highest-energy emission should occur at the
beginning of each pulse, when the greatest energy is
available for kinematic deposition, and should be followed by
emission indicating that energy losses have occurred.
GRB pulse shape asymmetries further support these models,
by indicating either that energy has been lost in the emission process
or that an abrupt change in Doppler shift has occurred.

We explore these models below in more detail.

\subsubsection{Model 1: Reflected Motion of Impactor}

\begin{figure}
\epsscale{0.50}
\plotone{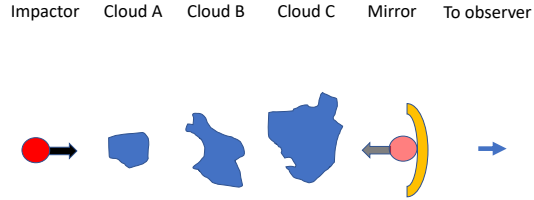}
\caption{A model of temporally-reversed pulse structure. An impactor (red circle) produces variable radiation as it travels through clouds (blue), and has its motion physically reversed by a mirror (yellow), so that it travels through the same material in reverse order (pink circle). \label{fig:model1}}\end{figure}

We first consider a clump of moving particles (presumably electrons and ions) or some 
condensed wave phenomenon such as a soliton traveling at high velocity outward
from the source that emitted it, in concordance with the standard model of GRB jet outflow. 
We hypothesize that this {\em impactor} produces emission as it passes through clouds
of material.
The time over which light is emitted depends on 
the size and density of each cloud as well as on the velocity and size of the clump/soliton.
The mirrored temporal pulse light curve features can be explained if 
the clump first emits as it travels through clouds A, B, and C,
then emits again as it travels through the clouds in reverse order (C B A). 
This model is shown in Figure \ref{fig:model1}.

Note that this model produces the proper temporal signature if one impactor travels through 
multiple clouds, but not if multiple impactors (1 and 2) travel through
these same clouds and are reflected to travel back through the clouds
in reverse order. This scenario produces events in a non-reversible 
chronological order (A1 A2 B1 B2 C1 C2 C1 C2 B1 B2 A1 A2) rather than in the
reverse chronological order (A1 A2 B1 B2 C1 C2 C2 C1 B2 B1 A2 A1) and is thus not viable.
A non-reversible chronological order is obtained for any number of impactors
greater than one.

The mechanism for reversing the impactor motion is unknown, but
might be due to a soliton reflecting off of a region characterized by higher
opacity ({\em e.g.,} \cite{sin06}) or to charged particles reflecting off a
region associated with a high magnetic field density, such as
a magnetic mirror or magnetic bottle ({\em e.g.,} \cite{bar18}).

Our kinematic model can be related to the standard GRB model if
we assume that the mirror results from increased density
({\em e.g.,} \cite{lev13}) and/or tightly-wound
magnetic fields in the head of a GRB jet
({\em e.g.,} \cite{lyn96}),
and that the clouds are density or magnetic
structures formed within the jet. These clouds would move toward
the head of the jet as the jet head
plows into an external medium
({\em e.g.,} \cite{har98,har00,agu01,kat04,por13,mar15,mar16}).
The impactor could be a superluminal 
event generated by the central engine. 

Interactions between the impactor and the clouds cannot be
seen in emitting regions that move away from the observer,
due to relativistic motion. However, all 
forward and reverse interactions can be observed if the emitting regions
(clouds and mirror) are moving towards the observer.
Thus, the impactor could change its direction of motion
in the mirror's reference frame while still be moving towards
the observer. Clouds catching up with the impactor would
cause another round of radiation to be emitted. The temporal compression of the residuals
on the pulse rise would indicate relativistic blueshifted motion toward the observer
prior to reflection while the stretching observed during the pulse decay 
would indicate less-pronounced blueshifted motion toward the observer following reflection.

A gradient of cloud structure, density, and/or magnetic field close to the mirror
might also explain increased residual amplitudes close to the time of reflection. 
Increased interactions should lead to increased residual amplitudes.

In a GRB with multiple pulses, this kinematic model requires either that two or more impactors
strike a single mirror at different times, or that a single impactor somehow
interacts with two or more mirrors as it moves through the jet. The single mirror/multiple impactor
hypothesis can be explained in hypernova models of long GRBs,
where multiple impactors might be emitted during the stellar collapse
process ({\em e.g., }\cite{mac99}), or in the cannonball GRB model ({\em e.g.,} \citep{dar06}),
but it is not consistent with merging neutron
star models of short GRBs, where repeated explosions from the
central engine seem less likely.
The single impactor/multiple mirror hypothesis seems unlikely in both long and short GRB scenarios, 
as the number of pulses is limited by the presence of a mirror that completely reflects ejecta. 
This model only works if multiple mirrors exist within the jet and if these mirrors are accessible
to the impactor. We do not know how this situation might arise;
perhaps isolated bubbles within the jet  exist as a result of magnetic
reconnection \citep{kat04}, and perhaps part of the impactor is somehow transmitted 
forward through each mirror in addition to the reflected part.

\subsubsection{Model 2: Unaltered Motion of Impactor}

\begin{figure}
\epsscale{0.50}
\plotone{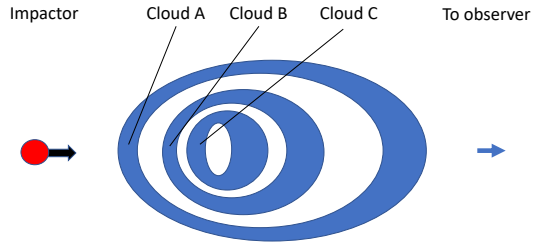}
\caption{A model of temporally-reversed pulse structure. An impactor (red) 
produces variable radiation as it travels through axisymmetric, stretched 
clouds (blue). \label{fig:model2}}\end{figure}

A physical mirror is not required to explain the stretched and time-reversible pulse residuals.
A more direct explanation is that
a simple impactor moves constantly in one direction (presumably toward the observer)
rather than one undergoing reflection and changing its direction of motion.
In this case, interactions between the impactor and the clouds might produce
the observed light curve structures if the clouds have some sort of
radial bilateral symmetry
({\em e.g.,} if they are distributed with cylindrical or spherical symmetry along the 
particle path). The increased residual amplitudes near the time of reflection
might indicate an increased density or cloud size toward the center of the cloud
distribution. This model, like the previous one, thus presupposes that structure exists within the jet.
The model is shown in Figure \ref{fig:model2}.

Non-uniform flow is expected in an expanding jet due to instabilities in the accretion disk 
as well as to dynamics of the spout through which the jet flows \citep{mac99}.
These instabilities likely depend 
on the angular momentum of the star, the accretion disk characteristics,
the stellar mass, and the magnetic field. 
Variability in jet outflow can be caused both by instabilities
at the central engine and those generated as the material
passes through the star \citep{mor10}.

Without knowing the form of GRB
jet structure, it is not obvious why this
proposed structure might cause the residual emission to stretch
following the time of reflection (which corresponds to the center of the radial cloud distribution).
It is possible that the motion of the impactor might compress the 
structure as it moves through it. However, any corresponding
deceleration of the impactor should be gradual rather than abrupt as observed.

Note that again two or more impactors cannot produce the residuals
seen in a single pulse, because the order of events is then irreversible.
Suppose that interactions between
the first impactor and symmetrically-distributed clouds A and B produce 
emission A1 and B1, respectively, and that interactions between
the second impactor (moving closely behind the first) and the clouds 
similarly produce emission A2 and B2. 
The order of the episodes will not reverse once the impactors pass the
center of the cloud distribution, and is thus not viable. In other words,
the order of emission will be A1 A2 B1 B2 B1 B2 A1 A2
or something equally irreversible depending on the difference in
time when each impactor interacts with the clouds.

Multi-pulsed bursts can be explained if each impactor produces its own pulse.
In other words, two or more separate impactors can pass through the same set of clouds
and produce different pulse/residual properties if the time between impactor
arrivals at the cloud distribution is much longer than the time it takes the impactor
to travel from cloud to cloud. Similarly, the reversible emission can be 
produced in separate pulses if one impactor passes through
two isolated cloud distributions that are located long distances from one another.

\subsubsection{Model 3: Bilaterally-Symmetric-Shaped Impactor}

It is possible that the bilaterally-symmetric physical structure
belongs to the impactor rather than to the clouds. In order to produce
the time-reversed structure this model requires the interaction to take
place within a single cloud, rather than with a distribution of clouds. 
The impactor could be a clump of material containing a
bilaterally-symmetric density distribution, a bilaterally-symmetric 
distribution of streaming clumps of material, 
or a stream of bilaterally-symmetric distributed solitons. 
The mechanism that produces this impactor structure 
could again relate to instabilities in the accretion disk
and/or in the jet spout \citep{mac99}.

\begin{figure}
\epsscale{0.50}
\plotone{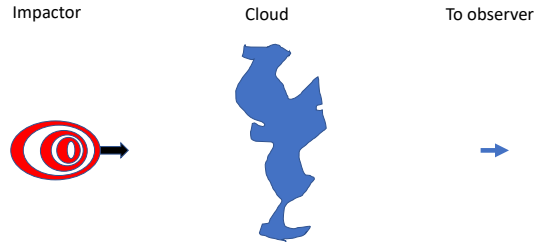}
\caption{A model of temporally-reversed pulse structure. A bilaterally-symmetric impactor (red) produces variable radiation as it travels through a cloud (blue). \label{fig:model3}}\end{figure}

The impactor distribution might also be produced magnetically
through the linkage of a series of magnetic bottles or sheets
in a process tied to jet collimation. 
Symmetry in the physical structure of an impactor might have to be
matched by a uniformity of structure in the impacted material
in order to produce the time-reversible pulse structure.
Stretching of the wave would almost require an asymmetric
impactor structure, with clumps being compressed on the 
leading side of the moving impactor and stretched on the 
trailing side (perhaps indicating compression of the impactor's
magnetic field in the direction of motion). This model is demonstrated
in Figure \ref{fig:model3}.

This model provides a simple explanation for the observation of
multiple GRB pulses. Either multiple impactors are ejected from the
central engine (which is still a potential problem for merging neutron stars
and short GRBs) or one impactor produces multiple pulses as it travels through multiple clouds.

Kinematic models 2 and 3 are similar to one another,
differing primarily in whether the physical structure responsible for producing
the time-reversible and stretched residuals is found in the
lower-velocity material (Model 2) or the higher-velocity material (Model 3).
This is a minor difference from our perspective, since we do
not presently know the origins of either the clouds or the impactor. 

\subsubsection{Model 4: Wavelike Instabilities Caused by Impactor}

A time-reversible wave-like residual structure might also be
produced from oscillations created when an impactor passes through a
cloud medium. If the characteristics of the medium are
spherically or cylindrically symmetric (for example, if the cloud
density decreases towards the edges), so that they appear to be
bilaterally symmetric to the observer, then the characteristics
of the oscillations ({\em e.g.,} frequency, amplitude) could
change as the impactor passes
through the cloud. 
Stretching of the
residual wave might be caused by a factor such as a change in the
cloud's density gradient. {\em Note that this model, like the previous two,
preferentially favors interactions of a single impactor
with one or more clouds.} This appears to be a constraint
imposed by the time-reversed residual structure. We
do not know how these waves could be formed.

\subsubsection{Other Models}

Other astrophysical models seem improbable given the data. 
For example, variable high-energy lightcurves are often associated with
flaring, orbital motion ({\em e.g.} compact binary stars), or rotation ({\em e.g.} pulsars).
The wavelike residual structures do not seem to represent any sort of
periodic motion (especially since the periods of the wavelike residual structures
decrease before the time of reflection then increase after it). Nor do the
patterned temporal sequences found in the residuals seem indicative
of any sort of flaring process that would not involve orbital motion or rotation.

\subsection{Comments on Radiation Mechanisms: Are the Residuals ``Waves" or ``Sub-Pulses?"}

The residuals produced by subtracting the monotonic pulse
model from the observed light curve produce a wavelike function that
exceeds the monotonic light curve half the time and is less than it
otherwise. If we want to model this behavior using
standard radiation mechanisms, then half the time this behavior must
represent emission in excess of the monotonic pulse
whereas otherwise it must represent some process
that also removes photons from the line-of-sight ({\em e.g.,} absorption,
scattering, polarization).

It is simpler to imagine a GRB physical process
that only produces photons than one that rapidly
and systematically alternates between producing photons and removing them
from the line-of-sight. 
Our Occam's Razor assumption
that GRB emission should be explained by the fewest number of
pulses led to the characterization of ``pulses'' as monotonic components
overlaid by structure, which we found to be time-reversed and stretched. 
However, applying Occam's Razor instead to GRB radiation mechanisms
would suggest that ``pulses'' should be explained by emission processes alone
rather than by emission processes combined with absorption,
scattering, and/or polarization; 
this assumption requires each pulse to be composed of many smaller sub-pulses
in order to account for the large and rapid intensity variations
found in GRB emission episodes while always representing positive flux. Using this 
hypothesis that GRB light curves are formed only from monotonic sub-pulses might not
allow us to find the time-reversed and stretched light curve structures.
Is there some way to determine
whether time-reversed and stretched residuals are an inherent characteristic
of GRB pulse light curves or whether they are simply artifacts of assumptions
and therefore of the fitting process?

In order to answer this question, we need to revert to simple monotonic pulse
models rather than a ``structured'' pulse model. The literature is full of monotonic
pulse models, each designed only to emit photons. 
In addition to the \cite{nor05} model used in this study, these models have taken
many functional forms, including pulses with power law rises and decays 
({\em e.g.} \cite{nor96, lee00a, lee00b}), 
power-law rises and exponential decays ({\em e.g.,} \cite{jia05}), 
Gaussian rises and exponential decays ({\em e.g.,} \cite{ste96}), 
Gaussian rises and decays \citep{bha12}, and 
more complex rise and decay functions ({\em e.g.,} \cite{koc03}). 
Each of these models by definition treats 
any structure that cannot be due to noise as a separate pulse.
As indicated in Section \ref{sec:analyze}, we have previously applied
this approach ourselves, and have found that making this assumption forces each 
GRB in our sample to be composed of dozens of individual sub-pulses.

One of our sampled GRBs exhibits a light curve that is nearly time-reversible without
any pulse-fitting or stretching. The decay portion of BATSE 249's
light curve (right panel of Figure \ref{fig:249}) contains a time-reversed memory
of events that happened during the rise portion of the light curve (we previously
described this as being like a series of cutout paper dolls). 
This reversibility is present regardless of whether the emission episode
is composed of a single structured pulse or many monotonic sub-pulses.
However, the features responsible for making the time-reversibility recognizable
are themselves structured because a strict monotonic interpretation would
require almost every recognizable feature to be subdivided 
into multiple sub-pulses.
The recognizable structures are not perfect mirrored images of one another, but
they are more recognizable as reversible structures than they would be
as sub-pulses, as a structure on the pulse rise will likely be deconstructed into
a different number of sub-pulses than the corresponding structure on the pulse decay,
due to the slightly different amplitudes and shapes of these non-identical features.
Since time-reversibility is present in 249 regardless of whether or not
the rise and decay portions contain equal numbers of matching sub-pulses,
we conclude that the assumption of monotonicity is not required
for time-reversibility to be present.

Time-reversibility is not as obvious in the light curves
of the other five sample GRBs as it is in 249. It would be somewhat visible in 143p2 if the
pulse's rise was stretched to have a similar duration as the decay. In other words, asymmetry plays a role
in disguising 143p2's time-reversibility, and stretching the rise side
by a fixed amount aligns rise side features with corresponding
decay side features. However, even after stretching the rise side, 
some pulses do not exhibit temporal symmetry and require the
time of reflection to be moved away from the pulse peak. These pulses also need to have
flux subtrracted from the continuum underneath them until they match. 
A combination of stretching,
flux-subtraction, and temporally offsetting removed flux from the center
of the temporally-matched features seems generally sufficient
to identify time-reversed structures.
It is interesting to note that the amount of flux that should be removed corresponds to the flux
found in a monotonic pulse, that the asymmetry of this pulse corresponds
to the amount of stretching needed in the residuals, and that the alignment
of the mirrored residuals is close to the peak of the fitted pulse.

We made a choice to fit emission episodes using a single monotonic pulse model
based on an Occam's Razor argument. Our choice was not based on
the expectation that the residuals would produce a 
time-reversed and stretched structure. The time-reversed residuals
belong to the structured component of the emission episode rather than to the monotonic component;
it is almost impossible that we could have fine-tuned any monotonic function
to produce these residuals if they had not already been present. 

Perhaps these features represent some combination of
structured sub-pulses rather than a wave (this model
splits the difference between simple monotonic and
structured monotonic pulse models).
Demonstrating that structured sub-pulses somehow
combine to form a time-reversible pulse still requires
some pulse continuum to be excluded and removed. 
Going through the same sort of detailed flux balancing and removal process 
described above for simple monotonic pulses might
make it possible for us to recover a time-reversible pulse composed of 
many structured monotonic sub-pulses, but the process would be complex, burst-dependent, and would
involve carefully balancing all the sub-pulse properties as they are being extracted.
The current process is much simpler. Occam's Razor states that we should
accept our original simple assumption of one pulse per emission episode,
rather than something more complicated.
{\em We thus believe that the time-reversed and stretched residual
structure is not an artifact of the fitting procedure, and that the
radiation mechanisms required to explain GRB pulse physics require
some combination of emission, absorption, scattering, and/or polarization.}

\section{Conclusions}

We have shown that instrumental characteristics are quite capable of smearing out inherent GRB pulse structure,
and of reducing complex structures to both the triple-peaked pulse structure found at moderate $S/N$ and the simple monotonic pulse structure found at low $S/N$. It is possible that all GRB pulses
have complex structures, but these structures can only be studied in GRB pulses observed at high $S/N$.
Basic properties of the monotonic part of the pulse such as duration, asymmetry and lag can still be reliably extracted from
faint pulses using the \cite{nor05} pulse model. These properties do not entirely or adequately characterize GRB pulses without the addition of the \cite{hak14} residual model. 
However, as resilient to $S/N$ variations as these monotonic pulse measurements are,
they do not entirely or adequately characterize GRB pulses,
which can have faint emission extending from long before the monotonic portion of the pulse
begins until long after it ends. 

We have chosen to avoid the smearing of intrinsic pulse structural properties caused by 
instrumental noise by studying a sample of bright, complex BATSE GRBs. Despite their complexity,
we find that these pulses share behaviors that can potentially be used to characterize
all GRB pulses. The most surprising result is that the residual structure, defined as the 
non-monotonic pulse component, is characterized by complex yet {\em time-reversible} behavior. 
This behavior can encompass multiple structures, can span a time considerably longer than
the duration of the monotonic component, and can have amplitudes similar to that of the monotonic component.
Each pulse appears to have only one time-reversible residual structure,
indicating that it may be a defining characteristic of a GRB pulse.

Pulses generally evolve in a hard-to-soft manner. However, they undergo spectral re-hardening
at the times of each residual peak. Some pulses undergo an extended re-hardening that
lasts through much of the monotonic component. In at least one case, we 
demonstrate how this re-hardening is responsible for the phenomenon of negative spectral lags,
even though the pulse containing it evolves from hard to soft.

The predictable behaviors of structure in GRB pulses allow us to more clearly
identify pulses in GRB light curves. As a result, we find that many GRBs 
likely contain only a few pulses, even if their structural appearance
makes it seem otherwise. We suggest that caution must be applied
when interpreting the number of pulses and the characteristics of
these pulses as they are used in pulse population studies.
For example, \cite{rrm00} have concluded that internal shocks rather
than external shocks are responsible for GRB pulses based on the 
relatively constant pulse durations found in multi-pulsed GRBs.
From our perspective, the \cite{rrm00} pulse sample is predominately
composed of complex single- or double-pulsed GRBs, rather than
multi-pulsed GRBs, invalidating the reasons for their conclusions.
Short GRBs \citep{hak18} have also been
found to contain only small numbers of structured pulses, and many of these
sub-structures contain temporal symmetries
(u-shaped, twin-peaked, etc. - similar
to what has been found here). The light curves of short GRBs and long GRBs
seem to differ primarily by temporal scale and hardness, but are similar
in many other ways.

The existence of time-reversed pulse structure leads us to believe
that physical models of GRB pulses must contain strong physical symmetries
and an interaction with a single impactor. We have
explored a number of simple kinematic models, and find that 
either the distribution of impacted material in a GRB jet must be
bilateral-symmetrically distributed and impacted by a single impactor,
a physical phenomenon is responsible for reversing the course
of a single impactor, or a single impactor creates emission in
bilateral-symmetrically distributed material as it passes through it. 

The literature contains many descriptions of
how structure might form in GRB light curves. This is often described as ``variability''
(although this term convolves variability due to multiple pulses with 
variability due to structure). The models explain variability
in prompt emission as being due to
internal shocks
({\em e.g.,} \cite{pac94,ree94,kob97,dai98,bos09}),
relativistic turbulence
({\em e.g.,} \cite{kum09,laz09,nar09}),
mini-jets
({\em e.g.,} \cite{lyu03,yam04,zha14}),
fluctuations in black-hole hyper-accretion disks
({\em e.g.,} \cite{lin16}) or fluctuations in 
accretion disk magnetic fields
\citep{llo16}.
None of these models predicts time-symmetric and 
stretched temporal structures, and few associate variability
specifically with spectral re-hardening. Thus, 
the observations reported here provide new and 
potentially useful constraints on GRB central engines
and jet structure.

Application of Occam's Razor leads us to believe that 
the time-reversed and stretched residual structure is 
more appropriately described by a single monotonic pulse 
overlaid by a time-reversible and stretched residual wave structure
than it is by a large number of appropriately-distributed monotonic sub-pulses.
From this we conclude that the radiation mechanism responsible for
producing GRB pulses is not solely due to emission, but is instead due
to some combination of emission and photon removal along the
line-of-sight.


Two of the six GRBs in our study exhibit multiple pulses. 
Some GRB pulses overlap temporally (see Figure \ref{fig:148});
this implies either that distributions of impacted material 
can lie close together spatially or that multiple impactors can interact
with a single distribution of material on timescales shorter than the 
interaction times.

The results derived here have been obtained from a small sample of bright GRB pulses.
High $S/N$ and good time resolution seem to be necessary instrumental requirements
for identifying  and extracting time-reversible and stretched features. 
As many as 100 BATSE pulses with 64-ms resolution and having $S/N > 25$ might be characterized by
complex light curves potentially capable of exhibiting these features. We are 
currently working to increase the size of our sample by studying more bright
BATSE bursts.
GRBs observed by other experiments may also exhibit these features,
although our potential success in observing them depends on their instrumental properties. 
For example, Fermi's GBM has a smaller surface area than BATSE
but has good $S/N$ and excellent time resolution. Swift's BAT has a similar surface area 
to BATSE and excellent $S/N$, but operates in a noisier, lower-energy regime.
We recognize that not all bright GRB pulses may exhibit these behaviors, and
thus make no claims about how generic our results are. However,
the consistency of our results suggests that we are one step
closer to understanding the physics of cosmic gamma-ray bursts.

\acknowledgments

We express our great appreciation
to Narayanan Kuthirummal, whose patience and courteous
attention to discussions of this issue have helped it 
evolve to it present state, and to our referee,
whose detailed recommendations greatly improved
the quality of the final manuscript.
We also acknowledge discussions during the
development of this project with Rebecca Brnich, Bailey Williamson,
and Drew Ayers. As always, we appreciate helpful
comments from Rob Preece. This project has been supported in part
by NASA SC Space Grant NNX15AL49H.

\end{document}